\documentclass[11pt]{article}

\usepackage[margin=1in]{geometry}
\usepackage{pstricks,pst-node,pst-text}
\usepackage{amsthm}
\usepackage{amssymb}
\usepackage{amsmath}
\usepackage{amsfonts}
\usepackage{enumerate}
\usepackage[english]{babel}
\usepackage{graphicx}
\usepackage[cp1250]{inputenc}
\usepackage[round, authoryear]{natbib}
\usepackage{float}
\usepackage{chngcntr}
\usepackage{dsfont}
\usepackage{geometry}
\usepackage[title]{appendix}
\usepackage{multicol}
\usepackage{enumitem}
\usepackage{multirow}
\usepackage{booktabs}
\usepackage{setspace}
\usepackage{makecell}
\usepackage{tabularx}
\usepackage[colorlinks=false, urlbordercolor=green, linkbordercolor=green]{hyperref}
\usepackage{cleveref}

\counterwithout{equation}{section}
\counterwithout{figure}{section} 
\counterwithout{table}{section}

\newcommand{\beqo}{\begin{eqnarray*}}
\newcommand{\eeqo}{\end{eqnarray*}\noindent}
\newcommand{\beq}{\begin{eqnarray}}
\newcommand{\eeq}{\end{eqnarray}\noindent}

\newtheorem{thm}{Theorem}
\newtheorem{lem}{Lemma}
\newtheorem{rem}{Remark}
\newtheorem{cor}{Corollary}
\newtheorem{df}{Definition}
\newtheorem{prop}{Proposition}

\def\be{{\mathbb{E}}}

\def\bp{{\mathbb{P}}}

\def\br{{\mathbb{R}}}

\def\R{{\mathbb R}}  

\begin{document}

\author{{\L}ukasz Delong\footnote{University of Warsaw, Department of Statistics and Econometrics,
l.delong@uw.edu.pl} \and
Mario W\"{u}thrich\footnote{ETH Z\"{u}rich, Department of Mathematics, mario.wuethrich@math.ethz.ch}}

\date{\today}

\title{Measures of predictive accuracy, miscalibration and discrimination}

\maketitle

\begin{abstract}
\noindent
We study the evaluation of real-valued point predictors under the decision-theoretic framework of mean-consistent loss functions given by the Bregman divergences. We first derive a new version of Murphy's decomposition of the expected loss which does not directly include the response itself but only its predictors. We then relate the miscalibration and the discrimination component of Murphy's decomposition to Lorenz-curve-based accuracy measures that are widely used in practice. Besides the usual area between the concentration and Lorenz curves, ABC, we introduce a mean-squared version ABC$^2$ that mitigates some of the weaknesses of the original ABC in identifying mean-calibration. More importantly, both ABC and ABC$^2$ are shown to rely on predictor-dependent weights, so they fail to align with the class of mean-consistent scoring functions. In the same spirit, we derive a similar result for the widely used Gini score. These results indicate that ABC, ABC$^2$ and Gini scores may lead to misleading evaluation of point predictions when used for model selection; this gives support to use mean-consistent loss functions as well as the miscalibration and the discrimination measures from Murphy's decomposition of the expected loss for model evaluation. Finally, we study forecast dominance when Lorenz curves intersect. We show that Lorenz and Murphy's curves have the same number of crossings and, in the one-crossing case, we establish weaker dominance criteria for subclasses of Bregman divergences through third-degree stochastic dominance.

\medskip


\end{abstract}

\section{Introduction}
In his pioneering work, 
\cite{Gneiting} initiated a discussion about the specific choice of the loss function and its role in evaluating point predictions. Assume there are $K$ competing point predictions $(x^{(k)}_1,\ldots, x^{(k)}_n)_{k=1}^K$ for forecasting the observations $y_1,\ldots, y_n$. Select a loss function $(y,x)\mapsto L(y,x)$ that measures the discrepancy between the prediction $x$ and the outcome $y$ of the response. To evaluate the point predictions, consider the following performance criterion
\beq\label{empirical_score}
\bar{S}_k = \frac{1}{n}\sum_{i=1}^nL\big(y_i,x^{(k)}_i\big),\qquad \text{ for $ k=1,\ldots, K$,}
\eeq 
and compare $\bar{S}_1,\ldots, \bar{S}_K$ to select the best prediction, i.e., the one with the smallest average loss \eqref{empirical_score}. The key question is what loss function $L$ should be selected and which specific choices have theoretical support. \cite{Gneiting} develops a decision-theoretic approach to the evaluation of point predictions and concludes that one should always choose so-called \emph{consistent loss functions}. The idea promoted by \cite{Gneiting} is that one should first decide on a functional of the predictive distribution in which one is interested in -- such as the mean or a quantile -- and then select a loss function $L$ that is \emph{consistent} with this functional choice. It turns out that the class of loss functions consistent for a certain functional is identical to the class of loss functions under which the functional is the optimal point forecast. Hence, if a forecaster chooses the best prediction by minimizing the empirical score \eqref{empirical_score}, they have to choose a loss function under which the functional of the predictive distribution under consideration is the optimal point forecast. Other choices of loss functions may lead to misguided conclusions. The notion of consistent loss functions can be extended to \emph{proper scoring rules}, see \cite{GneitingRaftery}.

In this paper we are interested in the estimation of mean values of observations. Hence, by evaluating point predictions we always think of evaluating mean estimates. The class of loss functions consistent for mean values has been characterized by \cite{Savage}, \cite{Gneiting} and \cite{ehm}, and it includes so-called \emph{Bregman divergences}; in fact, under mild conditions this is an if-and-only-if statement. The most popular examples include the mean squared loss, Kullback--Leibler divergences and loss functions related to deviance functions in the exponential dispersion family. 

The main criterion for evaluating point predictions is prediction accuracy characterized by closeness between predictions and observations measured with a consistent loss function. However, good predictions should also have other desirable properties. \cite{murphy} and \cite{MurphyWinkler} investigate decompositions of loss functions. \cite{pohle_decomp} proves that any consistent loss function can be decomposed into measures of \emph{miscalibration} and \emph{discrimination (resolution)}. This decomposition, called Murphy's decomposition, allows one to look deeper into predictive accuracy of predictors. Murphy's decomposition can identify strengths and weaknesses of predictions and guide forecasters to improve their predictive models. The measures of miscalibration and discrimination from Murphy's decomposition are in agreement with the decision-theoretic approach to the evaluation of point predictions from \cite{Gneiting} since they are related to consistent loss functions, see \cite{GneitingResin}.

The first contribution of this paper is to derive a new version of Murphy's decomposition. In this new version, the measures of miscalibration and discrimination only depend on the predictor, and not on the responses. The main benefit is that this provides one with a more intuitive interpretation of theses measures. One can now directly measure, with a consistent loss function, how close the predictor is to its calibrated version (the miscalibration statistics) and how far the predictor is from a constant prediction (the discrimination statistics), and this does not involve the responses which, in particular, makes it robust towards different tail properties of the responses. 

As discussed above, the loss functions consistent for mean values can be decomposed to get the measures of miscalibration and discrimination of mean estimates. At the same time, there are other measures of miscalibration and discrimination which have gained popularity in practical applications. For actuarial applications, \cite{DenuitSznajderTrufin}, \cite{DenuitBook2020}, \cite{DenuitTrufinTweedie} introduce and promote the area between the Lorenz and the concentration curve (the so-called ABC statistics) to measure the miscalibration of predictions. Recently, \cite{DenuitTrufinABC} investigate properties and failures of this measure. Yet, we believe that a deeper investigation of the ABC measure is still missing in the literature. In particular, \cite{DenuitSznajderTrufin} remark that there is a connection between the ABC and the miscalibration statistics linked with consistent loss functions. However, the relation between the ABC measure and the miscalibration statistics from Murphy's decomposition is not fully understood yet. The second contribution of this paper is to develop this relation by deriving a new representation of the ABC measure. This new result presented in this paper allows us to assess if the ABC measure fits the framework of the decision-theoretic approach to the evaluation of point predictions from \cite{Gneiting}. In addition to the ABC measure, we also introduce a new miscalibration statistics, the area between the Lorenz and the concentration curve measured in a mean squared sense, and study its properties. We also prove a new representation of this measure in the spirit of the new representation for the ${\rm ABC}$.

As far as measures of discrimination are concerned, the Gini index, introduced by \cite{Gini}, is the most popular statistics used in applications, also in actuarial applications. \cite{DenuitSznajderTrufin}, \cite{DenuitBook2020} and \cite{DenuitTrufinTweedie} use the classical Gini index based on the Lorenz curve to evaluate predictors. \cite{wuthrich_gini_2} investigate a machine learning version of the Gini index in an actuarial context where the treatment of case weights may pose some challenges. \cite{Frees_1} and \cite{Frees_2} promote a version of the Gini index based on a so-called ordered Lorenz curve. These papers mainly focus on the advantages of using the Gini index in evaluating the discriminatory power of predictors, however, the Gini index also faces serious disadvantages that should be emphasized. The Gini index has been well-studied in the statistical literature, see the main reference by \cite{Gini_book}. At the same time, the link between the Gini index and the discrimination statistics from Murphy's decomposition is not complete to the best of our knowledge. \cite{hand} and \cite{hand_2} have made a first step in this area of research and considered the misclassification error for binomial responses and distributions of misclassification costs. These authors conclude that the loss function, implied by the Gini index and the Area Under the Curve (AUC), which measures the closeness between predictions and a constant mean value, depends on the predictor. They conclude that: "the AUC is equivalent to averaging the misclassification loss over a cost ratio distribution which depends on the score distributions. Since the score distributions depend on the classifier, this means that, when evaluating classifier performance, the AUC evaluates a classifier using a metric which depends on the classifier itself. That is, the AUC evaluates different classifiers using different metrics. It is in that sense that the AUC is an incoherent measure of classifier performance". Henceforth, this statement indicates that the Gini index does not fit the decision-theoretic approach to the evaluation of point predictions from \cite{Gneiting}. Our third contribution is that we explore the ideas from \cite{hand} and \cite{hand_2} and we investigate real-valued predictors. We prove a new representation of the Gini index and relate the Gini index with the discrimination statistics from Murphy's decomposition. With this representation we confirm the findings from \cite{hand} and \cite{hand_2} in a more general setting. 

In both cases of the ABC statistics and the Gini index, we show that these measures of miscalibration and discrimination cannot be reconciled with the framework from \cite{Gneiting}. We believe that our results are of theoretical interest, but they also have practical implications. In practice, we rank predictors and select the best one for prediction. We characterize some special cases when the ranking of predictors based on the ABC measure/the Gini index and the miscalibration/the discrimination statistics from Murphy's decomposition coincide. Yet, in most practical cases the measures would yield different rankings of predictors in terms of their miscalibration and discrimination statistics. Specific examples are provided in the paper. Given that the ABC measure and the Gini index are not the scores in the sense of \eqref{empirical_score} and \cite{Gneiting}, we favor the miscalibration and discrimination statistics from Murphy's decomposition. As a by-product of our results, we also show that minimizing expected weighted Bregman divergences with weights depending on the predictor may lead to misleading point predictions, which complements the results from \cite{GneitingRanjan} and \cite{taagart} where weights depending on the responses are considered. 

Finally, we investigate \emph{forecast dominance} between predictors in the spirit of \cite{kruger_ziegel} and \cite{DenuitTrufinTweedie}, as well as other equivalent formulations such as \emph{Murphy's dominance} with \emph{Murphy's curves} developed by \cite{ehm}. Forecast dominance refers to a situation where one predictor outperforms all other predictors in terms of the performance criterion \eqref{empirical_score} for all relevant loss functions. In the context of mean estimates, \cite{kruger_ziegel} show that a mean-calibrated $X_1$ dominates a mean-calibrated $X_2$ for all Bregman divergences if the Lorenz curves of the predictors do not cross. Non-intersecting Lorenz curves are very rare in practice. We generalize the result of \cite{tryptych} from classification to regression and we show that the number of crossing points of the Lorenz curves is equal to the number of crossing points of Murphy's curves of two mean-calibrated predictors. Most importantly, the fourth contribution of this paper is to establish  forecast dominance between two mean-calibrated predictors for a subclass of Bregman divergences in the case when the Lorenz curves of the predictors cross once. In particular, we extend the results from \cite{role_of_variance}, \cite{muliere}, \cite{zoli}, \cite{dominance_book} which link \emph{third-degree stochastic dominance} for distributions with ranking of distributions in terms of expected utility, and we apply our extension in the context of forecast dominance. We also introduce a weaker form of local forecast dominance with Murphy's curves between predictors with Lorenz curves that cross once, inspired by the literature on wealth inequalities, see \cite{second_lorenz} and \cite{second_lorenz_2}.

This paper is structured as follows. In Sections \ref{sec_bregman}, \ref{sec_cal_disc} and \ref{sec_lorenz}, we recall and extend key concepts needed to evaluate point predictions in our setting. We prove a new version of Murphy's decomposition in Section \ref{sec_murphy}. In Section \ref{sec_cal}, we study miscalibrations measures, and in Section \ref{sec_disc}, we investigate discrimination measures. Section \ref{sec_disc_intersect} focuses on forecast dominance between predictors with Lorenz curves that cross once. Our results are illustrated with examples throughout the paper. Section \ref{conclusions section} concludes. A real data set in studied in Appendix \ref{app_real}. All proofs are presented in Appendix \ref{app_proofs}.

\section{Bregman divergence, Bregman and Murphy's dominance}\label{sec_bregman}

We start by presenting the statistical set-up, as well as introducing the main notation and definitions.
Let $Y$ denote a real-valued response variable with finite mean value $\be[|Y|]<\infty$. The goal is to predict $Y$ by estimating its mean value using a real-valued predictor $X$. We also assume that $X$ has finite mean value $\be[|X|]<\infty$. In a decision-theoretical setting, see, e.g., Section 3.1 in \cite{ehm}, we consider a random pair $(Y,X)$ defined on a so-called point prediction space $\big(\Omega,\mathcal{A}_0,\bp\big)$, where $\bp$ describes the joint distribution of $(Y,X)$. The predictor $X$ represents a point prediction which utilizes the information set (the forecaster's basis) $\mathcal{A}\subset \mathcal{A}_0$, hence, $X$ is assumed to be $\mathcal{A}$-measurable. A pair $(Y,X)$ is usually interpreted as a random observation from a test set. We denote the marginal distributions $Y\sim F_Y$ and $X\sim F_X$. We assume:
\begin{itemize}
\item[(A)] $X$ is globally unbiased for $Y$, meaning that $\be[Y]=\be[X]<\infty$;
\item[(B)] The distributions $F_X$ and $F_Y$ are supported on $[0,\infty)$;
\item[(C)] The distribution $F_X$ is continuous, strictly increasing on $[0,\infty)$ and $F_X(0)=0$.
\end{itemize} 
Select a loss function $(y,x)\mapsto L(y,x)$ that measures the discrepancy between a prediction value $x$ of $X$ and an outcome $y$ of $Y$. The loss function $L$ allows one to measure the accuracy of predictor $X$ for predicting $Y$ by the expected loss (subject to existence)
\begin{equation}\label{score}
S_L(Y,X) = \be\big[L(Y,X)\big],
\end{equation}
also called the score. We only consider loss functions $L$ for which the expected loss is finite $\be[|L(Y,X)|]<\infty$. It is standard to assume that $L(y,x)\geq 0$ and $L(y,x)=0$ if $y=x$. Other technical conditions, not necessary for the presentation of our results, can be found in \cite{Gneiting}. The score \eqref{empirical_score} is the empirical version of the expected loss \eqref{score}. In this paper we only consider loss functions consistent for mean estimation; see \cite{Savage}, \cite{Gneiting}, \cite{ehm} and \cite{GneitingResin}. This choice of loss functions is justified by the following definition.

\begin{df}\label{def_consistent_loss}
A loss function $L$ is consistent for the estimation of the mean value of $Y$ if
\beq\label{consistent_loss}
\be\Big[L\big(Y,\be[Y|\mathcal{A}]\big)\Big|\mathcal{A}\Big] \leq \be\Big[L\big(Y,X\big)\Big|\mathcal{A}\Big], \qquad a.s.,
\eeq
for all $\mathcal{A}$-measurable predictors $X$, supposed that all expected values in \eqref{consistent_loss} are finite. The loss function $L$ is strictly consistent for the estimation of the mean value of $Y$ if it is consistent and if the equality in \eqref{consistent_loss} holds if and only if $X=\be[Y|\mathcal{A}]$, a.s.
\end{df}

\begin{rem}[conditional mean estimation]\label{conditional_mean_estimation} \normalfont
In Definition \ref{def_consistent_loss} we consider loss functions consistent for the estimation of the conditional mean value of $Y$, given $\mathcal{A}$. E.g., let $X=h(\mathbf{Z})$, with $h:\br^d\to\br$ denoting a measurable regression function, $\mathbf{Z}:\Omega\to \br ^d$ be a vector of covariates, and the sub-$\sigma$-field $\mathcal{A}=\sigma(\mathbf{Z})$ denotes the information set generated by the covariates $\mathbf{Z}$. Then, finding the optimal ${\cal A}$-measurable predictor $X$ is equivalent in trying to find the optimal regression function $h$ on $\mathbf{Z}$.
In the sequel, we do not explicitly state the regression problem and
the information set $\mathcal{A}$ generated by $\mathbf{Z}$, but one should keep this in mind as the underlying motivation of the problem studied.
\end{rem}

In view of Definition \ref{def_consistent_loss}, using a strictly consistent loss function $L$ for mean estimation in the expected loss \eqref{score} allows us to score predictors $X_1$ and $X_2$, and we give preference to the one with the lower expected loss (lower score). The minimizer of this expected loss gives us the true (conditional) mean value of $Y$. This turns mean estimation into a minimization problem. This is particularly useful in fitting mean values with statistical methods and machine learning methods, i.e., by empirically minimizing strictly consistent loss functions for mean estimation.

Following \cite{Savage} and \cite{Gneiting}, the consistent loss functions for mean estimation are given by Bregman divergences; under mild conditions this is an if-and-only-if statement; see \cite[Theorem 7]{Gneiting}. There are two different characterizations of Bregman divergences. First, Bregman divergences are characterized by the representation
\beq\label{bregman_representation_2}
L_\phi(y,x)=\phi(y)-\phi(x)-\phi'(x)(y-x),
\eeq
for a convex function $\phi$ with (sub-)derivative $\phi'$. For a strictly convex function $\phi$, we obtain a strictly consistent loss function for mean estimation, supposed that $\be[\phi(Y)]$ is finite; see \cite[Theorem 7]{Gneiting}. Examples of strictly convex choices are given by deviance losses coming from the exponential dispersion family (EDF); see \cite[Section 2.3]{WM2023}. As a result of strict convexity of $\phi$, we have $L_\phi(x,y)=0$ if and only if $x=y$. Furthermore, from representation \eqref{bregman_representation_2} we can conclude that $t\mapsto L_\phi(x+t,x)$ is convex in $t$ with a minimum at $t=0$, and the larger the discrepancy between $y=x+t$ and $x$, the larger the Bregman divergence between $y$ and $x$. Second, Bregman divergences are characterized by the representation
\beq\label{bregman_representation}
L_H(y,x)=\int_{\R} L_\theta(y,x)\,dH(\theta),    
\eeq
for a non-negative measure $H$ on $\br$, and $L_\theta$ is an {\it elementary Bregman divergence}, for $\theta \in \R$, defined by
\begin{equation}\label{elementary Bregman loss}
L_\theta(y,x)=(y-\theta)^+-(x-\theta)^+-\mathbf{1}\{x>\theta\}(y-x),
\end{equation}
which itself is a Bregman divergence, we refer to \cite[Theorem 1b]{ehm}. Since $X$ and $Y$ are supported on $[0,\infty)$, the measure $H$, which characterizes the Bregman divergence $L_H$, is chosen also supported on $[0,\infty)$. W.l.o.g.~we can also assume that $H(0)=0$ by assumptions (B)-(C). Indeed, if $H$ had a jump at zero, then the expected loss $\mathbb{E}[L_H(Y,X)]$ would include a term of the form $\mathbb{E}\big[(Y - X)\mathbf{1}\{X = 0\}\big]$, arising from $L_{\theta=0}(y,x)$, but $(Y-X)\mathbf{1}\{X=0\}=0, \ a.s.$. 

Furthermore, following \cite{ehm}, we introduce  Murphy's curve (Murphy's diagram). 

\begin{df}\label{def Murphy curve}
Choose $\theta\in\br$. Let $M_\theta(Y,X)=\be[L_\theta(Y,X)]$ denote the expected loss \eqref{score} w.r.t.~the elementary Bregman divergence \eqref{elementary Bregman loss}. The function
\begin{equation}\label{definition Murphy curve}
\theta\mapsto M_\theta(Y,X),\quad \theta\in\br.
\end{equation}
is called {\it Murphy's curve}. Under assumption (A) the Murphy's curve is well-defined. 
\end{df}

Using Fubini's theorem, the score and Murphy's curve satisfy the relation
\beq\label{m_s}
S_{L_H}(Y,X)=\be\big[L_H(Y,X)\big]=\int_\br M_\theta(Y,X)\,dH(\theta).
\eeq
This motivates the following result which appears as Corollary 1b in \cite{ehm}; the proof is straightforward
from \eqref{m_s}.
\begin{prop}\label{prop_murphy_dominance}
For two predictors $X_1$ and $X_2$, it holds that
\beq\label{murphy_dominance}
M_\theta(Y,X_1)\leq M_\theta(Y,X_2) \quad \text{for all $\theta \in\br$} \quad \Longleftrightarrow \quad S_{L_H}(Y,X_1)\leq S_{L_H}(Y,X_2) \quad \text{for all $L_H$},
\eeq
where for all $L_H$ means for all Bregman divergences.
\end{prop}

The property on the right-hand side of \eqref{murphy_dominance} is called {\it forecast dominance}; see \cite[Definition 2.1]{kruger_ziegel} and \cite[Definition 2b]{ehm}; or {\it Bregman dominance}; see \cite{DenuitTrufin2025}. The property on the left-hand side of \eqref{murphy_dominance} is called {\it Murphy's dominance} in the sequel.

In general, the ranking of predictors based on the expected loss \eqref{score} and Definition \ref{def_consistent_loss} depends on the specific choice of the Bregman divergence $L_\phi$ and $L_H$, respectively. Ideally, we would like to choose an optimal predictor $X$ which is preferred under any Bregman divergence. The main value of Proposition \ref{prop_murphy_dominance} lies in the fact that  Murphy's dominance for the elementary Bregman divergence is easier to verify than the forecast dominance for any Bregman divergence. 

Since $X$ and $Y$ are supported on $[0,\infty)$ and the measure $H$ which characterizes the Bregman divergence $L_H$ is supported on $[0,\infty)$, we can consider $\theta\geq 0$ in Definition \ref{def Murphy curve} and Proposition \ref{prop_murphy_dominance}.

\section{The notions of miscalibration and discrimination}\label{sec_cal_disc}

The main criterion for evaluating mean estimates is prediction accuracy characterized by closeness between predictions and observations measured by the score \eqref{score} and its empirical counterpart. However, good predictions should also have other desirable properties. In particular, they should be mean-calibrated and have high discriminatory power. We use the following definitions.

\begin{df}\label{def_mean_calibration}~
\begin{itemize}
\item[(i)] The predictor $X$ is mean-calibrated for response variable $Y$ if $\be[Y|X]=X$, a.s.
\item[(ii)] The predictor $X$ does not discriminate response variable $Y$ if $X=\be[Y]$.
\end{itemize}
\end{df}

Intuitively, calibration of $X$ is measured by the closeness between $X$ and $\be[Y|X]$ -- the smaller the discrepancy, the better. The discrimination power of $X$ is measured by the closeness between $X$ and $\be[Y]$ -- the larger the discrepancy, the better. One possibility is to use the score \eqref{score} and Bregman divergences to measure these discrepancies, this choice is justified in the next section. Other choices are also possible, and investigated in the sequel.

From a practical point of view, it might be challenging to estimate $\be[Y|X]$. \cite{corp} and \cite{tryptych} propose isotonic regression to estimate the conditional mean of the response given the prediction and promote the so-called CORP (Consistent, Optimally binned, Reproducible and PAV Algorithm based) reliability diagram as a tool for the graphical assessment of the calibration of a predictor. The CORP reliability diagram visualizes the calibration property of mean estimates by plotting estimates of $\be[Y|X_i]$ against the predictions $X_i$ for all instances $i$ in a data set. In line with the interpretation above, large deviations of the CORP reliability diagram from the diagonal suggest a lack of calibration. We use CORP reliability diagrams in Appendix \ref{app_real}.

We complete the results from the previous section for mean-calibrated predictors. Under the assumption that $X$ is mean-calibrated for $Y$, the expected loss w.r.t.~the elementary Bregman divergence simplifies to
\beq\label{score_calibrated}
M_\theta(Y,X)&=&\be\Big[\be\Big[(Y-\theta)^+-(X-\theta)^+-\mathbf{1}\{X>\theta\}(Y-X)\big|X\Big]\Big]\nonumber\\
&=&\be\big[(Y-\theta)^+-(X-\theta)^+\big],\qquad \theta\in\br,
\eeq
where the last term on the first line vanishes by the mean-calibration property of $X$ for $Y$. We use \eqref{score_calibrated} in the sequel when we focus on mean-calibrated predictors.

\begin{rem}[stop-loss order]\label{stop-loss order} \normalfont
Assume we have two mean-calibrated predictors $X_1$ and $X_2$ supported on $[0,\infty)$. By \eqref{score_calibrated} we have the following equivalence to Murphy's dominance (under mean-calibration)
\beq\label{murphy_dominance 1}
M_\theta(Y,X_1)\leq M_\theta(Y,X_2)  \text{ for all $\theta \geq 0$} ~ \Longleftrightarrow ~
\be\big[(X_2-\theta)^+\big] \leq \be \big[(X_1-\theta)^+\big] \text{ for all $\theta \geq 0$}.
\eeq
The right-hand side gives us the {\it stop-loss order} $X_2 \preceq_{sl} X_1$ for the two predictors $X_1$ and $X_2$. Furthermore, because $\be[X_1]=\be[X_2]$ and our predictors are positively supported, by assumptions (A)-(B), the right-hand side means that $X_1$ dominates $X_2$ in {\it convex order}, denoted by $X_2\preceq _{cx} X_1$; see \cite[Theorem 3.6]{Muller}. 
\end{rem}

\section{Murphy's decomposition}\label{sec_murphy}

In this section we prove a new decomposition of expected losses w.r.t.~consistent loss functions. The new decomposition of the score \eqref{score} is interesting in its own, and is used in the next sections to prove the main results concerning the discrimination and the miscalibrations statistics.

Referring to \cite{pohle_decomp}, we consider the following decomposition of the 
expected loss \eqref{score} w.r.t.~a selected Bregman divergence, known as {\it Murphy's decomposition} for consistent loss functions $L$ for mean estimaton.
Murphy's decomposition splits the expected loss $S_L$ into the uncertainty (UNC), discrimination (DSC) and miscalibration (MCB) terms as follows
\begin{equation}\label{Murphy score_1}
S_L(Y,X)={\rm UNC}_L(Y)-{\rm DSC}_L(Y,X)+{\rm MCB}_L(Y,X),
\end{equation}
where the three terms are defined by
\begin{eqnarray*}
{\rm UNC}_L(Y)
&=& S_L(Y,\be[Y]),  \\
{\rm DSC}_L(Y,X)
&=& S_L(Y,\be[Y])-S_L(Y, \be[Y|X])~\ge ~0,\\
{\rm MCB}_L(Y,X)&=& S_L(Y,X)-S_L(Y,\be[Y|X])~\ge ~0.
\end{eqnarray*}
The later two inequalities on the discrimination and the miscalibration terms follow since $L$ is a consistent loss function for mean estimation, with equalities if $\be[Y|X]=\be[Y]$ and $\be[Y|X]=X$, respectively, see also Definition \ref{def_mean_calibration}. If the loss function $L$ is strictly consistent for mean estimation, then ${\rm DSC}_L(Y,X)=0$ iff $\be[Y|X]=\be[Y]$, a.s., and ${\rm MCB}_L(Y,X)=0$ iff $\be[Y|X]=X$, a.s.; see \cite[Theorem 2.23]{GneitingResin}.

Murphy's decomposition has been vividly used in meteorology for many years as a forecast evaluation method. It has recently also spread to other disciplines like psychology, statistics and actuarial science.
In the proposition below we prove a new version of Murphy's decomposition, which is our first main result.
\begin{thm}\label{prop_murphy_new_decomp}
Let $S_L$ denote the expected loss \eqref{score} w.r.t.~a Bregman divergence $L$ and compute Murphy's decomposition \eqref{Murphy score_1}. The discrimination and the miscalibration statistics are given by
\begin{eqnarray}
\label{discrimination X}
{\rm DSC}_L(Y,X)
&=& S_L(\be[Y|X],\be[Y])~\ge ~0,\\
{\rm MCB}_L(Y,X)&=& S_L(\be[Y|X],X)~\ge ~0.
\label{miscalibration X}
\end{eqnarray}
\end{thm}

As a consequence, we also have the decomposition in terms of Murphy's curves.

\begin{cor}
Let $\theta\mapsto M_\theta(Y,X)$ denote Murphy's curve defined by \eqref{m_s} as the expected loss \eqref{score} w.r.t.~the elementary Bregman divergence. We have the decomposition
\beq\label{murphy_sciore_3}
M_\theta(Y,X)=M_\theta(Y,\be[Y])-M_\theta(\be[Y|X],\be[Y])+M_\theta(\be[Y|X],X),\qquad \text{for all $\theta\in\br$}.
\eeq
\end{cor}

For the square loss $L(y,x)=(y-x)^2$, our decomposition yields the well-known decomposition of the mean squared loss (MSE)
\beqo
\be\big[(Y-X)^2\big]={\rm Var}[Y]-{\rm Var}\big[\be[Y|X]\big]+\be\big[\big(\be[Y|X]-X\big)^2\big];
\eeqo
see \cite{ehm}, and Theorem  \ref{prop_murphy_new_decomp} says that this decomposition generalizes to any Bregman divergence.

The first interest in this new formulation of the discrimination and the miscalibration terms is
that the explicit realization of the response $Y$ does no longer appear in the formulas, i.e., all involved terms are (directly) measurable w.r.t.~the predictor $X$.
This may suggest to replace the score \eqref{score} by a new score that measures the improvement of using the predictor $X$ over the (deterministic) mean predictor $\be[Y]$, that is,
\begin{equation}\label{Murphy score 2}
{\rm UNC}_L(Y,X)-S_L(Y,X)={\rm DSC}_L(Y,X)-{\rm MCB}_L(Y,X),
\end{equation}
where the right-hand side only involves the $\sigma(X)$-measurable terms
\eqref{discrimination X}-\eqref{miscalibration X}. The bigger this score, the more associated is the predictor $X$ and the response $Y$, hence, the more accurate the prediction will be. Thus, we aim to make the expected loss \eqref{score} small and the score \eqref{Murphy score 2} large, respectively, for receiving good predictors $X$ for predicting response $Y$.
Interestingly, the specific law of $Y$ does {\it not} appear on the right-hand side of \eqref{Murphy score 2} when using Theorem \ref{prop_murphy_new_decomp}, but only the conditional expectation $\be[Y|X]$. This may matter if we work with finite (empirical) samples, and it is related to the question of an optimal choice of the Bregman divergence $L$. Typically, such an optimal choice is done w.r.t.~the properties of the response $Y$; see, e.g., \cite{GourierouxMontfortTrogon}. However, if we validate predictors on the level of Theorem \ref{prop_murphy_new_decomp}, the responses $Y$ do not enter the consideration at all but only their conditional expectations. This may give raise to consider different necessary properties of $L$ for finite sample validation. 

The second conclusion from Theorem \ref{prop_murphy_new_decomp} is that once we decide on a Bregman divergence $L$ for the evaluation of point predictions, we shall quantify the measures
\beqo
S_L(Y,X),\quad S_L(\be[Y|X],\be[Y]),\quad S_L(\be[Y|X],X),
\eeqo
and rank the predictors based on the above values. In particular, the latter two $S_L(\be[Y|X],\be[Y])$ and $S_L(\be[Y|X],X)$ are intuitive measures of discrimination and miscalibration, respectively. We can use the score \eqref{score} with Bregman divergences to measure the discrepancies between $\be[Y|X]$ and $\be[Y]$, and $\be[Y|X]$ and $X$, for discrimination and micalibration, and turn these discrepancies into rankings. A predictor $X$ with a smaller $S_L(\be[Y|X],\be[Y])$ has a lower discriminatory power, in the sense that $X$ is closer to $\be[Y]$, and a predictor $X$ with a smaller $S_L(\be[Y|X],X)$ has a better calibration, in the sense that $\be[Y|X]$ is closer to $X$. The discrimination and the miscalibration measures from Murphy's decomposition have thus a clear interpretation.

\section{Lorenz curve and concentration curve}\label{sec_lorenz}

We have introduced the miscalibration and the discrimination statistics from Murphy's decomposition. We now introduce other measures of miscalibration and discrimination which have been considered in the literature. We study the area between the Lorenz and the concentration curve and the Gini index. The Gini index is already well-established in many fields of science, whereas the former one is gaining fast popularity in the field of actuarial science. The results presented in this section are used in the next two sections, where we relate the area between the Lorenz and the concentration curve to the miscalibration statistics and the Gini index to the discrimination statistics from Murphy's decomposition.

For $p \in [0,1]$, the Lorenz curve and the concentration curve are given, respectively, by
\beqo
LC_p(X)&=&\frac{\be\big[X\mathbf{1}\{X\leq F_X^{-1}(p)\}\big]}{\be[X]},\\
CC_p(Y,X)&=&\frac{\be\big[Y\mathbf{1}\{X\leq F_X^{-1}(p)\}\big]}{\be[Y]}=\frac{\be\big[\be[Y|X]\mathbf{1}\{X\leq F_X^{-1}(p)\}\big]}{\be[Y]}.
\eeqo
The Lorenz and concentration curves are well-defined under assumptions (A)-(B). If $X$ denotes an estimated insurance premium and $\be[Y|X]$ is the true premium, then $LC_p(X)$ represents the proportion of the total estimated
premium income corresponding to the sub-portfolio with the $100p\%$ of contracts with the smallest estimated premium, and $CC_p(Y,X)$ represents the proportion of the total true
premium income corresponding to the sub-portfolio with the $100p\%$ of contracts with the smallest estimated premium. We refer to \cite[Chapter 5]{Gini_book} for mathematical results on the Lorenz and the concentration curve. These curves have been investigated and promoted, e.g., by \cite{Frees_1, Frees_2},\cite{DenuitSznajderTrufin}, \cite{DenuitBook2020}, \cite{DenuitTrufinTweedie}, \cite{Denuitautocalibration}, \cite{wuthrich_gini}, \cite{wuthrich_gini_2} in the context of insurance pricing. These papers study two important measures for the evaluation of point predictions. The first one is the \emph{area between the curves} (ABC) defined as
\beq\label{ABC_measure}
{\rm ABC}(Y,X) &=& \int_0^1 \big(CC_p(Y,X) - LC_p(X)\big) \ dp \nonumber\\
&=& \frac{{\rm Cov}(Y-X, F_X(X))}{\be[Y]} = \frac{\be\big[(Y-X)F_X(X)\big]}{\be[Y]}.
\eeq
The ABC statistics was introduced by \cite{DenuitSznajderTrufin} as the integrated difference between the two curves. The second equality is proved in \cite[Section 4.2.3]{DenuitSznajderTrufin}, under slightly weaker assumptions than (A)-(C), and the third equality follows from the second one by the global unbiasedness property $\be[X]=\be[Y]$. \cite{DenuitTrufinTweedie} notice that the Lorenz curve and the concentration curve coincide for mean-calibrated predictors, see their Property 3.1. Let us recall a short proof of that remark, using the tower property of conditional expectation and mean-calibration, for any $p\in[0,1]$ we have
\begin{eqnarray}\label{mean-calibration ABC}
\be[Y\mathbf{1}\{X\leq F_X^{-1}(p)\}]&=&\be[\be[Y\mathbf{1}\{X\leq F_X^{-1}(p)\}|X]]\nonumber\\
&=&\be[\be[Y|X]\mathbf{1}\{X\leq F_X^{-1}(p)\}]]=\be[X\mathbf{1}\{X\leq F_X^{-1}(p)\}].
\end{eqnarray}
This motivates to measure the miscalibration of $X$ by the ABC statistics. Clearly, the ${\rm ABC}$ is a different measure of calibration than ${\rm MCB}$. The second measure studied in the above mentioned papers is the Gini index, which measures the dispersion of the distribution of $X$ based on the Lorenz curve
\beq\label{gini}
{\rm Gini}(X) &=&1 - 2 \int_0^1 LC_p(X)\, dp \nonumber\\
&=& \frac{2}{\be[X]}{\rm Cov}\left(X, F_X(X)\right)=\frac{\be[|X-X'|]}{2\be[X]}=\frac{\be[(X-X')^+]}{\be[X]}\nonumber\\
&=&\frac{1}{\be[X]}\int_0^\infty F_X(z)(1-F_X(z))dz,
\eeq
where $X'$ is an independent copy of $X$, see equations (2.1), (2.9), (2.15) and Chapter 2.1.4 in \cite{Gini_book}. The Gini index was introduced by \cite{Gini} and seems to be the most popular measure used for evaluation of point predictions in all disciplines. Let us emphazise that we consider the classical version of the Gini index, as later we use the Gini index only for mean-calibrated predictors. There is a machine learning version of the Gini index, tailored to the evaluation of point predictions. Both versions of the Gini index coincide, up to a constant depending only on the response, for mean-calibrated predictors, see \cite[Corollary 4.2]{wuthrich_gini}. It is also worth pointing out that one can find other definitions of the Gini index in the literature on predictive modelling, e.g., \cite{Frees_1, Frees_2} define the Gini index as twice the area between the so-called ordered Lorenz curve and the 45-degree line. It turns out that the Gini index in the classical version \eqref{gini} may serve as an alternative measure of a discriminatory power of a mean-calibrated predictor. By \cite[Definition 4.1]{DenuitSznajderTrufin}, let $X_1$ and $X_2$ be non-negative with equal expected values, the mean-calibrated $X_1$ is more discriminatory than the mean-calibrated $X_2$, if
\beq\label{lorenz_dominance}
X_2\preceq_{cx} X_1 \quad \Longleftrightarrow \quad LC_p(X_2)\geq LC_p(X_1), \quad \text{for all $p\in[0,1]$.}
\eeq
Assuming (A), $\be[Y]=\be[X]\preceq_{cx} X$ for any $X$. The Lorenz curve with the diagonal line arises for $X=\be[Y]$, hence, we expect that the closer the Lorenz curve of $X$ to the diagonal line, then the smaller the discrepancy between $X$ and $\be[Y]$, and the lower the discriminatory power of $X$. Thus, \eqref{lorenz_dominance} agrees with our definition of discriminatory power given in Section \ref{sec_cal_disc}. The relation between the Lorenz curves presented on the right-hand side of \eqref{lorenz_dominance} is called {\it Lorenz dominance}. It is denoted by $X_2\preceq _{L} X_1$, which says that $X_1$ Lorenz dominates $X_2$;  see \cite[Theorem 3.A.10]{shaked} for the equivalence between the convex order and the Lorenz order. The condition that $LC_p(X_1)\leq LC_p(X_2)$ for all $p$ implies that ${\rm Gini}(X_1)\geq {\rm Gini}(X_2)$, hence this motivates to measure a discriminatory power of $X$ by the Gini index. Clearly, Gini is a different measure of discrimination than ${\rm DSC}$.

In addition to the ABC statistics \eqref{ABC_measure}, we also study another measure of miscalibration related to the ABC, but based on the difference between the Lorenz and the concentration curves in the mean squared sense
\beq\label{ABC_measure_sq}
{\rm ABC}^2 (Y,X) &=& \int_0^1 \big(CC_p(Y,X) - LC_p(X)\big)^2 \ dp.
\eeq
A sum of squared differences between the curves appears in \cite{wuthrich_calibration_tests} in the context of testing mean-calibration of discrete point predictions. To the best of our knowledge, the miscalibration statistics \eqref{ABC_measure_sq} is new and it has not been investigated in the form of \eqref{ABC_measure_sq} in the literature.  In the next section we conclude that ${\rm ABC}^2$ is more intuitive than ${\rm ABC}$ for decision-making about mean-calibration of predictors. First, we prove a representation of ${\rm ABC}^2$ similar to \eqref{ABC_measure}.

\begin{prop}\label{prop_abc_sq}
Let (A)-(C) hold. We have the identities
\beq\label{ABC_measure_sq_rep}
{\rm ABC}^2(Y,X)&=&\frac{1}{\be[Y]^2}\be\Big[\big(Y-X\big)\big(Y'-X'\big)\Big(1-\max\big\{F_X(X),F_X(X')\big\}\Big)\Big]\nonumber\\
&=&\frac{\be\Big[\big(Y-X\big)\mathcal{Q}\Big(F_X(X)\Big)\Big]}{\be[Y]}=\frac{{\rm Cov}\big(Y-X, \mathcal{Q}(F_X(X))\big)}{\be[Y]},
\eeq
where 
\beqo 
z\in[0,1]\mapsto\mathcal{Q}\big(z\big)=\frac{\be\Big[\big(Y'-X'\big)\Big(1-\max\big\{z,F_X(X')\big\}\Big)\Big]}{\be[Y']},
\eeqo
and $(Y',X')$ is an independent copy of $(Y,X)$. Moreover, $z\mapsto\mathcal{Q}\big(z\big)$ is continuous on $[0,1]$ with $\mathcal{Q}(0)=-{\rm ABC}(Y,X)$ and $\mathcal{Q}(1)=0$.
\end{prop}

It is also interesting to investigate the derivative of $\mathcal{Q}$.

\begin{lem}\label{lemma_abc_sq}
Let (A)-(C) hold and let the distribution $F_X$ be absolutely continuous. Consider the ${\rm ABC}^2$ statistics \eqref{ABC_measure_sq} with representation \eqref{ABC_measure_sq_rep}. The function $x\mapsto \mathcal{Q}(x)$ is continuously differentiable on $[0,1]$ and we have
\beqo
\mathcal{Q}'(x)=LC_{x}(X)-CC_{x}(Y,X), \qquad \textit{for all $x\in[0,1]$}.
\eeqo
\end{lem}

From Lemma \ref{lemma_abc_sq} we deduce that the function $x\mapsto\mathcal{Q}(x)$ increases at points $x$ at which $CC_{x}(Y,X)\leq LC_{x}(X)$, and decreases at points $x$ at which $CC_{x}(Y,X)\geq LC_{x}(X)$. The ABC measure \eqref{ABC_measure} is in insurance applications interpreted as the covariance between the profits and the ranks of the estimated premiums; see \cite[Section 6.3.5.4]{DenuitBook2020}. The ${\rm ABC}^2$ measure \eqref{ABC_measure_sq_rep} can now be interpreted as the covariance between the profits and transformed ranks of the estimated premiums. The covariance, in particular its sign and absolute value, between the profits and strictly increasing or strictly decreasing transformations of the ranks of the estimated premiums can still have a business interpretation, but the transformation $\mathcal{Q}$ of the ranks of the predicted values used for ${\rm ABC}^2$ may be flat or not be monotonic in general.

\section{The miscalibration statistics}\label{sec_cal}

We have introduced three miscalibration statistics \eqref{miscalibration X}, \eqref{ABC_measure} and \eqref{ABC_measure_sq}. The key question we answer in the section is how these measures are related to each other and which miscalibration measure should be preferred for ranking predictors based on their miscalibration.

By the tower property of conditional expectations, we modify \eqref{ABC_measure} and \eqref{ABC_measure_sq} as follows
\beq\label{ABC_measure_misc}
{\rm ABC}(Y,X) &=& \frac{{\rm Cov}\big(\be[Y|X]-X, F_X(X)\big)}{\be[Y]} = \frac{\be\big[\big(\be[Y|X]-X\big)F_X(X)\big]}{\be[Y]},
\eeq
and
\beq\label{ABC_measure_sq_misc}
{\rm ABC}^2 (Y,X)=\frac{{\rm Cov}\big(\be[Y|X]-X, \mathcal{Q}(F_X(X))\big)}{\be[Y]}=\frac{\be\Big[\big(\be[Y|X]-X\big)\mathcal{Q}\Big(F_X(X)\Big)\Big]}{\be[Y]}.
\eeq
We can now see that the miscalibration measures \eqref{ABC_measure_misc}-\eqref{ABC_measure_sq_misc} compare the conditional mean $\be[Y|X]$ against $X$. The miscalibration statistics ${\rm MCB}$ from Murphy's decomposition also compares $\be[Y|X]$ against $X$. Clearly, the loss functions implicitly and explicitly used in comparing $\be[Y|X]$ against $X$ are different for the three miscalibration measures \eqref{miscalibration X}, \eqref{ABC_measure} and \eqref{ABC_measure_sq}. ${\rm MCB}$ explicitly uses a Bregman divergence. We recover the loss functions, and their properties, behind ${\rm ABC}$ and ${\rm ABC^2}$ in the sequel.

First, we observe that ${\rm ABC}$ takes real values, whereas ${\rm ABC}^2$ and ${\rm MCB}$ take non-negative values. We expect that negative values of ${\rm ABC}$ may be problematic in investigating mean-calibration of predictors. This disadvantage of ${\rm ABC}$ has recently been noticed by \cite{DenuitTrufinABC}.

\begin{prop}\label{prop_abc_zero}
Let (A)-(C) hold. (i) $\be[Y|X]=X$ implies that ${\rm ABC}(X,Y)={\rm ABC}^2(X,Y)=0$. (ii) ${\rm ABC}(X,Y)=0$ does not imply that $\be[Y|X]=X$, it only implies that $\be\Big[\be[Y|X]F_X(X)\Big]=\be\Big[XF_X(X)\Big]$. (iii) ${\rm ABC}^2(X,Y)=0$ implies that $\be[Y|X]=X$.
\end{prop}

Recall from Section \ref{sec_murphy} that ${\rm MCB}_L(Y,X)=0$ is a necessary and a sufficient condition for mean-calibration of $X$ if we choose a strictly consistent Bregman divergence $L$ to measure the discrepancy between $\be[Y|X]$ and $X$, see \cite[Theorem 2.23]{GneitingResin}. Proposition \ref{prop_abc_zero} now establishes that having ABC$(Y,X)=0$ is only a necessary condition for having mean-calibration of $X$, but it is not sufficient. Whereas ${\rm ABC}^2(X,Y)=0$ gives both a necessary and a sufficient condition for having mean-calibration of $X$. Hence, the interpretation that a predictor $X$ with a smaller miscalibration statistics has a better calibration, in the sense that $\be[Y|X]$ is closer to $X$, can be used for ${\rm ABC^2}$, but it cannot be used for ${\rm ABC}$, see \hyperref[ex_1]{Example 1} below.

The assertion (i) of Proposition \ref{prop_abc_zero} follows immediately from the definitions of the miscalibrations measures. The assertion (ii) is intuitively clear. We construct a counter-example below. The proof of (iii) is presented in the appendix. 

\medskip

\noindent \phantomsection\label{ex_1}\textit{Example 1}. Let $Z\sim {\rm Unif}(0,1)$. We set $\be[Y|Z]=Z$ and $X=Z+q\cos(2\pi Z)$. We can easily verify that $\be[\cos(2\pi Z)]=0$. We have $\be[Y]=\be[X]=0.5$. In this example we assume that $0<q<\frac{1}{2\pi}$. $X$ is positive and increasing in $Z$. Hence, $\be[Y|X]=Z$ and $F_X(X)=Z$. We derive 
$${\rm ABC}(X,Y)=\frac{\be[(\be[Y|X]-X)F_X(X)]}{\be[Y]}=\frac{-q\be[Z\cos(2\pi Z)]}{\be[Y]}=0.$$ 
In addition, we can also derive the Lorenz and the concentration curve
\beqo
LC_p(X)=p^2+q\frac{\sin(2\pi p)}{\pi},\quad CC_p(Y,X)=p^2,\qquad p\in[0,1].
\eeqo
We observe that the Lorenz and the concentration curve intersect (clearly this must hold if we have ${\rm ABC}(Y,X)=0$). In particular, $LC_p(X)>CC_p(Y,X)$ for $p\in[0,0.5)$ and $LC_p(X)<CC_p(Y,X)$ for $p\in(0.5,1]$. We remark that \cite{DenuitTrufinABC} construct a different example, but the qualitative conclusion is the same. \qed

\medskip 

\begin{prop}\label{prop proof two examples}
Let (A)-(C) hold. (i) The ranking of predictors based on their calibration as measured by the miscalibration statistics ${\rm ABC}$ or ${\rm ABC}^2$ generally differs from the ranking based on the miscalibration statistics ${\rm MCB}_L$, in the sense that for specific examples there exist Bregman divergences $L$ that provide rankings with ${\rm MCB}_L$ different from the rankings with ${\rm ABC}$ or ${\rm ABC}^2$. (ii) The rankings of predictors based on their calibration as measured by the miscalibration statistics ${\rm ABC}$ and ${\rm ABC}^2$ generally differ from each other.
\end{prop}

We prove this proposition by giving counter-examples based on the mean squared loss.

\medskip

\noindent \phantomsection\label{ex_2}\textit{Example 2}. We select the mean squared loss ${\rm MSE}(y,x)=(y-x)^2$.
Let $Z\sim {\rm Unif}(0,1)$. We set $\be[Y|Z]=Z$ and choose two predictors: $X_1=\frac{1}{2}(1-b)+bZ$ and $X_2=Z+q\cos(2\pi Z)$. We have $\be[Y]=\be[X_1]=\be[X_2]=0.5$. In this example we assume that $0<b<1$ and $0<q<\frac{1}{2\pi}$. $X_1$ and $X_2$ are positive and increasing in $Z$. Hence, $\be[Y|X_1]=\be[Y|X_2]=Z$ and $F_{X_1}(X_1)=Z$ and $F_{X_2}(X_2)=Z$. We get
\beqo
\be[Y|X_1]-X_1=(1-b)\big(Z-\frac{1}{2}\big),\quad \be[Y|X_2]-X_2=-q\cos(2\pi Z).
\eeqo
One verifies that
\beqo
&{\rm ABC}(Y,X_1) = \frac{1-b}{6},\quad  {\rm ABC}(Y,X_2)=0, \\
&{\rm MSE}\big(\be[Y|X_1],X_1\big)=\frac{(1-b)^2}{12},\quad  {\rm MSE}\big(\be[Y|X_2],X_2\big)=\frac{q^2}{2}.
\eeqo
Under the assumption that $0<b<1$, ${\rm ABC}(Y,X_1) > {\rm ABC}(Y,X_2)$. If we consider $q>\frac{1-b}{\sqrt{6}}$, and $0<q<\frac{1}{2\pi}$ by our assumption, then
${\rm MSE}\big(\be[Y|X_1],X_1\big)<{\rm MSE}\big(\be[Y|X_2],X_2\big)$. Hence, the rankings of the predictors with ABC and MSE may be different; e.g. if we choose a large $b$ and a small (but not too small) $q$.

We also derive the Lorenz and the concentration curves
\beqo
&LC_p(X_1)= (1-b)p+bp^2,\quad LC_p(X_2)=p^2+q\frac{\sin(2\pi p)}{\pi},\qquad p\in[0,1], \\
&CC_p(Y,X_1)=CC_p(Y,X_2)=p^2,\qquad p\in[0,1].
\eeqo
We already know that $LC_p(X_2)$ intersects with $CC_p(Y,X_2)$ at $p=0.5$. It is immediate to observe that $LC_p(X_1)>CC_p(Y,X_1)$. We are left with proving the relation between the Lorenz curves for $X_1$ and $X_2$. In order to have $LC_p(X_1)>LC_p(X_2)$ for some $p$, we require that $(1-b)(p-p^2)>q\frac{\sin(2\pi p)}{\pi}$, or equivalently
\beqo
\frac{\sin(2\pi p)}{p-p^2}<\frac{1-b}{q}\pi.
\eeqo
We investigate the function $p\mapsto {\sin(2\pi p)}/{(p-p^2)}$ on $[0,1]$. We can conclude it is strictly decreasing from $2\pi$ to $-2\pi$ with the root at $p=0.5$. If $q\leq \frac{1-b}{2}$, then $LC_p(X_1)\geq LC_p(X_2)$ for all $p\in[0,1]$, with the equality only at $p=0$ for $q=\frac{1-b}{2}$. If $q>\frac{1-b}{2}$, then $LC_p(X_1)<LC_p(X_2)$ for $p\in[0,p^*)$ and $LC_p(X_1)>LC_p(X_2)$ for $p\in(p^*,1]$, with $p^*\in(0,0.5)$. In particular, we observe that the Lorenz order between the predictors is not equivalent to the common rankings of the predictors with ABC and MSE. This completes the example.\qed

\medskip

\noindent \phantomsection\label{ex_3}\textit{Example 3}. We continue \hyperref[ex_2]{Example 2}. First, let us calculate 
$${\cal Q}(z)=\frac{\be\big[(\be[Y|X]-X)\big(1-\max\{z,F_X(X)\}\big)\big]}{\be[Y]},\qquad z\in[0,1].$$ 
We have for the two considered cases
\beqo
{\cal Q}_1(z)&=&2(1-b)\big(-\frac{1}{12}+\frac{z^2}{4}-\frac{z^3}{6}\big),\qquad z\in[0,1],\\
{\cal Q}_2(z)&=&-2q\frac{\cos(2\pi z)-1}{4\pi^2},\qquad z\in[0,1].
\eeqo
We notice that $z\mapsto \mathcal{Q}_1(z)$ is strictly increasing on $[0,1]$, but $z\mapsto \mathcal{Q}_2(z)$ is strictly increasing on $[0,0.5]$ and strictly decreasing on $[0.5,1]$. Next, we calculate 
\beqo
{\rm ABC}^2(Y,X)=\frac{\be\big[\big(\be[Y|X]-X\big){\cal Q}(F_X(X))\big]}{\be[Y]}=\frac{\be\big[f(Z){\cal Q}(Z)\big]}{\be[Z]},
\eeqo 
for $f_1(z)=(1-b)(z-\frac{1}{2})$ and $f_2(z)=-q\cos(2\pi z)$. We derive
\beqo
{\rm ABC}^2(Y,X_1) = \frac{(1-b)^2}{30},\quad  {\rm ABC}^2(Y,X_2)=\frac{q^2}{2\pi^2}, \\
{\rm MSE}\big(\be[Y|X_1],X_1\big)=\frac{(1-b)^2}{12},\quad  {\rm MSE}\big(\be[Y|X_2],X_2\big)=\frac{q^2}{2}.
\eeqo
In order to have ${\rm ABC}^2(Y,X_1) > {\rm ABC}^2(Y,X_2)$ and ${\rm MSE}\big(\be[Y|X_1],X_1\big)<{\rm MSE}\big(\be[Y|X_2],X_2\big)$ we choose
\beqo
\frac{1-b}{\sqrt{6}}<q<\frac{(1-b)\pi}{\sqrt{15}},
\eeqo
and we recall that we have assumed $0<b<1$ and $0<q<\frac{1}{2\pi}$ in \hyperref[ex_2]{Example 2}. Hence, the rankings of the predictors with ${\rm ABC}^2$ and MSE may be different; e.g. if we choose a large $b$ ($X_1$ is slightly miscalibrated, and the Lorenz curve dominates the concentration curve) and a sufficiently small (but not too small) $q$ ($X_2$ is moderately miscalibrated, but the Lorenz and the concentration curve cross).

Finally, in order to have ${\rm ABC}(Y,X_1) > {\rm ABC}(Y,X_2)$ and ${\rm ABC}^2(Y,X_1) < {\rm ABC}^2(Y,X_2)$, we choose $q>\frac{(1-b)\pi}{\sqrt{15}}$.

\begin{figure}[htb!]
\begin{center}
\includegraphics[width=0.6\textwidth]{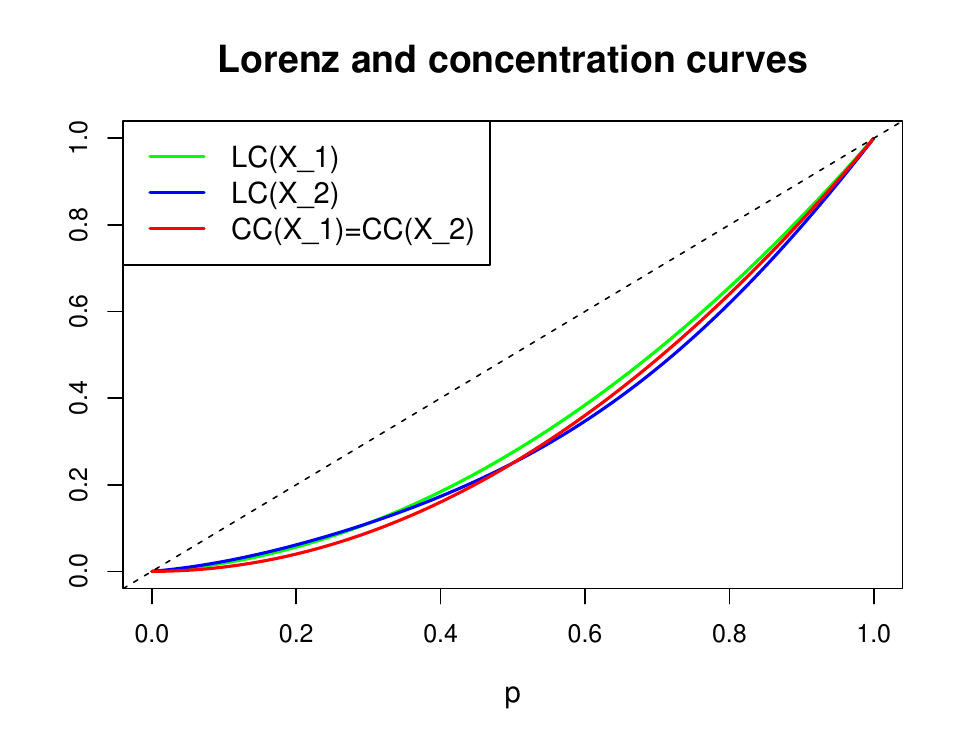}
\end{center}
\vspace{-.7cm}
\caption{The Lorenz curves and the concentrations curves of the predictors $X_1$ and $X_2$ from \hyperref[ex_1]{Example 1}-\hyperref[ex_3]{Example 3}.}
\label{fig_1}
\end{figure}

Let us choose $b=0.9$ and $q=0.07$. The Lorenz curves and the concentrations curves for $X_1$ and $X_2$ are presented in Figure \ref{fig_1}. In this case the Lorenz curves cross. We have the miscalibration statistics
\beqo
{\rm ABC}(Y,X_1) = 0.0166 \quad &>& \quad  {\rm ABC}(Y,X_2)=0, \\
{\rm ABC}^2(Y,X_1) = 0.00033 \quad &>& \quad  {\rm ABC}^2(Y,X_2)=0.00024, \\
{\rm MSE}\big(\be[Y|X_1],X_1\big)=0.00083 \quad &<& \quad  {\rm MSE}\big(\be[Y|X_2],X_2\big)=0.0049.
\eeqo

This completes the example and it also completes the proof of Proposition \ref{prop proof two examples}.\qed

\medskip

Let us now investigate the very special case when all three miscalibration measures give the same ranking of predictors.

\begin{lem}
Let (A)-(B) hold. Assume $\be[Y|X]=(1-b)\be[Y]+bX$. We have
\beqo
CC_p(Y,X)&=&(1-b)p+bLC_p(X),\qquad p\in[0,1],\\
{\rm ABC}(Y,X)&=&(1-b)\int_0^1(p-LC_u(X))du=\frac{1-b}{2}{\rm Gini}(X),\\
{\rm ABC}^2(Y,X)&=&(1-b)^2\int_0^1(p-LC_u(X))^2du,\\
{\rm MSE}\big(\be[Y|X],X\big)&=&\be\big[|\be[Y|X]-X|^2\big]=(1-b)^2Var(X).
\eeqo
\end{lem}

The proof is straightforward by direct substitution.

\begin{prop}\label{prop_rankings_abc}
Let (A)-(B) hold.
\begin{itemize}
\item Assume $\be[Y|X]=(1-b)\be[Y]+bX$.
(i) ${\rm ABC}(Y,X)=0$ iff $\be[Y|X]=X$, i.e. $b=1$. (ii) If $b<1$, then $CC_p(Y,X)>LC_p(X)$ for all $p\in(0,1)$ and ${\rm ABC}(Y,X)>0$. (iii) If $b>1$, then $CC_p(Y,X)<LC_p(X)$ for all $p\in(0,1)$ and ${\rm ABC}(Y,X)<0$.
\item Consider $X_1$ and $X_2$ and assume $\be[Y|X_i]=(1-b_i)\be[Y]+b_iX_i$ for $i=1,2$. If $|1-b_2|\leq|1-b_1|$ and $LC_p(X_1)\leq LC_p(X_2)$ for all $p\in[0,1]$, then the rankings of $X_1$ and $X_2$ based on their calibration with the miscalibration statistics ${\rm ABC}, {\rm ABC}^2$ and ${\rm MCB}_L$ with the mean squared loss $L$ coincide. The rankings imply that $X_1$ has a worse calibration than $X_2$.
\end{itemize}
\end{prop}

The proof is again straightforward. By \eqref{lorenz_dominance}, we recall that $X_2\preceq_{cx} X_1$ and the Lorenz order imply that $Var(X_2)\leq Var(X_1)$, if $X_1$ and $X_2$ have equal expected values.

The first conclusion from Proposition \ref{prop_rankings_abc} is that for predictors with the miscalibration form $\be[Y|X]=(1-b)\be[Y]+bX$, a smaller absolute value of ${\rm ABC}(Y,X)$ for $X$ can now be interpreted as a better calibration of $X$ in the sense that $\be[Y|X]$ is closer to $X$. The second conclusion is that the three miscalibration measures can give a common view on the miscalibration form $\be[Y|X]=(1-b)\be[Y]+bX$ under the right Lorenz order between the predictors. Note that the larger deviation of $(1-b)\be[Y]+bX$ away from the diagonal, controlled by the slope $b$, is not enough to have a common ranking of miscalibration of the predictors with the miscalibration measures. One also has to investigate the distributions of predictors and require that their dispersion also increase as the deviation of a recalibrated predictor from the diagonal increases. Hence the Lorenz order is an important requirement to have a common ranking of miscalibration of predictors with our three miscalibration measures. However, the stand alone Lorenz order is not enough to guarantee the common ranking of miscalibration with ${\rm ABC}, {\rm ABC}^2, {\rm MCB}$ (in Proposition \ref{prop_rankings_abc}, \hyperref[ex_2]{Example 2} and \hyperref[ex_3]{Example 3} we also make assumptions about the parameters).

We now present our second main statement. We show how the miscalibration statistics ${\rm ABC}$ and ${\rm ABC}^2$ can be related to the miscalibration statistics ${\rm MCB}$, and how ${\rm ABC}$ and ${\rm ABC}^2$, based on the area between the Lorenz and the concentration curves, measure the discrepancy between $X$ and $\be[Y|X]$. In the view of the representations presented below, it is clear that the rankings with our three miscalibration measures are different in general.

\begin{thm}\label{main_thm_2}
Let (A)-(C) hold. Let $L_H$ denote a Bregman divergence with representation \eqref{bregman_representation}. Consider the miscalibration statistics \eqref{miscalibration X} from Murphy's decomposition of the expected loss \eqref{score} w.r.t. $L_H$. We have 
\beq\label{mcb_decomposition}
{\rm MCB}_{L_H}(Y,X)&=&\int_0^\infty M_\theta\big(\be[Y|X],X\big)dH(\theta)\nonumber\\
&=&{\rm Cov}\Big(X-\be[Y|X],H(X)\Big)+\int_0^\infty\big(F_X(\theta)-F_{\be[Y|X]}(\theta)\big)H(\theta)d\theta.
\eeq
In particular, if $H$ is linear, then
\beq
{\rm MCB}(Y,X)&=&\be\Big[\Big(\be[Y|X]-X\Big)^2\Big]\nonumber\\
&=&{\rm Cov}\Big(X-\be[Y|X],X\Big)+\int_0^\infty\big(F_X(\theta)-F_{\be[Y|X]}(\theta)\big)\theta d\theta.
\eeq
\end{thm}

\begin{cor}\label{main_cor_2}
Let (A)-(C) hold. Let $L_H$ denote a Bregman divergence with representation \eqref{bregman_representation}. Consider the miscalibration statistics \eqref{miscalibration X} from Murphy's decomposition of the expected loss \eqref{score} w.r.t. $L_H$.\\
(i) If we set $H(\theta)=F_X(\theta)$, then we have
\beqo
{\rm ABC}(X,Y)=-\frac{{\rm MCB}_{L_H}(Y,X)+\int_0^\infty\big(F_{\be[Y|X]}(\theta)-F_X(\theta)\big)H(\theta)d\theta}{\be[Y]}.
\eeqo
(ii) Let the function $x\mapsto {\cal Q}(x)$ defined in Proposition \ref{prop_abc_sq} be non-decreasing on $[0,1]$. If we set $H(\theta)=\mathcal{Q}(F_X(\theta))+{\rm ABC}(Y,X)$, then we have
\beqo
{\rm ABC}^2(X,Y)=-\frac{{\rm MCB}_{L_H}(Y,X)+\int_0^\infty\big(F_{\be[Y|X]}(\theta)-F_X(\theta)\big)H(\theta)d\theta}{\be[Y]}.
\eeqo
(iii) Let the function $x\mapsto {\cal Q}(x)$ defined in Proposition \ref{prop_abc_sq} be non-increasing on $[0,1]$. If we set $H(\theta)=-\mathcal{Q}(F_X(\theta))-{\rm ABC}(Y,X)$, then we have
\beqo
{\rm ABC}^2(X,Y)=\frac{{\rm MCB}_{L_H}(Y,X)+\int_0^\infty\big(F_{\be[Y|X]}(\theta)-F_X(\theta)\big)H(\theta)d\theta}{\be[Y]}.
\eeqo
\end{cor}

\begin{rem}[non-monotonic $\mathcal{Q}$]\normalfont
Recall that monotonicity of $\theta\mapsto \mathcal{Q}(F_X(\theta))$ is studied in Lemma \ref{lemma_abc_sq}. Assume that $\theta\mapsto \mathcal{Q}(F_X(\theta))$ is not monotonic on $[0,\infty)$. Assume next that $[0,\infty)$ can be split into two disjoint subintervals on which $\theta\mapsto \mathcal{Q}(F_X(\theta))$ is monotonic (non-decreasing and non-increasing). We notice that $\theta\mapsto \mathcal{Q}(F_X(\theta))$ is of bounded variation on $[0,\infty)$ by Proposition \ref{prop_abc_sq}. Hence, we can decompose $\mathcal{Q}(F_X(\theta))=H(\theta)=H_1(\theta)-H_2(\theta)$, and both $\theta\mapsto H_1(\theta)$ and $\theta\mapsto H_2(\theta)$ are non-negative and non-decreasing on $[0,\infty)$. We have 
\beqo
\lefteqn{{\rm MCB}_{L_{H_1}}(Y,X)-{\rm MCB}_{L_{H_2}}(Y,X)}\nonumber\\
&=&\int_0^\infty M_\theta\big(\be[Y|X],X\big)dH_1(\theta)-\int_0^\infty M_\theta\big(\be[Y|X],X\big)dH_2(\theta)\nonumber\\
&=&{\rm Cov}\Big(X-\be[Y|X],H(X)\Big)+\int_0^\infty\big(F_X(\theta)-F_{\be[Y|X]}(\theta)\big)H(\theta)d\theta.
\eeqo
The result can be generalized to multiple finite subintervals on which $\theta\mapsto \mathcal{Q}(F_X(\theta))$ is monotonic. Hence, the conclusion of Corollary \ref{main_cor_2} presented below still holds even if $\theta\mapsto \mathcal{Q}(F_X(\theta))$ is not monotonic on $[0,\infty)$.
\end{rem}

We observe that the miscalibration statistics ${\rm ABC}$ and ${\rm ABC}^2$ consist of two terms. The first term describes the discrepancy between the predictor $X$ and its recalibrated version $\be[Y|X]$ measured by a kind of miscalibration statistics ${\rm MCB}_{L_H}$ from Murphy's decomposition of the expected loss w.r.t.~$L_H$. The second term has a similar interpretation but it uses an integral of the simple difference between the distributions of $X$ and $\be[Y|X]$. As a result, ${\rm ABC}$ and ${\rm ABC}^2$ take a large value if ${\rm MCB}_{L_H}(Y,X)$ is large and the integrated distance between the distributions $F_{\be[Y|X]}(x)$ and $F_X(x)$ is large. As far as the first term is concerned, a critical point is that the miscalibration statistics 
${\rm MCB}_{L_H}(Y,X)=S_{L_H}(\be[Y|X],X)$ with $H$ defined in Corollary \ref{main_cor_2} is strictly speaking not coming
from an expected loss w.r.t.~a Bregman divergence in the sense of our definition \eqref{score} and Murphy's decomposition. The choice of $H$ in the corollary depends on the predictor $X$, and different predictors lead to different $H$'s. Thus, ${\rm MCB}_{L_{H_1}}(Y,X_1)$ and ${\rm MCB}_{L_{H_2}}(Y,X_2)$ cannot be benchmarked against each other as the evaluation of $X_1$ vs $X_2$ is based on different loss functions $L_{H_1}$ and $L_{H_2}$. The second term is a weighted integrated distance between the distributions of $X$ and $\be[Y|X]$, which is again a measure of miscalibration, but the weight also depends on the predictor $X$. Hence, the same objection applies to the second term as for ${\rm MCB}_{L_H}(Y,X)$. These representations imply that if we compare two predictors $X_1$ and $X_2$, the differences in ${\rm ABC}$ and ${\rm ABC}^2$ do not only reflect how close, or far, $X$ is from $\be[Y|X]$, but they also reflect the changes in the measure $H$ used for defining the expected loss $\be\big[\int_0^\infty L_\theta(\be[Y|X],X)dH(\theta)\big]$ and the changes in the weight $H$ used for defining the integrated loss $\int_0^\infty (F_{\be[Y|X]}(\theta)-F_X(\theta))H(\theta)d\theta$, respectively. As discussed in the introduction, this critical point (and disadvantage) of statistical measures used in forecast evaluation has been noticed by \cite{hand} and \cite{hand_2} in the case of the Gini index and binomial responses. These authors conclude that it is highly inappropriate to validate predictors with loss functions which depend on predictors. Clearly, the decision-theoretic approach to the evaluation of point predictions by \cite{Gneiting} is not fulfilled with ${\rm ABC}$ and ${\rm ABC ^2}$ since these measures cannot be represented as the expected losses w.r.t.~a Bregman divergence. We conclude that we prefer the miscalibration statistics ${\rm MCB}$ from Murphy's decomposition over the miscalibration statistics ${\rm ABC}$ and ${\rm ABC}^2$ derived from the difference between the Lorenz and the concentration curve.

Finally, we notice that the miscalibration statistics ${\rm ABC}$ and ${\rm ABC}^2$ given by \eqref{ABC_measure_misc}-\eqref{ABC_measure_sq_misc} are of the form
\beq\label{ABC_measure_weighted}
\be\Big[L\big(\be[Y|X],X\big)\mathcal{W}\big(F_X(X)\big)\Big],
\eeq
with a linear loss function $L$ and a weight $\mathcal{W}$ which depends on $F_X$. From this perspective, this representation does not match the expected loss \eqref{score} w.r.t.~a Bregman divergence. The loss $L$ involved in \eqref{ABC_measure_weighted} is not a Bregman divergence and the weight $\mathcal{W}$ depends on the predictor itself. More generally, let us now consider an expected weighted loss given by
\beq\label{weighted_abc}
S_{L,\mathcal{W}}(Y,X) = \mathbb{E}\Big[L(Y,X)\mathcal{W}\big(F_X(X)\big)\Big],
\eeq
where $L$ is now a Bregman divergence and the weight $\mathcal{W}$ still depends on $F_X$. This expected weighted loss still does not match the evaluation framework with the expected loss \eqref{score} in the sense of \cite{Gneiting} since the weighted loss $L(Y,X)\mathcal{W}\big(F_X(X)\big)$ cannot be written as a function of a prediction value of $X$ and an outcome of $Y$. It is also not a scoring rule in the sense of \cite{GneitingRaftery} since it cannot be written as a function of a predictive probability function of $X$ and an outcome of $Y$. Yet, such a type of loss function arises when we deal with ${\rm ABC}$ and ${\rm ABC^2}$. The next result shows that such a weighted scoring with weight depending on a predictor cannot be used for evaluation of predictors. 

\begin{prop}\label{prop_weighted loss}
Let $S_{L,\mathcal{W}}$ be defined by \eqref{weighted_abc} as the expected weighted loss w.r.t.~a Bregman divergence between $Y$ and $X$ with a weight which depends on the distribution of $X$. Such a score is in general not consistent for the estimation of the mean value of $Y$, in the sense that it may happen that
\beqo
S_{L,\mathcal{W}}\big(Y,\be[Y|\mathcal{A}]\big)>S_{L,\mathcal{W}}\big(Y,X\big),
\eeqo
for some $\mathcal{A}$-measurable predictor $X$.
\end{prop}

We give an example. 

\medskip

\noindent \phantomsection\label{ex_4}\textit{Example 4}. Let us consider the mean squared loss $L(y,x)=(y-x)^2$. We consider predictors of the form $X=h(Z)$, where $Z$ is a one-dimensional feature and $h$ is a regression function. Hence, $\mathcal{A}=\sigma(Z)$, see Remark \ref{conditional_mean_estimation}. There exists the true regression function such that $\be[Y|Z]=h^*(Z)$. Both $z\mapsto h(z)$ and $z\mapsto h^*(z)$ are strictly monotonic. We consider the expected weighted loss
\beqo
S_{L,\mathcal{W}}(Y,X) &=& \mathbb{E}\Big[\Big(Y-X\Big)^2F_X(X)\Big],
\eeqo
and derive the decomposition
\beqo
S_{L,\mathcal{W}}(Y,X) &=& \mathbb{E}\Big[{\rm Var}\big[Y|Z\big]F_X(X)\Big]+\mathbb{E}\Big[\Big(\be\big[Y|Z\big]-X\Big)^2F_X(X)\Big].
\eeqo
If ${\rm Var}\big[Y|Z\big]$ does not depend on $Z$, then 
\beqo
S_{L,\mathcal{W}}(Y,X) = \frac{1}{2}{\rm Var}\big[Y\big]+\mathbb{E}\Big[\Big(\be\big[Y|Z\big]-X\Big)^2F_X(X)\Big],
\eeqo
and we should choose $X=\be\big[Y|Z\big]=h^*(Z)$. However, if ${\rm Var}\big[Y|Z\big]$ depends on $Z$, then intuitively, we could sacrify the bias $\big(\be\big[Y|Z\big]-X\big)^2$ to reduce the variance ${\rm Var}\big[Y|Z\big]$ -- this is related to the variance-bias trade-off.

Assume that $Z\sim {\rm Unif}(0,1)$ and choose a Tweedie's distribution assumption $\be[Y|Z]=Z$ and ${\rm Var}[Y|Z]=\phi Z^p$. We consider two predictors $X_1=Z$ and $X_2=1-Z$. We calculate
\beqo
S_{L,\mathcal{W}}(Y,X_1)&=&\mathbb{E}\Big[\phi Z^pZ\Big]=\phi\frac{1}{p+2},\\
S_{L,\mathcal{W}}(Y,X_2)&=&\mathbb{E}\Big[\phi Z^p(1-Z)\Big]+\mathbb{E}\Big[\Big(2Z-1\Big)^2(1-Z)\Big]=\phi\Big(\frac{1}{p+1}-\frac{1}{p+2}\Big)+\frac{1}{6}.
\eeqo
We can obviously choose $\phi,p$ so that $S_{L,\mathcal{W}}(Y,X_2)<S_{L,\mathcal{W}}(Y,X_1)$, e.g., $\phi = 4, p = 1$. We get $S_{L,\mathcal{W}}(Y,X_1)=\frac{4}{3}$ and $S_{L,\mathcal{W}}(Y,X_2)=\frac{5}{6}<\frac{4}{3}$. We prefer $X_2$ over $X_1$ even though $X_1$ is the perfect predictor. This proves Proposition \ref{prop_weighted loss}.\qed

\medskip

\cite{GneitingRanjan} and \cite{taagart} show that expected weighted Bregman divergences where the weight depends on the outcome of a response variable cannot be used for forecasts evaluation. Our Proposition \ref{prop_weighted loss} shows that minimizing expected weighted Bregman divergences with weights depending on the predictor may also lead to misleading point predictions. Intuitively speaking, the weighting in \eqref{weighted_abc} can be interpreted as a change of measure in the Radon--Nikod\'ym sense, thus, the underlying probability measure $\mathbb{P}$ is changed to an $X$-dependent measure $\mathbb{P}_X$. Firstly, this changes the underlying expected values of the considered random variables and, secondly, for different predictors $X_1$ and $X_2$ it results in different changed measures and, thus, non-comparability.

\section{The discrimination statistics}\label{sec_disc}

Let us now investigate the two discrimination statistics \eqref{discrimination X} and \eqref{gini}. We show how these measures are related to each other and elaborate which discrimination measure should be preferred for ranking predictors based on their discrimination.

First, we note that the Gini index \eqref{gini} only depends on $X$, not on $Y$. At the same time, the discrimination statistics \eqref{discrimination X} from Murphy's decomposition depends on $(Y,X)$. Hence, it makes sense to consider the Gini index as a measure of discriminatory power of $X$ for predicting $Y$ only if $X$ is mean-calibrated for $Y$. \cite[Theorem 4.3]{wuthrich_gini} confirms that argument by the following theorem. Recall Definition \ref{def_consistent_loss} and Remark \ref{stop-loss order}.

\begin{thm}\label{Theorem mario?}
The predictor $X=\be[Y|\mathcal{A}]$ maximizes the Gini index \eqref{gini} on the class of ${\cal A}$-measurable and mean-calibrated predictors for $Y$.
\end{thm}

In this section we assume that 
\begin{itemize}
\item [(D)] $X$ is mean-calibrated for $Y$ in the sense of Definition \ref{def_mean_calibration}.
\end{itemize}
If $X$ is not mean-calibrated for $Y$, we recalibrate this predictor by $X\mapsto \be[Y|X]$ to get a mean-calibrated version of $X$. If $X$ is mean-calibrated for $Y$, then ${\rm MCB}(Y,X)=0$ and the expected loss $S(Y,X)$ depends on $X$ only through ${\rm DSC}(Y,X)=S(X,\be[Y])$; we have Murphy's decomposition for mean-calibrated predictors $S(Y,X)={\rm UNC}(Y)-{\rm DSC}(Y,X)$ by \eqref{Murphy score_1} and the predictive power is determined by the discriminatory power. Moreover, both the Gini index and ${\rm DSC}$ only depend on $X$ now. Intuitively, the ${\rm Gini}$ index and ${\rm DSC}$ should be related to some loss functions which compare $X$ against $\be[Y]$. By our new version of Murphy's decomposition, see Theorem \ref{prop_murphy_new_decomp}, ${\rm DSC}$ explicitly uses a Bregman divergence for this comparison. We recover the loss function behind the Gini index in the sequel.

We start with the following (trivial) result.

\begin{prop}
Let (A), (B), (D) hold. We have the equivalences
\beqo
X=\be[Y] \quad &\Longleftrightarrow& \quad {\rm Gini}(X) = 0 \\
\quad &\Longleftrightarrow& \quad {\rm DSC}_L(Y,X)=0 \qquad \text{for all strictly consistent $L$},
\eeqo
where for all strictly consistent $L$ means for all strictly consistent Bregman divergences.
\end{prop}

We need strict consistency of $L$ to get the equivalence between ${\rm DSC}_L(Y,X)=0$ and $X=\be[Y]$, recall \cite[Theorem 2.23]{GneitingResin}. Having the Gini index or the discrimination statistics from Murphy's decomposition equal to zero is a necessary and a sufficient condition for the lack of discriminatory power of a mean-calibrated predictor. By the representation of DSC with a Bregman divergence, the interpretation that $X$ with a smaller discrimination statistics has a lower discriminatory power, in the sense that $X$ is closer to $\be[Y]$, can be used for DSC. The question is whether this interpretation can  be used for the  Gini index as the loss function used for comparing $X$ with $\be[X]$ in the Gini index is not explicitly given.

The next proposition characterizes when the rankings of predictors based on their discriminatory power with the Gini index and the discrimination measure from Murphy's decomposition coincide.

\begin{prop}\label{prop_gini_dominance}
Let (A), (B), (D) hold. We have
\beq\label{gini_dominance_2}
LC_p(X_1)\leq LC_p(X_2) \quad \text{for all $p\in[0,1]$} &\Longleftrightarrow& \quad S_L(Y,X_1)\leq S_L(Y,X_2) \quad \text{for all $L$}\nonumber\\
\quad &\Longleftrightarrow& \quad {\rm DSC}_L(Y,X_1)\geq {\rm DSC}_L(Y,X_2) \quad \text{for all $L$},\qquad
\eeq
where for all $L$ means for all Bregman divergences.
\end{prop}

The first equivalence relation is proved by \cite[Theorem 3.1]{kruger_ziegel}. This also follows from \eqref{m_s}, Remark \ref{stop-loss order} and \eqref{lorenz_dominance}. Since $S_L(Y,X)={\rm UNC}_L(Y)-{\rm DSC}_L(Y,X)$, we immediately deduce the second equivalence relation.

Recall that the relation between the Lorenz curves presented on the left-hand side of \eqref{gini_dominance_2} is called Lorenz dominance, see \eqref{lorenz_dominance}. Lorenz dominance appears if the Lorenz curves of two predictors do not intersect. Proposition \ref{prop_gini_dominance} shows that Lorenz dominance is the key factor for common rankings of discrimination of mean-calibrated predictors with Gini and ${\rm DSC}$. The condition that $LC_p(X_1)\leq LC_p(X_2)$ for all $p\in[0,1]$ implies that ${\rm Gini}(X_1)\geq {\rm Gini}(X_2)$. However, the implication in the other direction is not generally true. The role of the Gini index in measuring the discrepancy between $X$ and $\be[X]$ and the discriminatory power of $X$ must still be understood in the general case when the Lorenz curves of competing predictors intersect. In the sequel we derive a useful representation of the Gini index.

\begin{cor}\label{cor_gini_dominance}
Let (A), (B), (D) hold. (i) Let the mean-calibrated predictors $X_1$ and $X_2$ be ordered in the sense of Lorenz dominance. The rankings of mean-calibrated predictors based on their discrimination with the discrimination statistics Gini and ${\rm DSC}_L$ with any Bregman divergence $L$ coincide. (ii) In general, the ranking of mean-calibrated predictors based on their discrimination with the discrimination statistics Gini is different from the ranking with the discrimination statistics ${\rm DSC}_L$, in the sense that there exist examples and Bregman divergences $L$ that provide a ranking with ${\rm DSC}_L$ different from the ranking with Gini.
\end{cor}

Assertion (i) follows immediately from Proposition \ref{prop_gini_dominance}. Assertion (ii) also follows from Proposition \ref{prop_gini_dominance}. Since the Lorenz dominance does not hold,  we can find $L_1$ and $L_2$ such that ${\rm DSC}_{L_1}(Y,X_1)\geq {\rm DSC}_{L_1}(Y,X_2) $ and ${\rm DSC}_{L_2}(Y,X_1)\leq {\rm DSC}_{L_2}(Y,X_2)$, and one of the ranking is different from the Gini ranking.

We shall remark that Proposition \ref{prop_gini_dominance} and Corollary \ref{cor_gini_dominance} are closely related to the fundamental result from \cite{inequality_measures}. By \eqref{score_calibrated} and assumptions (A), (B) and (D), for evaluating mean-calibrated predictors we use the utility $U(x)=x-(x-\theta)^+$ which is increasing and concave. \cite{inequality_measures} proved that the rankings of the distributions of $X_1$ and $X_2$ with equal expected values based on the criterion of the maximal expected utility $\be[U(X)]$ coincide for all increasing, concave and twice differentiable utilities $U$ iff $X_1$ and $X_2$ are ordered in the sense of Lorenz dominance.

\medskip

Let us illustrate assertion (ii) from Corollary \ref{cor_gini_dominance}.

\medskip

\noindent \phantomsection\label{ex_5}\textit{Example 5}.
Let us consider the mean squared loss $L(y,x)=(y-x)^2$. In this case, ${\rm DSC}_L(Y,X)={\rm Var}[X]$. We show that the ranking of $X_1$ and $X_2$ with equal expected values may be different with the Gini index and the variance. This phenomenon has been reported in the literature, see, e.g., \cite{inequality_measures} and \cite[Chapter 2.2]{Gini_book}, yet specific examples are hard to find.  \cite{example_lognormal} suggest that the Gini index and the variance may rank distributions differently in the class of shifted log-normal distributions. We construct an example, which is absent in \cite{example_lognormal}. Let $X=Z+a$ where $Z\sim {\rm LogN}(\mu, \sigma^2)$ and $a\geq 0$. Then, by \cite{example_lognormal} we have the measures
\beqo
\be[X]&=&a+e^{\mu+\frac{1}{2}\sigma^2},\\
{\rm Var}[X]&=&\big(\be[X]-a\big)^2\Big(e^{\sigma^2}-1\Big),\\
{\rm Gini}(X)&=&\frac{\be[X]-a}{\be[X]}\Big(2\Phi\Big({\sigma}/{\sqrt{2}}\Big)-1\Big).
\eeqo
We require that
\beqo
\be[X_1]=\be[X_2],\quad {\rm Var}[X_1]>{\rm Var}[X_2],\quad {\rm Gini}(X_1)<{\rm Gini}(X_2).
\eeqo
Hence, the parameters must satisfy
\beqo
\frac{\be[X]-a_1}{\be[X]-a_2}&>&\sqrt{\frac{e^{\sigma_2^2}-1}{e^{\sigma_1^2}-1}},\\
\frac{\be[X]-a_1}{\be[X]-a_2}&<&\frac{2\Phi\Big({\sigma_2}/{\sqrt{2}}\Big)-1}{2\Phi\Big({\sigma_1}/{\sqrt{2}}\Big)-1}.
\eeqo
All we have to do is to choose $0<a_2<a_1$ so that $c=\frac{\be[X]-a_1}{\be[X]-a_2}<1$, $\sigma_2>\Phi^{-1}\big(\frac{1+c}{2}\big)\sqrt{2}$ and $\sigma_1$ sufficiently larger than $\sigma_2$. At the same time we note that the ranking based on the Gini index and the variance always coincide for log-normal distributions without a shift ($a=0$).

Using classical results for log-normal distributions, the Lorenz curve for a shifted log-normal distribution is given by
\beqo
LC_p(X)=\frac{ap+\big(\be[X]-a\big)\Phi\Big(\Phi^{-1}(p)-\sigma\Big)}{\be[X]},\qquad p\in[0,1],
\eeqo
and one can show that $p\mapsto LC_p(X)$ is strictly convex on $[0,1]$ for any $a\geq0$, by calculating the second derivative. Let $X_{a,\sigma^2}$ denote the shifted log-normal distribution with parameters $(a,\sigma^2)$. Under our choice of the parameters for the two distributions, i.e. $\sigma_1>\sigma_2$, we have
\beqo
\Phi\Big(\Phi^{-1}(p)-\sigma_1\Big)<\Phi\Big(\Phi^{-1}(p)-\sigma_2\Big),\qquad p\in[0,1].
\eeqo
Hence, $LC_p(X_{0,\sigma_1^2})<LC_p(X_{0,\sigma_2^2})$ for all $p\in[0,1]$. By shifting the log-normal distributions by $a>0$ we move their Lorenz curves upward. Indeed, let $a\mapsto h_p(a)=ap+\big(1-a\big)\Phi\Big(\Phi^{-1}(p)-\sigma\Big)$ for $a\in[0,\infty)$, then $h_p'(a)=p-\Phi\Big(\Phi^{-1}(p)-\sigma\Big)>0$. First, we move the Lorenz curves for $X_{0,\sigma_1^2}$ and $X_{0,\sigma_2^2}$ upward by shifting the distributions by $a_2$. Since $\sigma_1>\sigma_2$ and $h_p'(a)$ is strictly increasing in $\sigma$ at each $a\in[0,\infty)$ and $p\in[0,1]$, we move the Lorenz curve for $X_{0,\sigma_1^2}$ more upward than the Lorenz curve for $X_{0,\sigma_2^2}$ at each $p\in[0,1]$. Next, we only move the Lorenz curve for $X_{a_2,\sigma_1^2}$ upward by shifting the distribution by $a_1-a_2$. All Lorenz curves considered are strictly convex. Since the ranking of $X_{a_1,\sigma_1^2}$ and $X_{a_2,\sigma_2^2}$ based on the Gini index and the variance is different, the Lorenz curves must intersect by Proposition \ref{prop_gini_dominance}. Our investigation above shows that they intersect only once. More precisely, there exists $p^*\in(0,1)$ such that $LC_p(X_{a_1,\sigma_1^2})>LC_p(X_{a_2,\sigma_2^2})$ for all $p\in[0,p^*)$, and $LC_p(X_{a_1,\sigma_1^2})<LC_p(X_{a_2,\sigma_2^2})$ for all $p\in(p^*,1]$.

\begin{figure}[htb!]
\begin{center}
\includegraphics[width=0.6\textwidth]{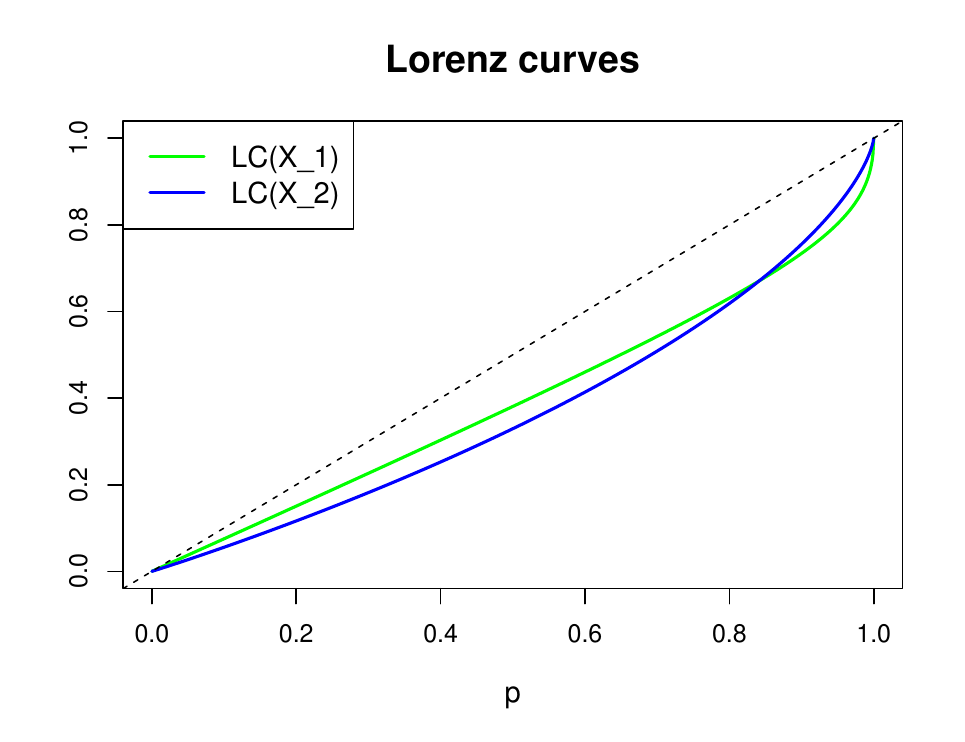}
\end{center}
\vspace{-.7cm}
\caption{Lorenz curves of the predictors $X_1$ and $X_2$ from \hyperref[ex_5]{Example 5}.}
\label{fig_2}
\end{figure}

Let us choose $\sigma_1=2, \sigma_2=1, a_1=7.5, a_2=5$ and $\be[X]=10$. The Lorenz curves for $X_1$ and $X_2$ are presented in Figure \ref{fig_2}. We have the discrimination statistics
\beqo
{\rm Gini}(X_1) = 0.2107 \quad &<& \quad  {\rm Gini}(X_2)=0.2603, \\
{\rm Var}(X_1)=334.9884 \quad &>& \quad  {\rm Var}(X_2)=42.9570.
\eeqo

This example also shows why $|{\rm ABC}(Y,X)|\sim {\rm Gini}(X)$ and ${\rm MSE}(\be[Y|X],X)\sim Var(X)$ considered in Proposition \ref{prop_rankings_abc} may lead to different rankings of the calibration of predictors if the Lorenz curves of the predictors intersect.
\qed

\medskip

Most importantly, we can identify the difference between the Gini index and the discrimination statistics from Murphy's decomposition for mean-calibrated predictors. We present the third main theorem of this paper. We show how the Gini index can be related to ${\rm DSC}_L$, and how the Gini index, based on the Lorenz curve, measures the discrepancy between $X$ and $\be[Y]$. In the view of the representation presented below, it is clear that the rankings with our two discrimination measures are different in general.

Based on \eqref{score_calibrated}, let us recall the relations
\beq\label{score_H}
S_{L_H}(Y,X)=\int_0^\infty\be\big[(Y-\theta)^+\big]dH(\theta)-\int_0^\infty\be\big[(X-\theta)^+]dH(\theta),\\
\label{discrimination_X_H}
{\rm DSC}_{L_H}(Y,X)=\int_0^\infty\be\big[(X-\theta)^+\big]dH(\theta)-\int_0^\infty\big(\be[Y]-\theta)^+dH(\theta).
\eeq

\begin{thm}\label{thm_main_3}
Let (A), (B), (D) hold. Let $L_H$ denote a Bregman divergence with representation \eqref{bregman_representation}. Consider the discrimination statistics \eqref{discrimination X} from Murphy's decomposition of the expected loss \eqref{score} w.r.t.~$L_H$. We choose $H(\theta)=F_X(\theta)$. The Gini index \eqref{gini} for $X$ has the following representations
\beq\label{gini_bregman_representation}
{\rm Gini}(X)&=&\frac{\be\big[\big(Y-X'\big)^+\big]-S_{L_H}(Y,X)}{\be[Y]}\nonumber\\
&=&\frac{{\rm DSC}_{L_H}(Y,X)+\frac{1}{2}\be\big[\big|X-\be[Y]\big|\big]}{\be[Y]}\nonumber\\
&=&\frac{{\rm DSC}_{L_H}(Y,X)}{\be[Y]}+\Big(F_X(\be[Y])-L_{F_X(\be[Y])}(X)\Big),
\eeq
where $X'$ is an independent copy of $X$.
\end{thm}

Let us first consider a special case of Theorem \ref{thm_main_3}. Let us assume that $Y$ has a Bernoulli distribution supported in $\{0,1\}$. The mean-calibrated predictor $X$ is supported on $(0,1)$. Let $\pi_0$ and $\pi_1$ denote the probabilities that an object randomly drawn from a population belongs to class $\{0\}$ and $\{1\}$, respectively. Based on \eqref{gini_bregman_representation} we derive
\beq\label{bregman_match_3}
S_{L_H}(Y,X)&=&-\pi_1{\rm Gini}(X)+\be\big[\big(0-X'\big)^+\big]\pi_0+\be\big[\big(1-X'\big)^+\big]\pi_1\nonumber\\
&=&-\pi_1{\rm Gini}(X)+(1-\be[X])\pi_1=-\pi_1{\rm Gini}(X)+\pi_0\pi_1\nonumber\\
&=&-\pi_1\big(2{\rm AUC}(X)-1\big)+\pi_0\pi_1=-2\pi_1{\rm AUC}(X)+(1+\pi_0)\pi_1,
\eeq
where ${\rm AUC}(X)$ defines the \emph{area-under-the-curve} for predictor $X$; 
see \cite{Tasche}. With formula \eqref{bregman_match_3} derived from Theorem \ref{thm_main_3} for binary responses, we recover the conclusion from \cite{hand} and \cite{hand_2}, where the authors consider the misclassification error and distributions of costs of misclassification. The main objection about applying the Gini index and the area-under-the-curve in classification problems raised by \cite{hand} and \cite{hand_2} is that these classification measures assume different measures $H$ for the Bregman divergences for the expected losses \eqref{bregman_representation} for different classifiers, or different cost distributions using the language from \cite{hand} and \cite{hand_2}. The authors conclude that it is highly inappropriate to validate predictors with measures which depend on predictors. The measure $H$ used for integrating the elementary Bregman divergence $L_H(y,x)=\int_0^\infty L_{\theta}(y,x)dH(\theta)$ for the expected loss $S_{L_H}=\be[L_H(Y,X)]$, which we use for the comparison of predictors, should be the same for all predictors we wish to compare with the score $S_{L_H}$. This is not the case in \eqref{bregman_match_3} since we choose $H(\theta)=F_X(\theta)$ if we compare the predictors with the Gini index. 

Theorem \ref{thm_main_3} goes beyond the binary case covered by \cite{hand} and \cite{hand_2} and presents the general case. The main conclusion is the same. The Gini index \eqref{gini_bregman_representation} is split into two terms. The first term in \eqref{gini_bregman_representation} is the discrimination statistics \eqref{discrimination X} from Murphy's decomposition. However, it is the expected loss w.r.t.~a loss calculated by integrating the elementary Bregman loss, which measures the discrepancy between $X$ and $\be[Y]$, with the measure $H=F_X$ which depends on the predictor under consideration. As in \cite{hand} and \cite{hand_2}, we conclude that if we use the Gini index for comparing the discriminatory power of two predictors, then the first term evaluates the discrimination of the predictors using different metrics. A fair comparison of the discriminatory power of mean-calibrated predictors is not possible with ${\rm DSC}_{L_H}$ as the loss function $L_H$, which compares $X$ against $\be[Y]$, depends on the evaluated predictor $X$ and its distribution $F_X$. The second term in \eqref{gini_bregman_representation} is the mean absolute deviation (MAD) of predictor $X$, which in particular can be related to the Lorenz curve of the predictor under consideration. The term $p-LC_p(X)$ is called the inequality gap in the literature on wealth inequalities. The key point here is that the loss $L(y,x)=|y-x|$ underlying MAD is not a Bregman divergence. Hence, the second term measures the discrepancy between $X$ and $\be[Y]$, but not with a Bregman divergence. For the shifted log-normal distribution used in \hyperref[ex_4]{Example 4}, the second term in \eqref{gini_bregman_representation} takes the form
\beqo
\Phi\big(\sigma/2\big)-LC_{p=\Phi(\sigma/2)}(X)&=&\Phi\big(\sigma/2\big)-\frac{a\Phi\big(\sigma/2\big)+\big(\be[X]-a\big)\Phi\big(-\sigma/2\big)}{\be[X]}\\
&=&2\Phi\big(\sigma/2\big)-1-\frac{a\Big(2\Phi\big(\sigma/2\big)-1\Big)}{\be[X]}.
\eeqo 

We observe that the Gini index is small if ${\rm DSC}_{L_H}=S_{L_H}(X,\be[Y])$ is small and the absolute mean deviation $\be[|X-\be[Y]|]$ is small. Hence, the interpretation that a mean-calibrated predictor $X$ with a smaller Gini index has a lower discriminatory power, in the sense that $X$ is closer to $\be[X]$, can fortunately be used for Gini in general. However, the decision-theoretic approach to the evaluation of point predictions by \cite{Gneiting} is not fulfilled with the Gini index since this measure cannot be represented as the expected loss w.r.t.~a Bregman divergence. The above characterization of the Gini index is not harmful if the Lorenz dominance/Murphy dominance/Bregman dominance hold for the mean-calibrated predictors under consideration, recall Proposition \ref{prop_gini_dominance}. In such a case, the Gini index \eqref{gini} and the discrimination statistics \eqref{discrimination X} from Murphy's decomposition give the common view on the discriminatory power of the predictors, even though the loss function behind the Gini index depends on the distribution of the predictor to be evaluated. Yet, in most cases we do not expect the Lorenz dominance/Murphy dominance/Bregman dominance to hold in practice. Taking into account all arguments presented above, we conclude that we prefer ${\rm DSC}$ over the Gini index. Let us recall that the same type of arguments were used in the previous section to question ${\rm ABC}$ and ${\rm ABC}^2$ in favor of ${\rm MCB}$.

\section{Discrimination measures with intersecting Lorenz and Murphy's curves}\label{sec_disc_intersect}

It is known that when Lorenz curves intersect, the standard Lorenz dominance criterion cannot provide a complete ranking of distributions, see, e.g., \cite{inequality_measures}, \cite{role_of_variance} and \cite{muliere}. Researchers in the field of inequality measurement have developed weaker ordering criteria than the Lorenz order. In our context of evaluation of point predictions, Proposition \ref{prop_gini_dominance} fails if the Lorenz curves or  Murphy's curves of two predictors intersect. In this section we propose a weaker version of Murphy's dominance \eqref{murphy_dominance} and derive a weaker version of Proposition \ref{prop_gini_dominance} in case when the Lorenz curves of two predictors cross once.

First, let us investigate the relation between the number of crossing points for Lorenz curves and Murphy's curves of two mean-calibrated predictors. The proposition below is an extension of Fact B from \cite{tryptych} where the authors consider receiver operating characteristics (ROC) curves for binary responses, instead of Lorenz curves for real-valued responses as we do in this paper; we also refer to \cite{Tasche} for the ROC curve. In the proposition below we state that Murphy's dominance is equivalent to the Lorenz dominance for mean-calibrated predictors, which can be immediately deduced from Propositions \ref{prop_murphy_dominance} and \ref{prop_gini_dominance}. More importantly, we show that if the Lorenz curves of two mean-calibrated predictors cross, their Murphy's curves cross as well, and the number of crossing points for the Lorenz curves and  Murphy's curves are equal.

\begin{prop}\label{prop_murphy_gini_dominance}
Let (A)-(D) hold. (i) We have
\beq\label{murphy_gini_dominance}
LC_p(X_1)\leq LC_p(X_2) \quad \text{for all $p\in[0,1]$} \quad \Longleftrightarrow \quad M_\theta(Y,X_1)\leq M_\theta(Y,X_2) \quad \text{for all $\theta\geq 0$}.
\eeq
(ii) Assume that $x\mapsto F_{X_1}(x)-F_{X_2}(x)$ has a finite number $n\geq 1$ of sign changes. The Murphy's curve difference $\theta \mapsto M_\theta(Y,X_1)- M_\theta(Y,X_2)$ and the Lorenz curve difference $p\mapsto LC_p(X_1)- LC_p(X_2)$ have the same number of sign changes, at most $n-1$. 
(iii) Assume that $x\mapsto F_{X_1}(x)-F_{X_2}(x)$ has two sign changes. If the Lorenz curves cross once, then Murphy's curve cross once, and vice-verse. It holds
\beqo
&LC_p(X_1)\geq LC_p(X_2), \ \textit{ for all $p\in[0,p^*]$} \quad \textit{and} \quad LC_p(X_1)\leq LC_p(X_2), \ \textit{ for all $p\in[p^*,1]$}\\
&\Longleftrightarrow\\
&M_\theta(Y,X_1)\geq M_\theta(Y,X_2), \ \textit{ for all $\theta\in[0,\theta^*]$} \quad \textit{and} \quad M_\theta(Y,X_1)\leq M_\theta(Y,X_2), \ \textit{ for all $\theta\in[\theta^*,\infty)$.}  
\eeqo
\end{prop}

By \cite{tryptych}, a function $x\mapsto h(x)$ on $[0,\infty)$ has $n$ sign changes if there exists a partition of $[0,\infty)$ with $n$ + 1 members, such that the function is non-negative (non-positive) with at least one non-zero value on the first and non-positive (non-negative) with at least one non-zero value on the second of any two consecutive members of the partition. 

\medskip

\noindent \phantomsection\label{ex_6}\textit{Example 6}. We continue \hyperref[ex_5]{Example 5}.
Using classical results for log-normal distributions, for $X$ which is a mean-calibrated predictor for $Y$ and has a shifted log-normal distribution, we derive the Murphy's curve
\beqo
\lefteqn{M_\theta(X,\be[Y])=\be\big[\big(X-\theta\big)^+-\big(\be[Y]-\theta\big)^+\big]}\\
&=&
\begin{cases}
  0, \quad \text{if } \theta\leq a, \\
  e^{\mu+\frac{1}{2}\sigma^2}\Phi\Big(\frac{\mu+\sigma^2-\log(\theta-a)}{\sigma}\Big)-(\theta-a)\Phi\Big(\frac{\mu-\log(\theta-a)}{\sigma}\Big) - \big(e^{\mu+\frac{1}{2}\sigma^2}+a-\theta\big)^+, \quad \text{if } \theta>a.
\end{cases}
\eeqo

\noindent We point out that $\theta\mapsto M_\theta(X,\be[X])$ focuses on the discriminatory power of $X$ in predicting $Y$.

\begin{figure}[htbp]
    \centering
    \hspace{-1.5cm}
    \begin{minipage}{0.45\textwidth}
        \centering
        \includegraphics[width=1.2\textwidth]{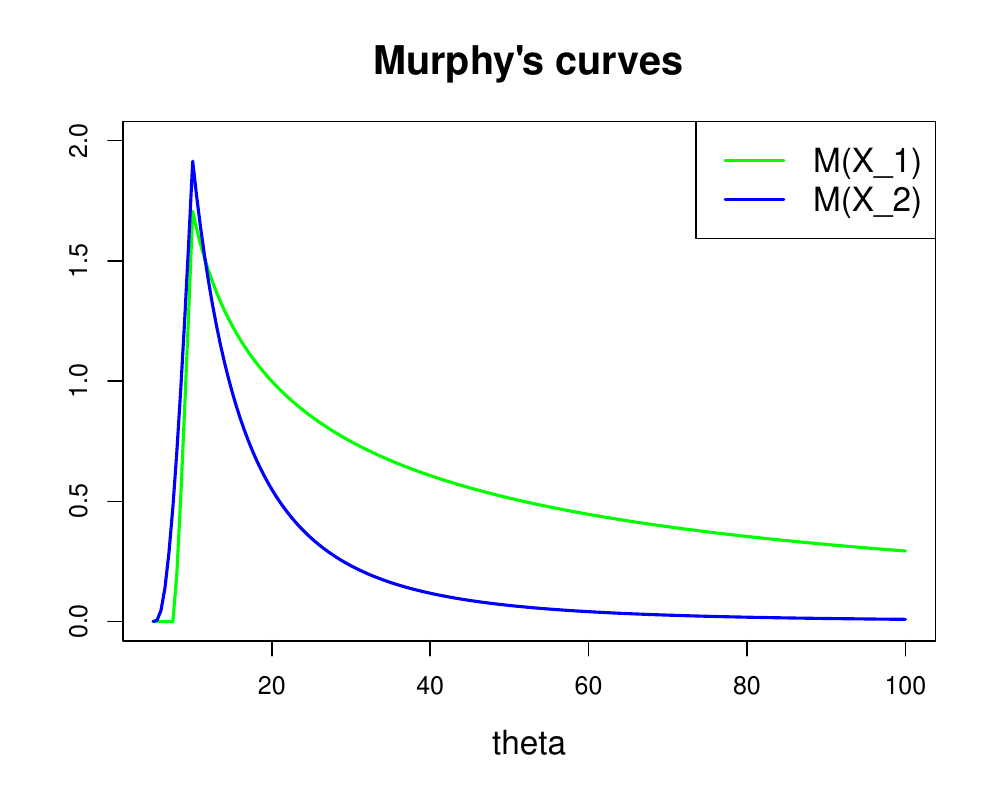}
    \end{minipage}
     \hspace{0.75cm}
    \begin{minipage}{0.45\textwidth}
        \centering
        \includegraphics[width=1.2\textwidth]{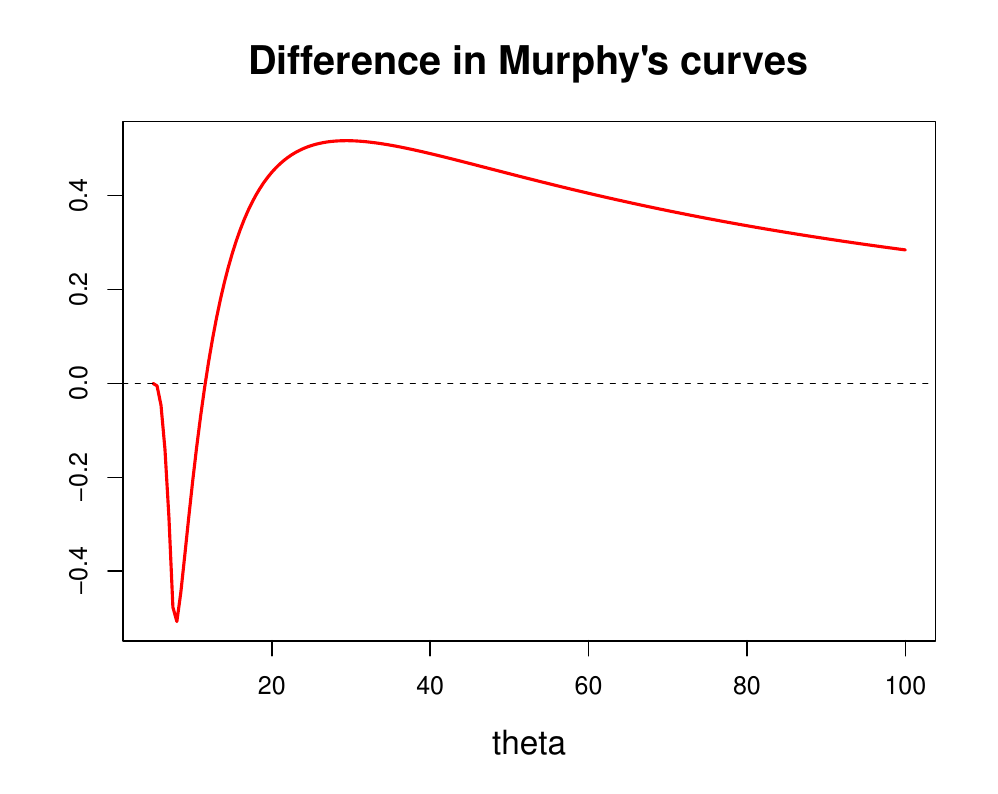}
    \end{minipage}
    \caption{Murphy's curves measuring the discriminatory power of the predictors $X_1$ and $X_2$ from \hyperref[ex_6]{Example 6}.}
    \label{fig_3_4}
\end{figure}

In \hyperref[ex_5]{Example 5}, $p\mapsto LC_p(X_1)$ and $p\mapsto LC_p(X_2)$ cross once from above. Proposition \ref{prop_murphy_gini_dominance} implies that the Murphy's curves $\theta\mapsto M_\theta(Y,X_1)$ and $\theta\mapsto M_\theta(Y,X_2)$ also cross once from above. The Murphy's curves $\theta\mapsto M_\theta(X_2, \be[Y])$ and $\theta\mapsto M_\theta(X_1, \be[Y])$, presented in Figure \ref{fig_3_4} cross once from above.
\qed

\medskip

Let us now focus on Lorenz curves and Murphy's curves that cross once. It turns out that the Lorenz curves of two predictors that cross once still offer an intuitive interpretation of discriminatory power of the predictors. For completeness, we present the following definition.

\begin{df}\label{def_lorenz_cross_once}
Let $X_1$ and $X_2$ denote two predictors. Their Lorenz curves $p\mapsto LC_p(X_1)$ and $p\mapsto LC_p(X_2)$ cross once from above, or $p\mapsto LC_p(X_1)$ crosses $p\mapsto LC_p(X_2)$ once from above, if there exists $p^*\in(0,1)$ such that
\beqo
LC_p(X_1)\geq LC_p(X_2), \ \textit{ for all $p\in[0,p^*]$} \quad \textit{and} \quad LC_p(X_1)\leq LC_p(X_2), \ \textit{ for all $p\in[p^*,1]$.} 
\eeqo
The same type of definition applies if we consider Murphy's curves that cross once from above.
\end{df}

Notice that $LC_p(X_1)\leq LC_p(X_2)$ for all $p\in[p^*,1]$ is equivalent to $\overline{LC}_p(X_2)=1-LC_{1-p}(X_2)\leq 1-LC_{1-p}(X_1)=\overline{LC}(X_1)$ for all $p\in[0,1-p^*]$, where $\overline{LC}_p(X)=\be[X\mathbf{1}\{X>F_X^{-1}(1-p)\}]/\be[X]$ is the descending order Lorenz curve above the diagonal. We postulate the following interpretation in the context of discriminatory power of predictors. Since the diagonal line in the space of Lorenz curves represents $X=\be[X]$, if $p\mapsto LC_p(X_1)$ and $p\mapsto LC_p(X_2)$ cross once from above for mean-calibrated $X_1$ and $X_2$, we may say that low values of $X_1$ have a lower discriminatory power than low values of $X_2$, and large values of $X_1$ have a larger discriminatory power than large values of $X_2$. This interpretation agrees with the interpretation presented in Section 3.

If Lorenz curves cross once, we have a so-called \emph{second-degree Lorenz dominance}, see, e.g., \cite{second_lorenz}, \cite{second_lorenz_2}.

\begin{lem}\label{lemma_2lorenz}
Let $X_1$ and $X_2$ denote two predictors. Assume $p\mapsto LC_p(X_1)$ and $p\mapsto LC_p(X_2)$ cross once from above. (i) ${\rm Gini}(X_1)\leq {\rm Gini}(X_2)$ iff
\beq\label{second_lorenz_up}
\int_0^pLC_u(X_1)du\geq \int_0^pLC_u(X_2)du,\quad \text{ for all } p\in[0,1].
\eeq
(ii) ${\rm Gini}(X_1)\geq {\rm Gini}(X_2)$ iff
\beq\label{second_lorenz_down}
\int_p^1(1-LC_u(X_2))du\leq \int_p^1(1-LC_u(X_1))du,\quad \text{ for all } p\in[0,1].
\eeq
\end{lem}

The relation $\Longrightarrow$ is proved in \cite[Theorem 2]{second_lorenz_2}. The relation $\Longleftarrow$ follows immediately if we choose $p=1$ in \eqref{second_lorenz_up}-\eqref{second_lorenz_down}. 

Following \cite{second_lorenz}, \cite{second_lorenz_2} and the interpretation of the Lorenz order \eqref{lorenz_dominance}, we say that $X_2$ \emph{second-degree upward Lorenz dominates} $X_1$, denoted by $X_1\preceq^2_{L} X_2$, if \eqref{second_lorenz_up} holds, and $X_1$ \emph{second-degree downward Lorenz dominates} $X_2$, denoted by $X_2\preceq^2_{\bar{L}} X_1$, if \eqref{second_lorenz_down} holds. The term upward dominance refers to the fact that the Lorenz curves are integrated from the left, and the term downward dominance refers to the fact that the Lorenz curves are integrated from the right. In the context of evaluating point predictions, we may use the following interpretation: if $X_2$ \emph{second-degree upward Lorenz dominates} $X_1$, then low predictions are of greater importance when evaluating discriminatory power of predictors, since low values of mean-calibrated $X_2$ have more discriminatory power than low values of mean-calibrated $X_1$ in the sense that $LC_p(X_1)\geq LC_p(X_2)$ for small $p$. If $X_1$ \emph{second-degree downward Lorenz dominates} $X_2$, then large predictions are of greater importance when evaluating of discriminatory power of predictors, since large values of mean-calibrated $X_1$ have more discriminatory power than large values of mean-calibrated $X_2$ in the sense that $1-LC_p(X_2)\leq 1-LC_p(X_1)$ for large $p$.

We now extend the above ideas and interpretations to Murphy's curves. We prove a result similar to Lemma \ref{lemma_2lorenz} formulated in terms of the discrimination statistics from Murphy's decomposition and Murphy's curves.

\begin{prop}\label{prop_2murphy}
Let (A), (B), (D) hold. Consider a Bregman divergence $L_H$ with representation \eqref{bregman_representation_2}, the discrimination statistics ${\rm DSC}_{L_H}$ from Murphy's decomposition of the expected loss w.r.t.~$L_H$ given with \eqref{discrimination X} and Murphy's curve \eqref{definition Murphy curve} for the elementary Bregman divergence. Assume that $\theta\mapsto M_\theta(Y,X_1)$ and $\theta\mapsto M_\theta(Y,X_2)$ cross once from above. (i) ${\rm DSC}_{L_H}(Y,X_1)\leq {\rm DSC}_{L_H}(Y,X_2)$ iff
\beq\label{second_murphy_up}
\int_0^pM_\theta(Y,X_1)dH(\theta)\geq \int_0^pM_\theta(Y,X_2)dH(\theta),\quad \text{ for all }p\geq 0,\nonumber\\ 
\int_0^pM_\theta(X_1,\be[Y])dH(\theta)\leq \int_0^pM_\theta(X_2, \be[Y])dH(\theta),\quad \text{ for all }p\geq 0.
\eeq
(ii) ${\rm DSC}_{L_H}(Y,X_1)\geq {\rm DSC}_{L_H}(Y,X_2)$ iff
\beq\label{second_murphy_down}
\int_p^\infty M_\theta(Y,X_1)dH(\theta)\leq \int_p^\infty M_\theta(Y,X_2)dH(\theta),\quad \text{ for all }p\geq 0,\nonumber\\
\int_p^\infty M_\theta(X_1,\be[Y])dH(\theta)\geq \int_p^\infty M_\theta(X_2, \be[Y])dH(\theta),\quad \text{ for all }p\geq 0.
\eeq
\end{prop}

We interpret Proposition \ref{prop_2murphy}. By assertion (iii) from Proposition \ref{prop_murphy_gini_dominance} and Definition \ref{def_lorenz_cross_once}, we conclude that $p\mapsto LC_p(X_1)$ and $p\mapsto LC_p(Y,X_2)$ cross once from above, consequently, low values of $X_1$ have less discriminatory power than low values of $X_2$, and large values of $X_1$ have more discriminatory power than large values of $X_2$. However, the second-degree Lorenz dominance cannot be established in the setting of Proposition \ref{prop_2murphy}. We proceed as follows. First, since $\theta\mapsto M_\theta(Y,X_1)$ and $\theta\mapsto M_\theta(Y,X_2)$ cross once from above, $\theta\mapsto M_\theta(X_2, \be[Y])$ and $\theta\mapsto M_\theta(X_1, \be[Y])$ cross once from above (recall $M_\theta(Y,X)=M_\theta(Y,\be[Y])-M_\theta(X,\be[Y])$ for mean-calibrated $X$). It holds
\beqo
M_\theta(X_1,\be[Y])&\leq& M_\theta(X_2, \be[Y]), \ \text{ for all $\theta\in[0,\theta^*]$,} \\
M_\theta(X_1,\be[Y])&\geq& M_\theta(X_2,\be[Y]), \ \text{ for all $\theta\in[\theta^*,\infty)$.} 
\eeqo
Next, by assumption (A), recall that
\beqo
M_\theta(X,\be[Y])=\be\Big[(X-\theta)^+-(\be[Y]-\theta)^+\Big]=\be\Big[\min(\be[Y],\theta)-\min(X,\theta)\Big].
\eeqo
Hence, $M_\theta(X_1, \be[Y])\leq M_\theta(X_2, \be[Y])$ for all $\theta\in[0,\theta^*]$ is equivalent to $\be\big[\min(X_1,\theta)\big]\geq \be\big[\min(X_2,\theta)\big]$ for all $\theta\in[0,\theta^*]$. The second inequality can be interpreted that $X_1$ has less dispersed values below $\theta\leq \theta^*$ than $X_2$, consequently low values of mean-calibrated $X_2$ have more discriminatory power than low values of mean-calibrated $X_1$. Finally, $M_\theta(X_1, \be[Y])\geq M_\theta(X_2, \be[Y])$ for all $\theta\in(\theta^*,\infty)$ is equivalent to $\be\big[(X_1-\theta)^+\big]\geq \be\big[(X_2-\theta)^+\big]$ for all $\theta\in(\theta^*,\infty)$. The second inequality can be interpreted that $X_1$ has more dispersed values above $\theta\geq\theta^*$ than $X_2$, consequently large values of mean-calibrated $X_1$ have more discriminatory power than large values of mean-calibrated $X_2$. These interpretations of the discriminatory power of the predictors agree with the interpretations based on the Lorenz curves. Following Lemma \ref{lemma_2lorenz}, we can now introduce {\it a second-degree Murphy's dominance}. We say that $X_2$ \emph{second-degree upward Murphy dominates} $X_1$ if \eqref{second_murphy_up} holds, and $X_1$ \emph{second-degree downward Murphy dominates} $X_2$ if \eqref{second_murphy_down} holds. The term upward dominance again refers to the fact that Murphy's curves are integrated from the left, and the term downward dominance refers to the fact that Murphy's curves are integrated from the right. If $X_2$ \emph{second-degree upward Murphy dominates} $X_1$, then low predictions are of greater importance when evaluating discriminatory power of predictors, since low values of mean-calibrated $X_2$ have more discriminatory power than low values of mean-calibrated $X_1$ in the sense that $M_\theta(X_1, \be[Y])\leq M_\theta(X_2, \be[Y])$ for small $\theta$. If $X_1$ \emph{second-degree downward Murphy dominates} $X_2$, then large predictions are of greater importance when evaluating discriminatory power of the predictors, since large values of mean-calibrated $X_1$ have more discriminatory power than large values of mean-calibrated $X_2$ in the sense that $M_\theta(X_1, \be[Y])\geq M_\theta(X_2, \be[Y])$ for large $\theta$. Clearly, the second-degree Murphy's dominance depends on the choice of $H$ used in measuring the performance of the predictors with Bregman divergences $L_H$. We state the following corollary deduced from Proposition \ref{prop_2murphy}. In contrast to the second-order Lorenz dominance, the second-order Murphy's dominance is in line with the decision-theoretic approach of the evaluation of point predictions from \cite{Gneiting}.

\begin{cor}\label{cor_2murphy_score}
Let (A), (B), (D) hold. Assume $p\mapsto LC_p(X_1)$ and $p\mapsto LC_p(X_2)$ cross once from above. Consider a class $\mathcal{L}$ of Bregman divergences $L$.
(i) $X_2$ second-degree upward Murphy dominates $X_1$ under all Bregman divergences $L\in\mathcal{L}$ iff $X_2$ is preferred over $X_1$ under all Bregman divergences $L\in\mathcal{L}$. (ii) 
$X_1$ second-degree downward Murphy dominates $X_2$ under all Bregman divergences $L\in\mathcal{L}$ iff $X_1$ is preferred over $X_2$ under all Bregman divergences $L\in\mathcal{L}$.
\end{cor}

Our final goal is to prove a weaker version of Proposition \ref{prop_gini_dominance} in the framework of Lorenz curves crossing once. First, we relate third-degree stochastic dominance for distributions with ranking of distributions in terms of expected utility of the payoff. Such a result is known; see \cite[Lemma 7]{role_of_variance} and \cite[Theorem 3.3]{dominance_book}. However, compared to the cited results from the literature, we consider classes of test functions which match our application and handle distributions with unbounded positive support, which requires handling limits at the end points of the distribution's support. Secondly, we use this result to modify Proposition \ref{prop_gini_dominance}.

Let us consider the classes of functions
\beq\label{utility_class}
\mathcal{U}&=&\big\{U:[0,\infty)\mapsto[-\infty,\infty];\nonumber\\ 
&& U \ \text{is piece-wise monotone on } [0,\infty) \ \text{and } \mathcal{C}^3((0,\infty)),\nonumber\\
&& U''(x)\geq 0, \ U'''(x)\leq 0 \ \text{ for all } x>0 \big\},
\eeq
and
\beq\label{utility_class_2}
\mathcal{V}&=&\big\{U:[0,\infty)\mapsto[-\infty,\infty]; \nonumber\\
&& U \ \text{is piece-wise monotone on } [0,\infty) \ \text{and } \mathcal{C}^3((0,\infty)),\nonumber\\
&& U''(x)\geq 0, \ U'''(x)\geq 0 \ \text{ for all } x>0 \big\}.
\eeq
We assume that $U(0)=\lim_{x\rightarrow 0}U(x)$ and we allow for infinite $|U(0)|$. In particular, $U(x)=x^2\in\mathcal{U}$ and $\in\mathcal{V}$. We point out that these two classes are more general than the class of utility functions considered in \cite{role_of_variance} and \cite{dominance_book}, they consider $U'\geq 0$, $U''\leq 0$, $U'''\geq 0, U\in\mathcal{C}^3$ and $U$ supported on a finite interval.

\begin{lem}\label{lem_order_dist_quant_utility}
Let (A)-(C) hold.  Assume that $\int_0^\infty |U(x)|dF_{X_1}(x)<\infty$ and $\int_0^\infty |U(x)|dF_{X_2}(x)<\infty$ for all $U\in\mathcal{U}$ and $U\in\mathcal{V}$. We have the equivalent relations
\beq\label{order_dist_quant_up_utility}
\lefteqn{\int_0^u\int_0^v\big(F_{X_1}(t)-F_{X_2}(t)\big)dtdv\geq 0,\quad \text{ for all } u\geq 0,}\nonumber\\
& \Longleftrightarrow & \be[U(X_1)]\geq \be[U(X_2)],\qquad \text{for all $U\in\mathcal{U}$,}
\eeq
and
\beq\label{order_dist_quant_downn_utility}
\lefteqn{\int_u^\infty \int_v^\infty\big(F_{X_1}(t)-F_{X_2}(t)\big)dtdv\geq 0,\quad \text{ for all } u\geq 0,}\nonumber\\
& \Longleftrightarrow & \be[U(X_1)]\leq \be[U(X_2)],\qquad \text{for all $U\in\mathcal{V}$.}
\eeq
\end{lem}

The relation \eqref{order_dist_quant_up_utility} characterizes  \emph{third-degree stochastic dominance} for distributions, it  is the adaptation of the results from \cite{role_of_variance} and \cite{dominance_book} into our setting. The result \eqref{order_dist_quant_downn_utility} has not been investigate by \cite{role_of_variance} and \cite{dominance_book}. By \eqref{variance_formula} in the appendix we observe the relation between \eqref{order_dist_quant_up_utility} and \eqref{order_dist_quant_downn_utility}
\beq\label{variance_formula_2}
\lefteqn{\int_u^\infty\int_v^\infty\big(F_{X_2}(t)-F_{X_1}(t)\big)dtdv}\nonumber\\
&=&\int_0^u\int_0^v\big(F_{X_2}(t)-F_{X_1}(t)\big)dtdv-\frac{1}{2}\big(Var[X_2]-Var[X_1]\big),\quad  u\geq 0.
\eeq

We present the fourth main result of this paper.

\begin{thm}\label{prop_gini_dominance_2nd}
Let (A)-(D) hold. Assume $p\mapsto LC_p(X_1)$ and $p\mapsto LC_p(X_2)$ cross once from above. We have
\beq\label{2lorenz_dominance}
\lefteqn{\int_0^u\int_0^v\big(F_{X_2}(t)-F_{X_1}(t)\big)dtdv\geq 0,\quad \text{ for all } u\geq 0} \nonumber\\
&\Longleftrightarrow& \quad Var(X_1)\leq Var(X_2)\nonumber\\
&\Longleftrightarrow& \quad S_{L_\phi}(Y,X_1)\geq S_{L_\phi}(Y,X_2) \quad \text{for all $L_\phi$ with $\phi\in\mathcal{U}$}\nonumber\\
&\Longleftrightarrow& \quad {\rm DSC}_{L_\phi}(Y,X_1)\leq {\rm DSC}_{L_\phi}(Y,X_2) \quad \text{for all $L_\phi$ with $\phi\in\mathcal{U}$},
\eeq
where for all $L_\phi$ with $\phi\in\mathcal{U}$ means for all Bregman divergences with representation \eqref{bregman_representation_2} with the function $x\mapsto\phi(x)$ which belongs to \eqref{utility_class}, and
\beq\label{2lorenz_dominance_2}
\lefteqn{\int_u^\infty\int_v^\infty\big(F_{X_2}(t)-F_{X_1}(t)\big)dtdv\geq 0,\quad \text{ for all } u\geq 0} \nonumber\\
&\Longleftrightarrow& \quad Var(X_1)\geq Var(X_2)\nonumber\\
&\Longleftrightarrow& \quad S_{L_\phi}(Y,X_1)\leq S_{L_\phi}(Y,X_2) \quad \text{for all $L_\phi$ with $\phi\in\mathcal{V}$}\nonumber\\
&\Longleftrightarrow& \quad {\rm DSC}_{L_\phi}(Y,X_1)\geq {\rm DSC}_{L_\phi}(Y,X_2) \quad \text{for all $L_\phi$ with $\phi\in\mathcal{V}$},
\eeq
where for all $L_\phi$ with $\phi\in\mathcal{V}$ means for all Bregman divergences with representation \eqref{bregman_representation_2} with the function $x\mapsto\phi(x)$ which belongs to \eqref{utility_class_2}.
\end{thm}

Compared to Proposition \ref{prop_gini_dominance} which concerns non-intersecting Lorenz curves and characterizes Bregman dominance between two mean-calibrated predictors measured with any Bregman divergence, Theorem \ref{prop_gini_dominance_2nd} establishes Bregman dominance between two mean-calibrated predictors within a class of Bregman divergences, in the case when the Lorenz curves cross once. By Corollary \ref{cor_2murphy_score}, Theorem \ref{prop_gini_dominance_2nd} also characterizes a class of Bregman divergences under which the second-degree Murphy's dominance holds. We point out that the variance, and not the Gini index, is decisive in determining the Bregman dominance in Theorem \ref{prop_gini_dominance_2nd}. This gives another argument for not using the Gini index as a discrimination measure of a predictor. This conclusion formulated by us in the context of evaluating point predictions agrees with \cite{role_of_variance} and \cite{zoli} in the context of the inequality evaluation. We remark that the second-degree Lorenz dominance does not agree with the third-degree stochastic dominance.

The question remains which Bregman divergences are in $\mathcal{U}$ and $\mathcal{V}$. In applications we usually consider Tweedie's deviance loss functions constructed with 
\beq\label{tweedie_loss}
\phi_p(x)=
\begin{cases}
  \frac{x^{2-p}}{(1-p)(2-p)}, & \qquad \text{Tweedie's loss with } p\in\br\setminus\{0,1,2\},\\
  x\log(x)-x, & \qquad \text{Poisson loss} \ (\text{Tweedie's loss with } p=1),\\
  -\log(x), &\qquad \text{Gamma loss} \ (\text{Tweedie's loss with } p=2),\\
  x^2,&\qquad \text{quadratic loss} \ (\text{Tweedie's loss with } p=0).
\end{cases}
\eeq
The family of Tweedie's deviance loss functions agrees with Patton's family of loss functions discussed by \cite{GneitingResin}. We can now verify that the Tweedie's deviance loss functions with $p\geq 0$ are in $\mathcal{U}$, and the Tweedie deviance loss functions with $p\leq 0$ are in $\mathcal{V}$. There is one point that deserves some attention, we can define Tweedie's deviance losses for all $p\in \R$, thus, Patton's family, however for $p\in (0,1)$ there is no proper distribution within the exponential dispersion family that induces this corresponding Tweedie's loss as a deviance loss.

\medskip

\begin{figure}[ht]
    \centering
    \hspace{-1.5cm}
    \begin{minipage}{0.45\textwidth}
        \centering
        \includegraphics[width=1.2\textwidth]{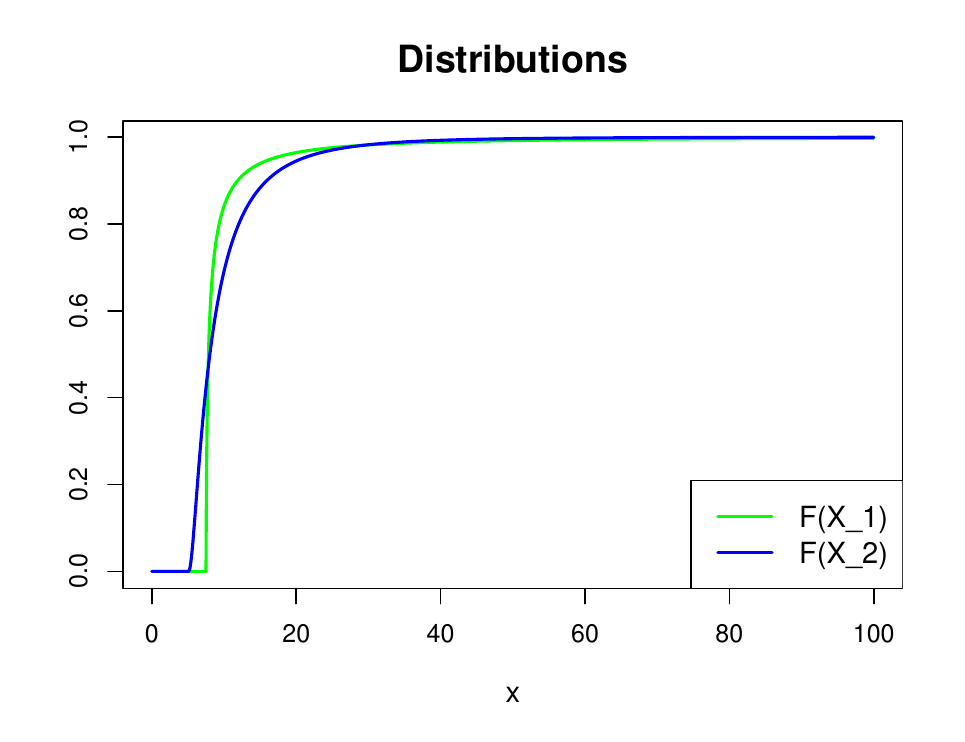}
    \end{minipage}
     \hspace{0.75cm}
    \begin{minipage}{0.45\textwidth}
        \centering
        \includegraphics[width=1.2\textwidth]{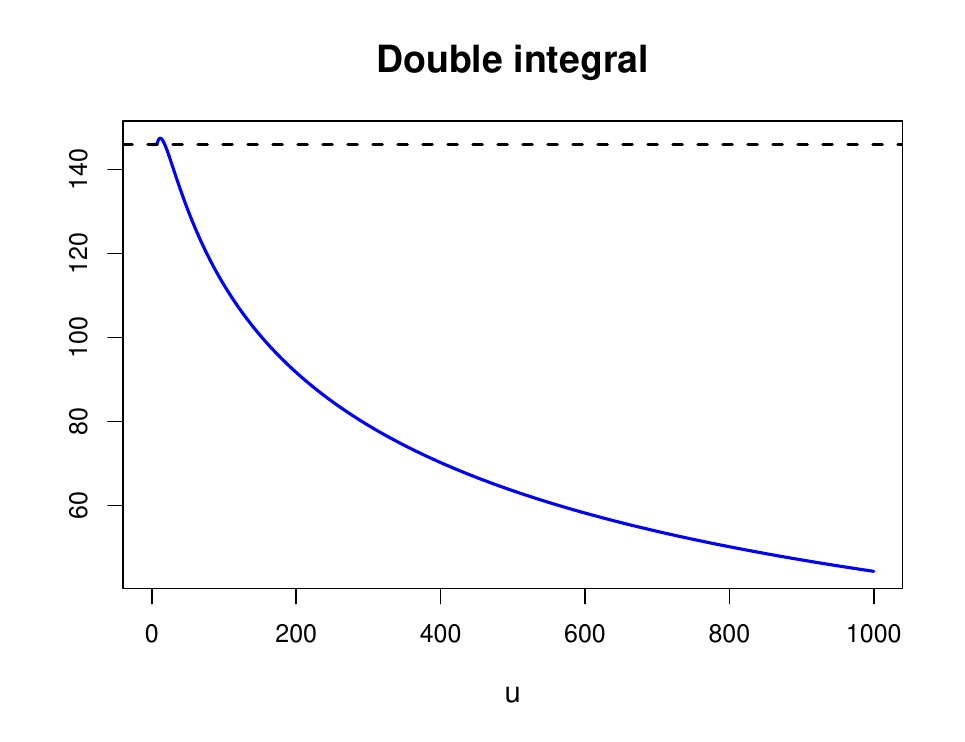}
    \end{minipage}
    \caption{LHS: The distributions $F_{X_1}$ and $F_{X_2}$ of the predictors $X_1$ and $X_2$ from \hyperref[ex_7]{Example 7}. RHS: The double integral $u\mapsto \int_u^\infty\int_v^\infty(F_{X_2}(t)-F_{X_1}(t))dtdv$ of the difference in the distributions of the predictors $X_1$ and $X_2$.}
    \label{fig_5_6}
\end{figure}

\noindent \phantomsection\label{ex_7}\textit{Example 7}. We continue \hyperref[ex_5]{Example 5} and \hyperref[ex_6]{Example 6}, and use the results presented before. In the left panel of Figure \ref{fig_5_6} we present the distributions of the predictors $X_1$ and $X_2$. The distributions cross twice, which agrees with Proposition \ref{prop_murphy_gini_dominance} (recall that we have shown that the Lorenz curves of $X_1$ and $X_2$ cross once). In this example the Lorenz curve of $X_1$ crosses the Lorenz curve of $X_2$ from above and $Var(X_1)>Var(X_2)$. Hence, we are in the framework of \eqref{2lorenz_dominance_2} of Theorem \ref{prop_gini_dominance_2nd}. In the right panel of Figure \ref{fig_5_6} we present the double integral \eqref{2lorenz_dominance_2} of the difference in the distributions of the predictors $X_1$ and $X_2$ calculated numerically. The dashed line shows $\frac{1}{2}(Var(X_1)-Var(X_2))=146.01$, which is $\int_0^\infty\int_v^\infty(F_{X_2}(t)-F_{X_1}(t))dtdv$ by \eqref{variance_formula_2}.

\begin{figure}[ht]
    \centering
    \hspace*{-0.52cm}
    \includegraphics[width=1.03\textwidth]{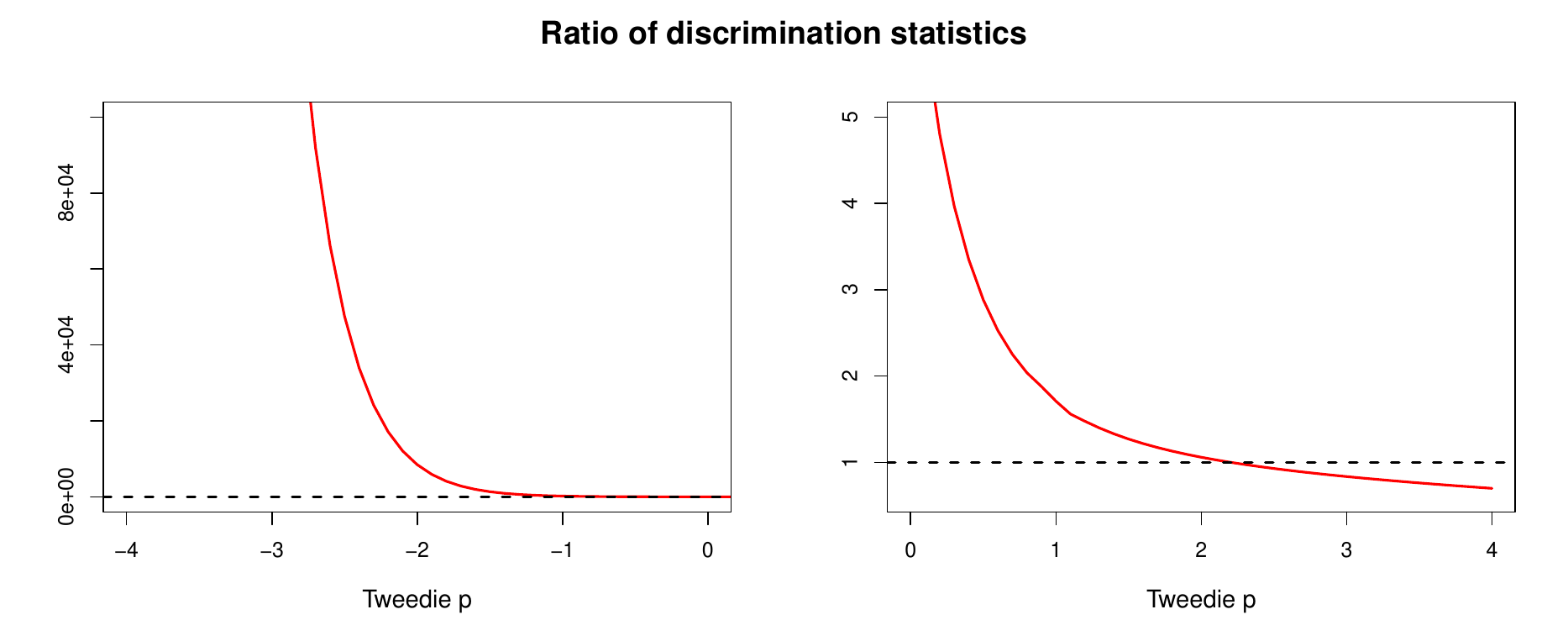}
    \caption{The ratio ${\rm DSC}_{L_{\phi_p}}(Y,X_1)/{\rm DSC}_{L_{\phi_p}}(Y,X_2)$ of the discrimination statistics of the predictors $X_1$ and $X_2$ from \hyperref[ex_7]{Example 7}.}
    \label{fig_7}
\end{figure}

In Figure \ref{fig_7} we estimate the discrimination measure ${\rm DSC}_{L_\phi}(Y,X)=S_{L_\phi}(X,\be[Y])=\be[\phi(X)]-\phi(\be[X])$ with Monte-Carlo simulations, assuming that $X$ is a mean-calibrated predictor for $Y$. We use the Tweedie's deviance loss functions constructed with $\phi_p$ specified in \eqref{tweedie_loss}. We present the relation between ${\rm DSC}_{L_{\phi_p}}(Y,X_1)$ and ${\rm DSC}_{L_{\phi_p}}(Y,X_2)$ as a function of the Tweedie parameter $p$. Since $Var(X_1)>Var(X_2)$, we confirm that ${\rm DSC}_{L_{\phi_p}}(Y,X_1)>{\rm DSC}_{L_{\phi_p}}(Y,X_2)$ for $p\leq 0$ which is expected in the view of Theorem \ref{prop_gini_dominance_2nd}. We conclude that the predictor $X_1$ dominates $X_2$ in the sense of Bregman dominance under the Tweedie's deviance loss functions with $p\leq 0$. We note that ${\rm DSC}_{L_{\phi_p}}(Y,X_1)>{\rm DSC}_{L_{\phi_p}}(Y,X_2)$ also holds for $p\in(0,2]$, but does not hold for $p>2$.\qed

\section{Conclusions}
\label{conclusions section}

This paper has studied the evaluation of point predictions of a response, or point estimates of a response's mean value, from a decision-theoretic perspective. Our starting point has been consistency of loss functions for mean estimation which coincides with the class of Bregman divergences. Within this framework, the predictive accuracy is naturally measured by the expected loss, while calibration and discrimination measures arise as structural properties of a predictor that should be analyzed jointly with the accuracy measure. The main message of the paper supports that Murphy's decomposition of the expected loss provides the suitable principle for this task, and that several popular alternatives based on Lorenz and concentration curves, although informative and often appealing in practice, should be interpreted with caution when they are used for ranking predictors.

Our first main contribution is a new representation of Murphy's decomposition. We have shown that the discrimination term (DSC) and the miscalibration term (MCB) can be represented without directly including the responses, but solely considering their predictors. This reformulation is conceptually important. It shows that discrimination is the discrepancy between the calibrated predictor and the unconditional mean, whereas miscalibration is the discrepancy between the calibrated predictor and the original predictor. As a consequence, Murphy's decomposition is not only an algebraic identity: it gives a direct and interpretable geometry of forecast evaluation.

Our second contribution concerns miscalibration measures. We have revisited the area between the concentration and Lorenz curves, ABC, and we have introduced the mean squared analogue ABC$^2$. The analysis reveals an important asymmetry. While ABC$^2=0$ is equivalent to mean-calibration, ABC$=0$ is only a necessary condition and may occur even when the predictor is not mean-calibrated. Thus, ABC does not support a desired interpretation of ``smaller value means better calibration'' in general. Moreover, we have also shown that both ABC and ABC$^2$ can be represented through predictor-dependent weights of expected losses, thus, they fail to be aligned with the decision-theoretic notion of mean consistent loss function. Therefore, these quantities cannot be treated as universal mean-consistent scoring functions in the sense of forecast evaluation. From this perspective, the miscalibration term from Murphy's decomposition is preferable because it compares predictors under a pre-specified Bregman divergence. Our counter-example for weighted Bregman divergences with predictor-dependent weights reinforces this conclusion by showing that such weights may even reverse the ordering between a truthful predictor and a distorted one.

Our third contribution concerns discrimination. For mean-calibrated predictors, we have related the Gini index to Murphy's discrimination term and shown that the Gini index can be decomposed into a Murphy-type discrimination component, built from a predictor-specific Bregman divergence, plus a mean absolute deviation term. This includes the predictor in the discrimination measure in an unwanted way, leading to the same conclusions as for ABC and ABC$^2$, namely, the discrimination term from Murphy's decomposition should be preferred over the Gini index to assess discrimination.

The final part of the paper addresses the practically relevant case where Lorenz curves intersect. In practice, strict Lorenz dominance is rare, so an evaluation theory that only covers non-intersecting curves is too restrictive. We have shown that, for mean-calibrated predictors, Lorenz curves and Murphy's curves have the same number of crossings. This creates a precise bridge between geometric comparisons in the Lorenz space and dominance comparisons in the Murphy-diagram space. When the curves cross once, we have introduced weaker notions of upward and downward local dominance and characterized the corresponding ranking conditions. Most importantly, we have proved a weaker version of forecast dominance for mean-calibrated predictors for subclasses of Bregman divergences. In this single-crossing setting, the direction of dominance is determined by a double integral of the difference in distributions of predictors and, equivalently, by the variance of predictors. This result complements the classical Lorenz theory for forecast dominance and shows that, once Lorenz curves intersect, the Gini index is no longer the decisive object for discrimination; rather, the admissible class of loss functions and the associated curvature properties become central.

Several directions for future research emerge from our analysis. A first direction concerns the choice of the Bregman divergence. Our new Murphy's decomposition suggests that the tail behavior of the response may play a different role for evaluation of miscalibration and discrimination than for selecting a loss function to achieve prediction accuracy. A systematic study of which Bregman divergences are most informative or robust in finite-sample validation remains open. A second direction is to extend the analysis beyond mean prediction, e.g., to quantiles, expectiles, or full predictive distributions. Finally, on the dominance side, the present paper gives a detailed treatment of the case of one crossing. A broader theory covering multiple crossings, richer local dominance concepts, and sharper characterizations of admissible loss classes would further deepen the connection between forecast evaluation and stochastic order theory.

\bibliographystyle{agsm}
\bibliography{bibliography_miscalibration_discrimination} 

\begin{appendices}

\section{Numerical example with real data set}
\label{app_real}

We illustrate our results on a real data set to show that the theoretical findings presented in this paper have important implications for real-world applications.

We study the French motor insurance data set {\tt freMTPL2freq} from the {\sf R} package CASdatasets \cite{dutang}, which is widely used in actuarial applications. The response variable is a claim frequency, which is supported by a range of explanatory variables. In this example, we use a sample of 100,000 observations. We fit four predictive models:
\begin{verbatim}
Model 1 : ClaimNb ~ offset(log(Exposure)) + AreaGLM + VehPowerGLM + Region +
           VehAgeGLM + DrivAgeGLM + BonusMalusGLM + DensityGLM +
           VehBrand + VehGas,
Model 2 : ClaimNb ~ offset(log(Exposure)) + VehAge + DrivAge,
Model 3 & 4 : ClaimNb ~ offset(log(Exposure)) 
                + Area + VehPower + Region + VehAge + DrivAge 
                + BonusMalus + Density + VehBrand + VehGas.
\end{verbatim}

\begin{figure}[htb!]
    \centering
    \hspace*{-0.52cm}
    \includegraphics[width=\textwidth]{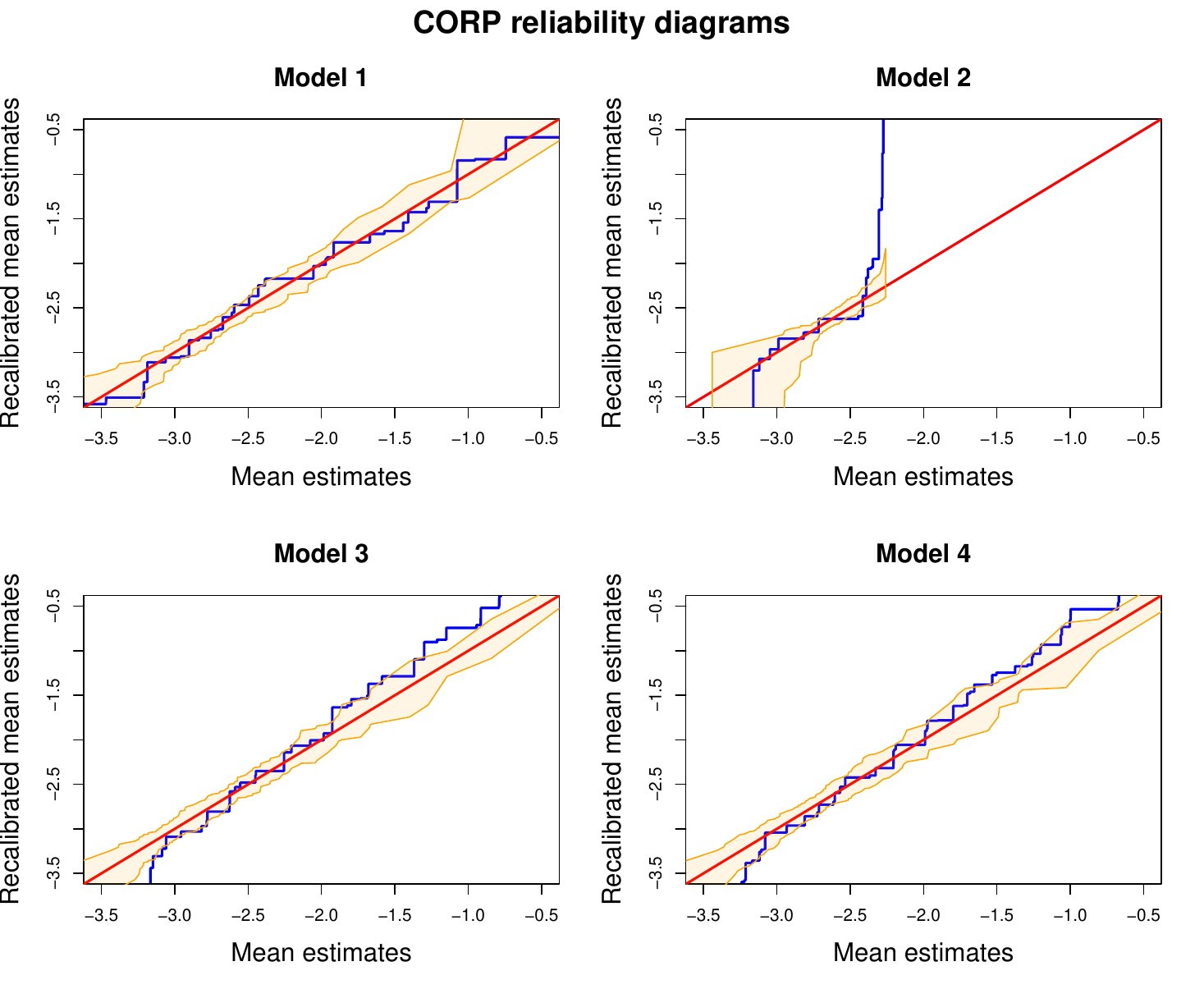}
    \caption{CORP reliability diagrams for the predictions from Models 1--4, together with 95\% point-wise consistency bands; all graphs are on the log-scale.}
    \label{fig_corp_4}
\end{figure}

For Models 1 and 2 we use generalized linear models being fitted by minimizing the Poisson deviance loss weighted with the exposures. The features with suffix 'GLM' are categorical features constructed based on the continuous features.  Model 3 is fitted using boosting trees, where the exposures are updated with the current predictions at each iteration, and the next weak learner is selected by minimizing the Poisson deviance loss weighted with the updated exposure, see \cite[Section 7.4.2]{WMBuser}. Model 4 is fitted using classical gradient boosting trees, where the model minimizes the mean squared error of the gradients, weighted with the exposures, at each iteration to determine the next weak learner. Thus, Models 3 and 4 differ in how the boosting trees are selected. In all cases we use the log-link functions. The predictions from Models 3 and 4 are rescaled so that their mean matches the mean of the response variable in the training data, ensuring global in-sample unbiasedness for all four models. The explanatory variables and the modelling of the claim frequency for this data set are discussed in detail in Sections 5.2.4 and 13.1 of \cite{WM2023}. The exposure corresponds to the duration over which each insurance policy is active. Since exposures are taken into account, we use the methods of \cite{wuthrich_gini_2} to construct the empirical Lorenz and concentration curves.

\begin{figure}[htb!]
    \centering
    \hspace*{-0.52cm}
    \includegraphics[width=\textwidth]{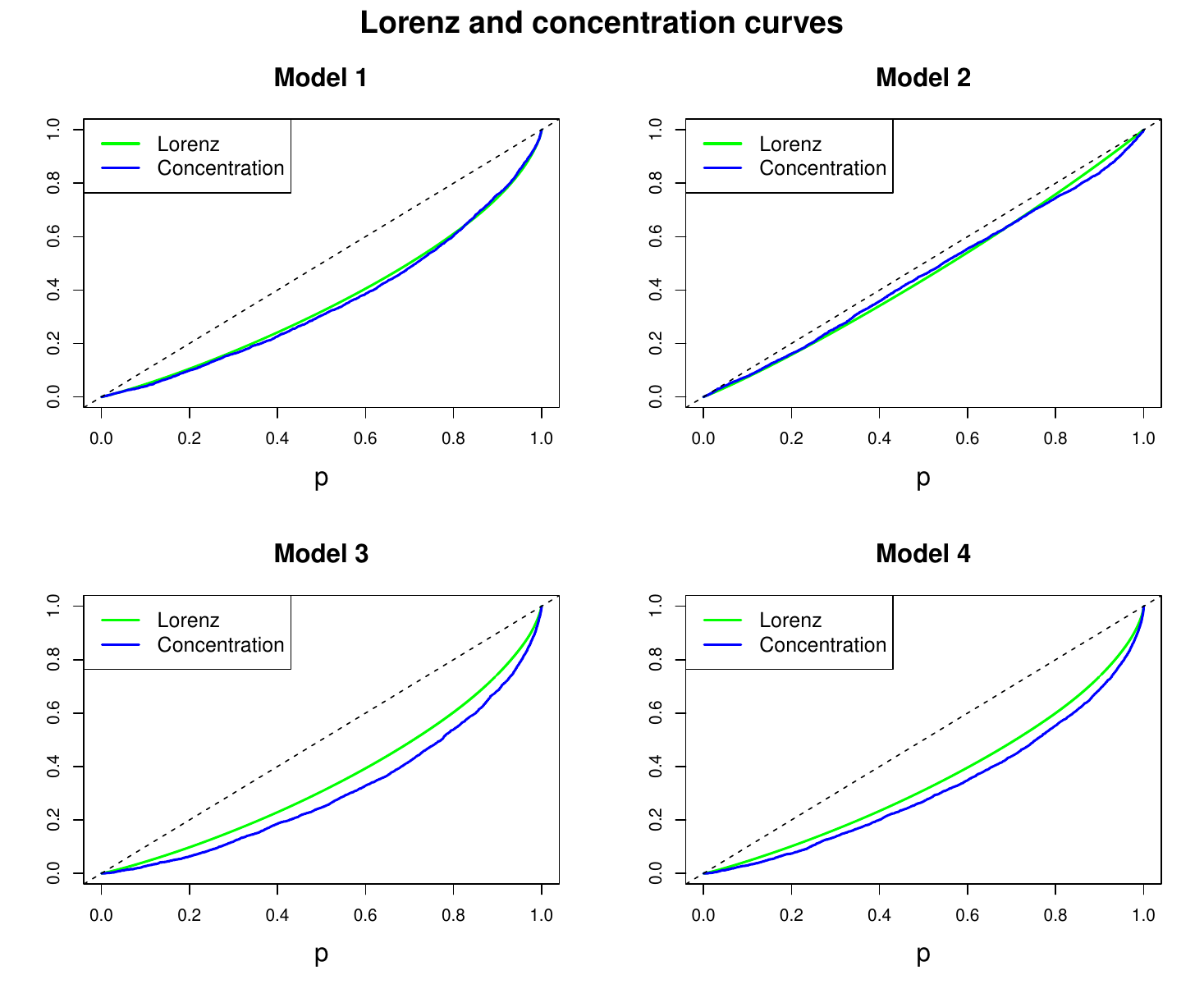}
    \caption{Lorenz and concentration curves for the predictions from Models 1--4.}
    \label{fig_lorenz_conc_4}
\end{figure}

First, we investigate whether the models are miscalibrated. Although CORP reliability diagrams are not the primary focus of this paper, we begin our case study by analysing them. The CORP reliability diagram serves as a valuable tool for graphically assessing the calibration of a predictor, see \cite{corp} and \cite{tryptych}, and it helps us interpret the numerical values of the miscalibration statistics presented later. In Figure \ref{fig_corp_4}, we present the CORP reliability diagrams for the predictions of the four models, together with 95\% point-wise consistency bands. The interpretation of CORP reliability diagrams is straightforward - large deviations from the diagonal suggest that a predictor is miscalibrated. To increase confidence in the decisions regarding the calibration of our predictors, we construct consistency bands using the parametric approach of \cite{delong_iso}. The consistency bands in Figure \ref{fig_corp_4} are generated by sampling the number of claims from negative binomial distributions with $Var[Y_i]=\phi\be[Y_i]$, where the dispersion parameter $\phi$ is estimated using Pearson's chi-square statistics. In Figure \ref{fig_lorenz_conc_4}, we display the Lorenz curve and the concentration curve; the discrepancy between the curves provides another graphical assessment of the calibration property.

\begin{figure}[htb!]
    \centering
    \hspace*{-0.52cm}
    \includegraphics[width=1.03\textwidth]{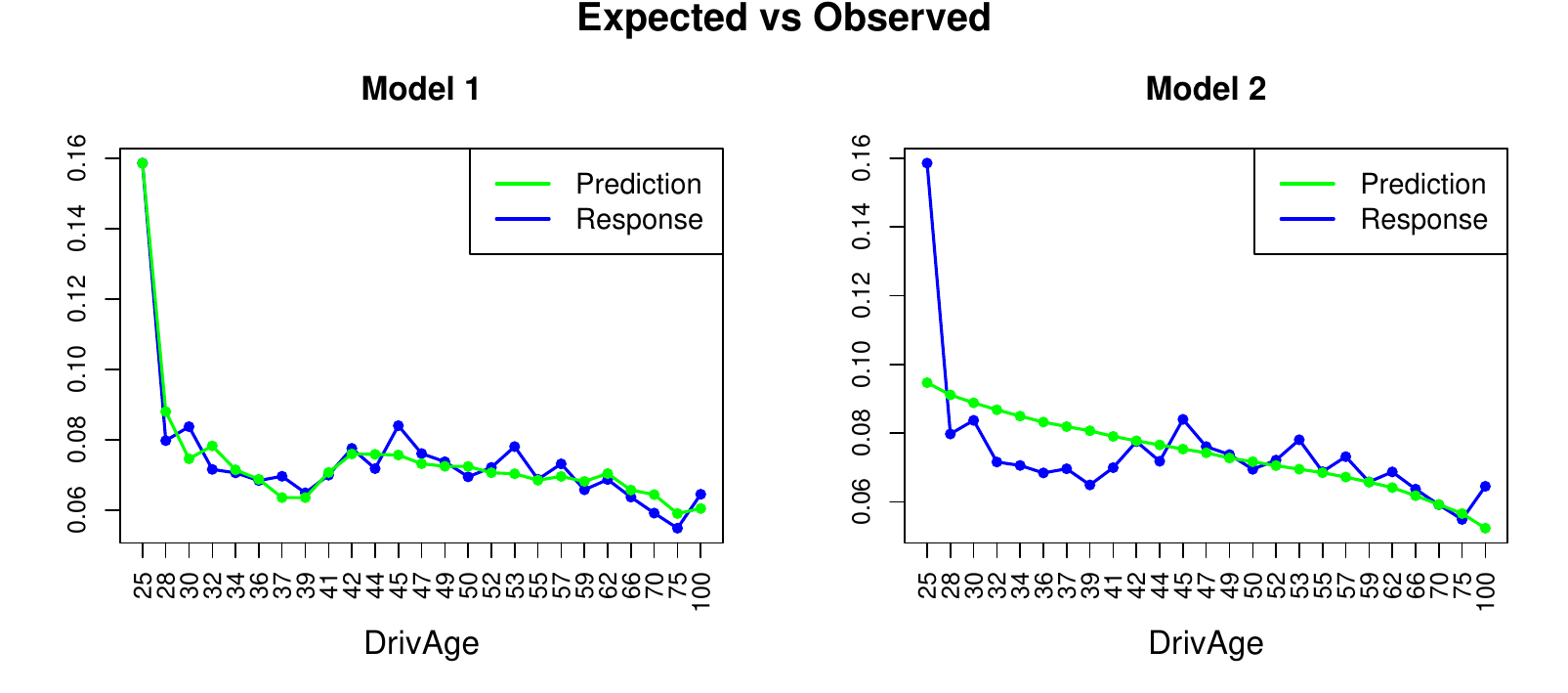}
    \caption{Mean predictions from Models 1 and 2 versus mean responses, stratified by VehAge.}
    \label{fig_exp_obs}
\end{figure}

We focus on Models 1 and 2. Based on Figures \ref{fig_corp_4}--\ref{fig_lorenz_conc_4}, we expect that Model 1 is well calibrated and Model 2 is miscalibrated. To confirm the poor fit of Model 2, in Figure \ref{fig_exp_obs} we also compare the mean predictions with the mean observations, weighted by exposures, stratified by DrivAge. We use 25 bins, to which we allocate our observations based on increasing values of the feature DrivAge. The same bins are used for both Model 1 and Model 2. It is now evident why Model 2 is miscalibrated. The log-linear functional form of the regression function assumed in Model 2, based on the original features VehAge and DrivAge, is inadequate. In contrast, in Model 1 these two variables are modelled in a categorical fashion by binning the variables into age bins. It turns out that Model 2 underestimates the claim frequency for the youngest drivers and overestimates the claim frequency for drivers aged 28--42. Please note that young drivers have a high predicted claim frequency in Model 2; hence, this miscalibration is observed in the right tail of the CORP reliability diagram, the Lorenz curve, and the concentration curve. Summing up, we are confident in stating that Model 1 is well calibrated and Model 2 is miscalibrated. 

\begin{table}[ht]
\small{
\centering
\begin{tabularx}{\textwidth}{>{\centering\arraybackslash}X 
                              >{\centering\arraybackslash}X 
                              >{\centering\arraybackslash}X 
                              >{\centering\arraybackslash}X 
                              >{\centering\arraybackslash}X}
  \toprule
  Model & Poisson MCB & MSE MCB & ABC & ${\rm ABC}^2$ \\ 
  \midrule
  Model 1 & 8.0435 & 1.3884 & 0.0081 & 0.0001 \\ 
  Model 2 & 12.3832 & 1.5163 & -0.0016 & 0.0002 \\ 
  Model 3 & 22.7809 & 5.8986 & 0.0477 & 0.0027 \\ 
  Model 4 & 15.5668 & 7.4663 & 0.0352 & 0.0014 \\
  \bottomrule
\end{tabularx}
\caption{Miscalibration measures MCB from Murphy's decomposition of the Poisson deviance loss and the mean squared loss ($10^{-4}$), the ABC, and the ${\rm ABC}^2$ measures for the predictions from Models 1--4.}
\label{tab_calibration}
}
\end{table}

We now turn to the proposed miscalibration measures. In Table \ref{tab_calibration}, we report the miscalibration statistics based on Murphy's decomposition of the Poisson deviance loss and the mean squared error loss, as well as the ABC and ${\rm ABC}^2$ statistics. The ABC measure gives a misleading result in this case, suggesting that Model 2 is better calibrated than Model 1. The reason for this is that very small positive differences between the concentration curve and the Lorenz curve observed over the range $p\in[0.2, 0.7]$ compensate larger negative differences observed for $p\in[0.7,1]$, exactly as in \hyperref[ex_1]{Example 1}. In contrast, the miscalibration measures derived from Murphy's decomposition and the ${\rm ABC}^2$ measure correctly indicate that Model 1 is better calibrated than Model 2, which agrees with our intuition and findings based on the validation plots. As in \hyperref[ex_2]{Example 2} and \hyperref[ex_3]{Example 3}, ABC gives a different ranking of the two predictors than the miscalibration statistics from Murphy's decomposition and ${\rm ABC}^2$, and it fails to provide the correct ranking.

Next, we study Models 3 and 4. By investigating Figures \ref{fig_corp_4}--\ref{fig_lorenz_conc_4}, in particular the magnitude of the deviation of the CORP reliability diagram from the diagonal and the discrepancy between the Lorenz curve and the concentration curve, we conclude that Models 3 and 4 are miscalibrated, with Model 3 appearing to be worse calibrated than Model 4. This expectation is confirmed by the miscalibration statistics derived from Murphy's decomposition of the Poisson deviance loss, as well as by the ABC and ${\rm ABC}^2$ measures. The ranking implied by the miscalibration statistics based on Murphy's decomposition of the mean squared error loss is different and suggests that Model 4 is worse calibrated than Model 3. More importantly, all measures reported in Table \ref{tab_calibration} for Models 3--4 are larger than those for Models 1--2, which fully agrees with our intuition and indicates the miscalibration of Models 3--4.

Next, we investigate the discrimination of the models. We apply isotonic regression to obtain calibrated versions of Models 2--4; that is we use the isotonic regression, which underlies the CORP reliability diagram, to estimate the mapping $X\mapsto\be[Y|X]$. To preserve continuity of the distributions of the recalibrated predictors, i.e. assumption (B), we replace the resulting step functions from the isotonic regression with piecewise linear approximations, where we connect linearly the midpoints of the flat regions of the step functions. This transformation leads to a loss of the global in-sample unbiasedness of the recalibrated predictors guaranteed by the isotonic regression; therefore, we rescale the predictions so that their means match the empirical means of the response variables. In fact, this scaling is negligible and is below 1\% for Models 2--4. We perform this global scaling to ensure assumption (A).

\begin{figure}[htb!]
    \centering
    \includegraphics[width=1.03\textwidth]{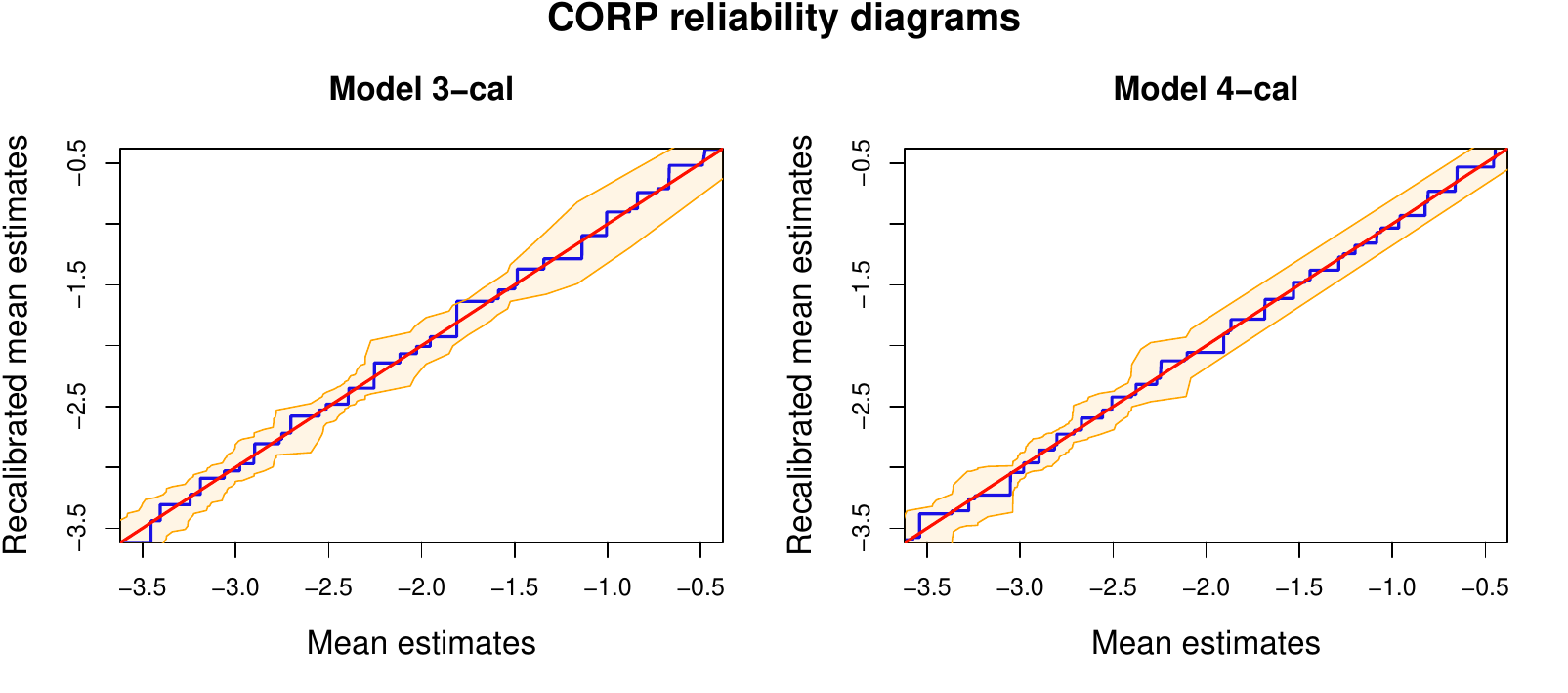}
    \includegraphics[width=1.03\textwidth]{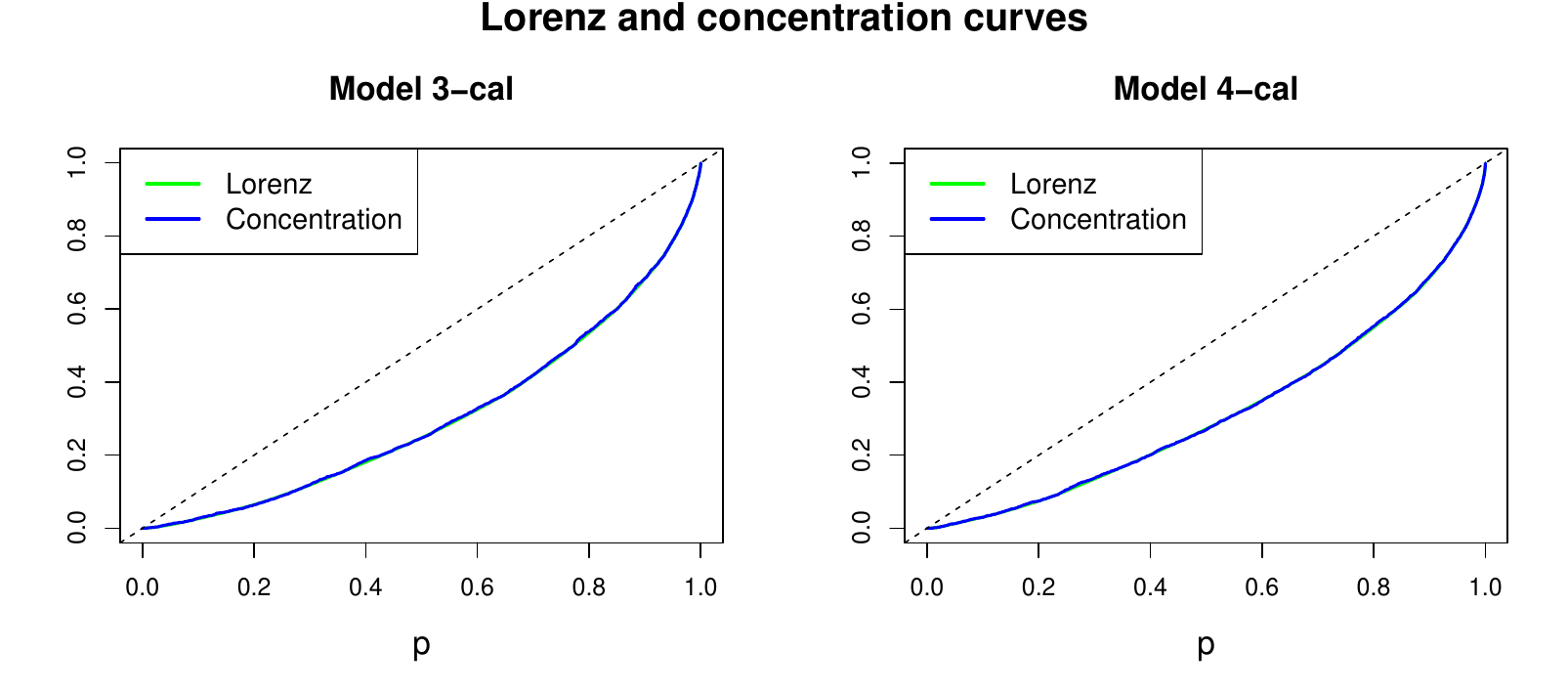}
    \caption{Top: CORP reliability diagrams for the predictions, together with 95\% point-wise consistency bands; the graphs are on the log-scale. Bottom: Lorenz and concentration curves for the predictions from Models 3-cal and 4-cal.}
    \label{fig_corp_2}
\end{figure}

\begin{figure}[htbp]
    \centering
    \includegraphics[width=1.03\textwidth]{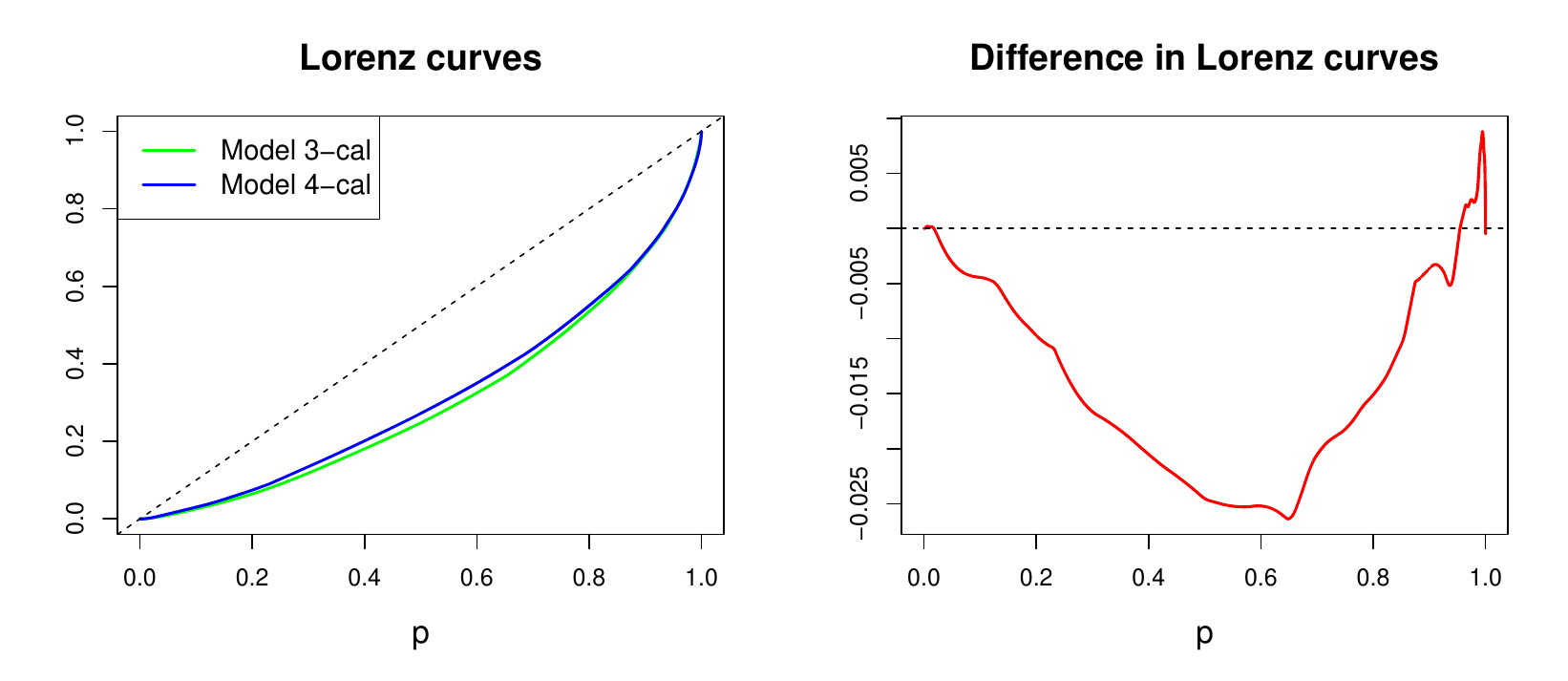}
    \includegraphics[width=1.03\textwidth]{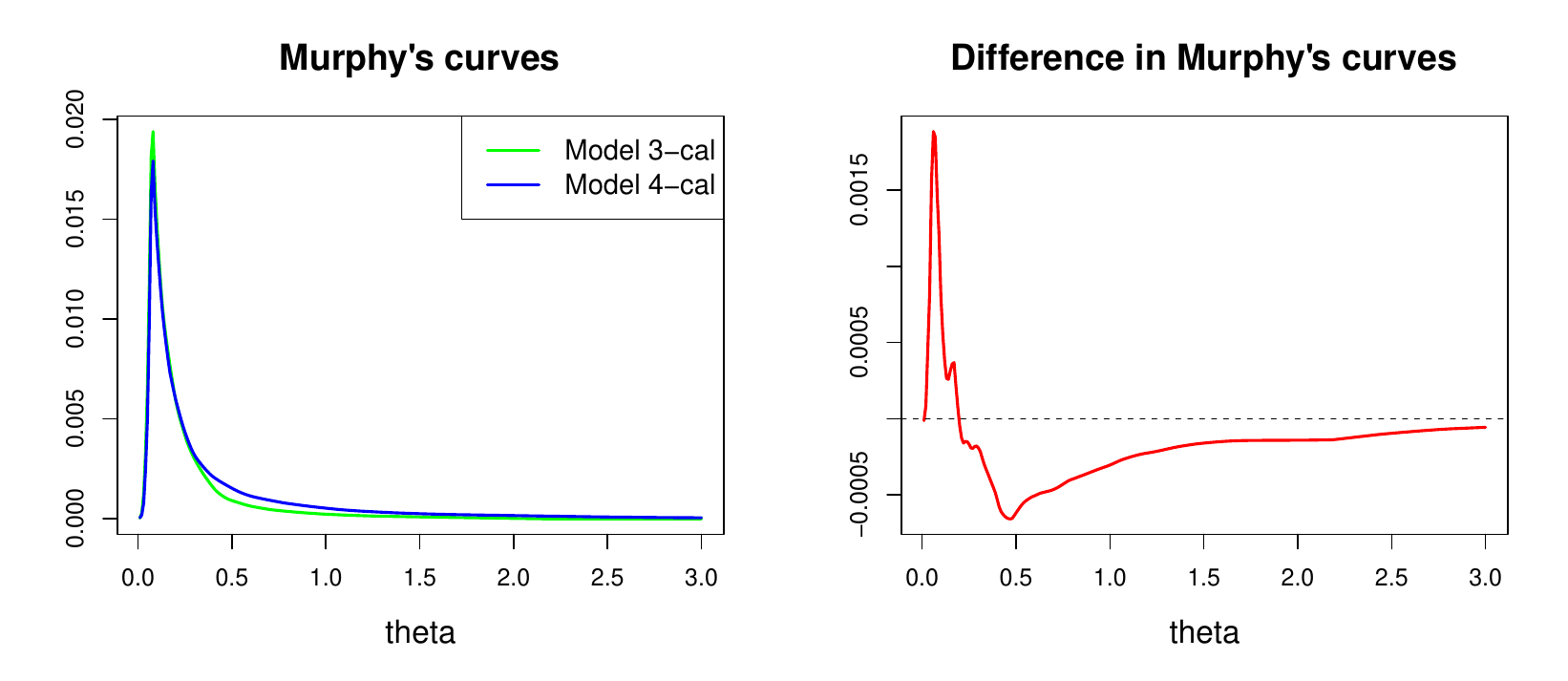}
    \caption{Top: Lorenz and concentration curves for the predictions. Bottom: Murphy's curves measuring the discriminatory power of the predictions from Models 3-cal and 4-cal.}
    \label{fig_murphy_2}
\end{figure}

We focus on the recalibrated Models 3 and 4, denoted by Models 3-cal and 4-cal. Recall that the ranking induced by a Bregman divergence coincides with the ranking induced by the discrimination component in Murphy's decomposition of the same divergence for mean-calibrated predictors. Figure \ref{fig_corp_2} confirms that Model 3-cal and Model 4-cal are indeed calibrated. The discrimination statistics from Murphy's decomposition are therefore estimated using ${\rm DSC}_L(Y,X)=S_L(X,\be[Y])$, assuming that the predictors are calibrated, i.e., $\be[Y | X]=X$. This implies that we assume that $S_L(Y,X)={\rm UNC}_L(Y,X)-{\rm DSC}_L(Y,X)$, even though the empirical estimate of $S_L(Y,X)$ has a non-zero component ${\rm MCB}_L(Y,X)$, since the models are not perfectly in-sample calibrated. We note that the MSE and the variance reported in Table \ref{tab_discrimination} are not identical. The discrimination measure is estimated as $n^{-1}\sum_{i=1}^n (X^{\text{cal}}_i-\overline{X^{\text{cal}}_i})^2 w_i$, whereas the variance is estimated as $\big(\sum_{i=1}^n w_i\big)^{-1}\sum_{i=1}^n (X^{\text{cal}}_i-\overline{X^{\text{cal}}_i})^2 w_i$, as is commonly done in practical applications. 

\begin{table}[ht]
\small{
\centering
\begin{tabularx}{\textwidth}{>{\centering\arraybackslash}X 
                              >{\centering\arraybackslash}X 
                              >{\centering\arraybackslash}X 
                              >{\centering\arraybackslash}X 
                              >{\centering\arraybackslash}X}
  \toprule
  Model & Poisson DSC & MSE DSC & Gini & Variance \\ 
  \midrule
  Model 1 & 121.9159 & 14.0719 & 0.2778 & 26.6280 \\ 
  Model 2-cal & 22.2522 & 2.3399 & 0.1001 & 4.4278 \\ 
  Model 3-cal & 236.9139 & 30.7684 & 0.3918 & 58.2222 \\ 
  Model 4-cal & 225.7111 & 36.6560 & 0.3636 & 69.3631 \\ 
  \bottomrule
\end{tabularx}
\caption{Discrimination measures DSC from Murphy's decomposition of the Poisson deviance loss and the mean squared loss ($10^{-4}$), the Gini index, and the variance for the predictions from Models 1, 2-cal--4-cal.}
\label{tab_discrimination}
}
\end{table}

\begin{figure}[htb!]
    \centering
    \includegraphics[width=0.6\textwidth]{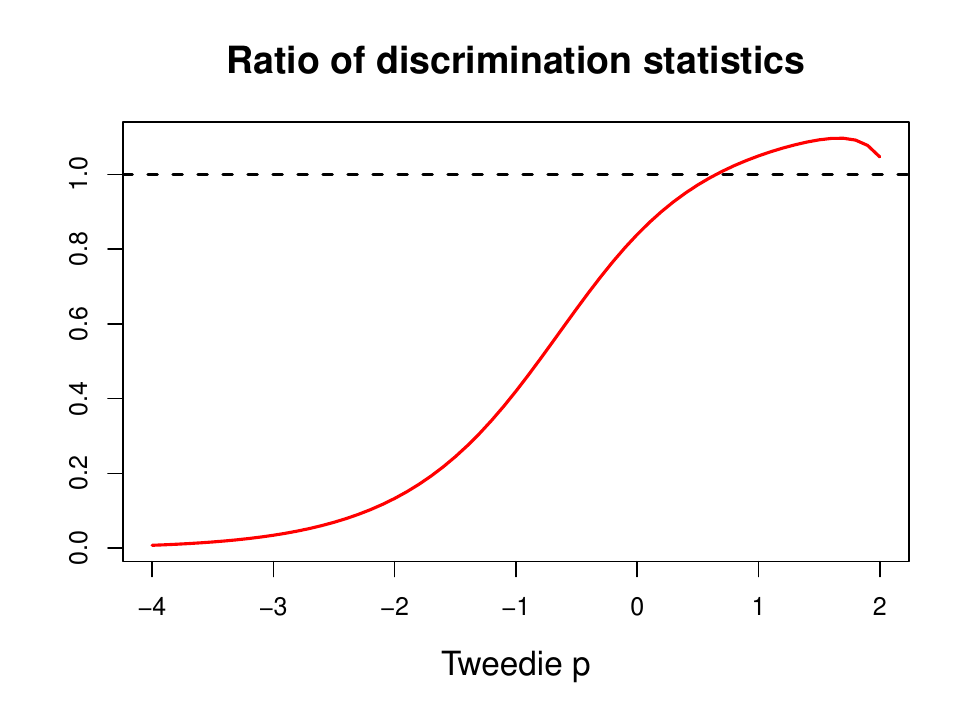}
    \caption{The ratio ${\rm DSC}_{L_{\phi_p}}(Y,X_{3\text{-cal}})/{\rm DSC}_{L_{\phi_p}}(Y,X_{4\text{-cal}})$ of the discrimination statistics for the predictions from Models 3-cal and 4-cal measured with Tweedie's deviance loss function with parameter $p$.}
    \label{fig_dsc}
\end{figure}

In Figure \ref{fig_murphy_2}, we observe that the Lorenz curve of Model 4-cal crosses the Lorenz curve of Model 3-cal once from above. Moreover, Table \ref{tab_discrimination} shows that the Gini index for Model 3-cal is greater than that for Model 4-cal, while the variance of the predictions from Model 4-cal exceeds that of Model 3-cal. As in \hyperref[ex_5]{Example 5}, we observe that the Gini index yields a different ranking of Models 3-cal and 4-cal compared to the variance-based ordering. The setting satisfies condition \eqref{2lorenz_dominance_2} of Theorem \ref{prop_gini_dominance_2nd}. Hence, as in \hyperref[ex_7]{Example 7}, in Figure \ref{fig_dsc} we display the discrimination measures from Murphy's decomposition of the Tweedie's deviance loss functions \eqref{tweedie_loss} with parameters $p<2$ and illustrate that Model 4-cal dominates Model 3-cal in the sense of Bregman dominance under the Tweedie's deviance loss functions with $p \leq 0$, i.e., we observe ${\rm DSC}_{L_{\phi_p}}(Y,X_{4\text{-cal}}) > {\rm DSC}_{L_{\phi_p}}(Y,X_{3\text{-cal}})$ for all $p \leq 0$, equivalently $S_{L_{\phi_p}}(Y,X_{4\text{-cal}}) < S_{L_{\phi_p}}(Y,X_{3\text{-cal}})$ for all $p \leq 0$. In particular, Table \ref{tab_discrimination} shows that the discrimination measure based on Murphy's decomposition of the mean squared error loss is larger for Model 4-cal than for Model 3-cal, i.e., ${\rm DSC}_{L_{\phi_p}}(Y,X_{4\text{-cal}}) > {\rm DSC}_{L_{\phi_p}}(Y,X_{3\text{-cal}})$ for $p=0$. However, the discrimination measure based on Murphy's decomposition of the Poisson deviance loss is smaller for Model 4-cal than for Model 3-cal, i.e., ${\rm DSC}_{L_{\phi_p}}(Y,X_{4\text{-cal}}) < {\rm DSC}_{L_{\phi_p}}(Y,X_{3\text{-cal}})$ for $p=1$. The explanation of this phenomenon is provided later. At this stage, the result regarding Tweedie's forecast dominance for $p\leq 0$ can be justified by Theorem \ref{prop_gini_dominance_2nd}. We remark that we only consider Tweedie parameters $p<2$ in Figure \ref{fig_dsc}, as the Tweedie's deviance loss $L_{\phi_p}(Y,X)$ is infinite for $Y=0$ if $p \geq 2$, even though the DSC empirically estimated using the formula ${\rm DSC}_{L{\phi_p}}(Y,X)=S_{L_{\phi_p}}(X,\be[Y])$ is finite for all $p\in\br$.

In line with our interpretations, since the Lorenz curve of Model 4-cal crosses the Lorenz curve of Model 3-cal once from above, we may conclude that low predictions from Model 4-cal have lower discriminatory power than low predictions from Model 3-cal, whereas large predictions from Model 4-cal have higher discriminatory power than large predictions from Model 3-cal. What is crucial for the final ranking of the predictors based on their discrimination is whether low or large predictions are more important for a forecaster. By Lemma \ref{lemma_2lorenz}, Model 3-cal second-degree upward Lorenz dominates Model 4-cal; hence, Model 3-cal is preferred over Model 4-cal. This ranking is implied by the Gini indices and the Lorenz curves of the predictors. However, we have demonstrated that the Gini index does not fit the decision-theoretic approach to the evaluation of point predictions. Using Theorem \ref{thm_main_3}, in Table \ref{tab_discrimination_2} we decompose the Gini indices of the predictors into two terms: the mean absolute deviation of the predictor and the discrimination statistics from Murphy's decomposition measured using the expected elementary Bregman loss integrated with respect to the distribution of the predictor. Apart from the fact that ${\rm DSC}_{L_{F_X}}$ depends on $F_X$ and cannot be compared across predictors, we note that the dominant term of the Gini index is the MAD in all cases, which is not a Bregman divergence. 

\begin{table}[ht]
\small{
\centering
\begin{tabularx}{\textwidth}{>{\centering\arraybackslash}X 
                              >{\centering\arraybackslash}X 
                              >{\centering\arraybackslash}X 
                              >{\centering\arraybackslash}X}
  \toprule
  Model & Gini & $\frac{1}{2}\be\big[|X-\be[Y]|\big]$ & ${\rm DSC}_{L_{F_X}}$ \\ 
  \midrule
  Model 1 & 0.2778 & 0.0148 & 0.0059\\ 
  Model 2-cal & 0.1001 & 0.0049 & 0.0026\\ 
  Model 3-cal & 0.3918 & 0.0212 & 0.0086\\ 
  Model 4-cal & 0.3636 & 0.0195 & 0.0082\\ 
  \bottomrule
\end{tabularx}
\caption{Decomposition \eqref{gini_bregman_representation} of the Gini index for the predictions from Models 1, 2-cal--4-cal.}
\label[table]{tab_discrimination_2}
}
\end{table}

We now turn to our new concept of Murphy's second-degree dominance. This dominance concept is based on the discrimination statistics from Murphy's decomposition and Murphy's curves. In Figure \ref{fig_murphy_2}, we present Murphy's curves which focus on the discriminatory power of the predictors, as in \hyperref[ex_6]{Example 6}. Recall that the Murphy's curve used for studying discriminatory power of a mean-calibrated predictor $X$ is defined by $\theta \mapsto M_\theta(X,\be[Y])$. The Murphy's curve for Model 3-cal crosses the Murphy's curve for Model 4-cal once from above, which is consistent with Proposition \ref{prop_murphy_gini_dominance}. By Corollary \ref{cor_2murphy_score} and condition \eqref{2lorenz_dominance_2} of Theorem \ref{prop_gini_dominance_2nd}, we conclude that $X_{4\text{-cal}}$ second-degree downward Murphy dominates $X_{3\text{-cal}}$ under the Tweedie's deviance loss functions with $p \leq 0$. Hence, Model 4-cal is preferred over Model 3-cal in terms of discrimination and, as already noted, also in terms of Bregman divergences under any Tweedie's deviance loss function with $p \leq 0$.

\begin{figure}[htb!]
    \centering
    \includegraphics[width=0.7\textwidth]{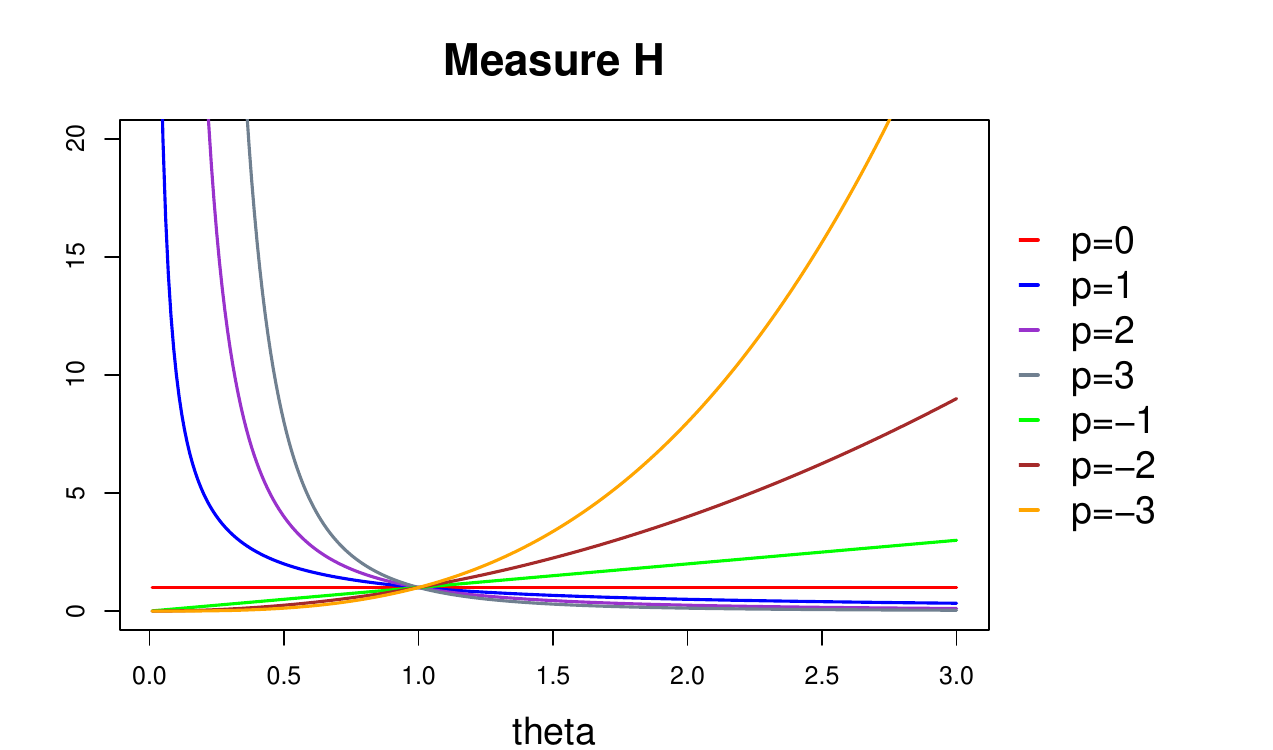}
    \caption{The measure $H(\theta)=\theta^{-p}$ arising from the Bregman representation \eqref{bregman_representation} of Tweedie's deviance loss function with parameter $p$.}
    \label{fig_h}
\end{figure}

The ranking of Model 3-cal and Model 4-cal under Tweedie's deviance loss functions with $p>0$ is more involved and goes beyond Theorem \ref{prop_gini_dominance_2nd}. In general, by Corollary \ref{cor_2murphy_score} and Proposition \ref{prop_2murphy}, we note that $X_{4\text{-cal}}$ second-degree downward Murphy dominates $X_{3\text{-cal}}$ under a Bregman divergence $L_H$ if $\theta \mapsto M_\theta(X_{3\text{-cal}},\be[Y])$ crosses $\theta \mapsto M_\theta(X_{4\text{-cal}},\be[Y])$ once from above and
\beq\label{contribution_positive_negative}
\int_{\theta_0}^\infty \Big(M_\theta(X_{3\text{-cal}},\be[Y])-M_\theta(X_{4\text{-cal}}, \be[Y])\Big) dH(\theta) \leq 0, \quad \text{for all } \theta_0 \geq 0,
\eeq
where we integrate with respect to the measure $H$ from the representation \eqref{bregman_representation} of $L_H$. We investigate the contributions of the discrepancies between the Murphy's curves, $M_\theta(X_{3\text{-cal}},\be[Y]) - M_\theta(X_{4\text{-cal}}, \be[Y])$, over $\theta \in [0,\infty)$ to the integral \eqref{contribution_positive_negative}. If we use Tweedie's deviance loss with $\phi_p$ defined in \eqref{tweedie_loss} as a Bregman divergence, then the function $H$ that weights the differences between the Murphy's curves is given by $H(\theta)=\phi_p''(\theta)=\theta^{-p}$, with the convention that $H(\theta)=1$ for $p=0$. These weighting functions are illustrated in Figure \ref{fig_h}. Tweedie's deviance loss functions with $p<0$ put more weight on the discrepancies between the Murphy's curves observed for $\theta>1$, and the larger $|p|$ is, the greater the contribution of these discrepancies to the integral \eqref{contribution_positive_negative}. In our case, the Murphy's curve of Model 4-cal lies above the Murphy's curve of Model 3-cal for $\theta>0.2$. Consequently, the differences between the Murphy's curves for $\theta>1$ are negative. These negative differences lead to a negative value of the integral \eqref{contribution_positive_negative} for any $\theta_0 \geq 0$ when we use the Tweedie's deviance loss function with $p=0$, for which $H(\theta)=1$ (the mean squared error loss). Indeed, if the mean squared error loss is used for evaluating the discriminatory power of predictors, then the discrimination measure is the variance, and the variance of the predictions from Model 4-cal exceeds that of Model 3-cal, as observed in Table \ref{tab_discrimination}. Hence, $X_{4\text{-cal}}$ second-degree downward Murphy's dominates $X_{3\text{-cal}}$ under the variance measure by Proposition \ref{prop_2murphy}. The negative differences in the Murphy's curves dominate the integral \eqref{contribution_positive_negative} for $p=0$ and become even more dominant as $|p|$ increases. This explains condition \eqref{2lorenz_dominance_2} of Theorem \ref{prop_gini_dominance_2nd}, which we apply here, and why $X_{4\text{-cal}}$ second-degree downward Murphy/Bregman dominates $X_{3\text{-cal}}$ under the Tweedie deviance loss functions for all $p \leq 0$. In contrast, Tweedie's deviance loss functions with $p>0$ put more weight on the discrepancies between the Murphy's curves observed for $\theta<1$, and the larger $p$ is, the greater the contribution of these discrepancies to the integral \eqref{contribution_positive_negative}. The integral \eqref{contribution_positive_negative} is negative for all $\theta_0 \geq 0$ if calculated with the Tweedie's deviance loss with $p=0$. As we increase $p$, more weight is placed on the positive differences between the Murphy's curves observed over the interval $\theta \in (0,0.2)$. Consequently, for sufficiently large $p>0$ (in our case, for $p>0.7$), the positive differences outweigh the negative ones, and the integral \eqref{contribution_positive_negative} becomes positive for some $\theta_0 \geq 0$. Consequently, in our example, $X_{4\text{-cal}}$ second-degree downward Murphy/Bregman dominates $X_{3\text{-cal}}$ under the Tweedie's deviance loss functions with $p \in (0,0.7]$, whereas $X_{3\text{-cal}}$ second-degree upward Murphy/Bregman dominates $X_{4\text{-cal}}$ under the Tweedie's deviance loss functions with $p \in (0.7,2)$. This offers a complete explanation of the results reported in Table \ref{tab_discrimination}.

Let us sum up the discussion of Models 3-cal and 4-cal and their ranking under Tweedie's deviance loss functions with parameter $p$. If we place more emphasis on large predictions, Model 4-cal is preferred over Model 3-cal in the sense of Bregman dominance for any $p \leq 0$. If we focus more on small predictions and choose $p \in (0,0.7)$, Model 4-cal is still preferred over Model 3-cal. If we wish to penalize prediction errors for low predictions more severely, we choose a larger $p$, in which case Model 3-cal is now preferred over Model 4-cal in the sense of Bregman dominance for $p \in [0.7,2)$. We note that in our cases approximately 95\% of predictions lie below $\theta = 0.2$, which is the point at which the Murphy's curves cross.

We point out that in the analysis above we use Tweedie's deviance loss function as the strictly consistent loss function for mean estimation. Hence, Tweedie's deviance loss function can be used with any $p \in \br$ within the framework of the decision-theoretic evaluation of point predictions. The choice of $p$ for Tweedie deviance loss should be determined by the forecaster's preferences regarding the types of prediction errors they consider most important. This choice has a different motivation than in classical statistical modelling, where Tweedie's deviance loss is typically selected to match the distribution of the response variable.

Finally, since Model 2-cal is the simplest model with only two features, the discrimination statistics presented in Table \ref{tab_discrimination} are the lowest for this model. Model 1 is a reasonable GLM with engineered features, but, as expected, it should have lower discriminatory power than modern machine learning predictive models such as regression trees. In line with our intuition, all discrimination statistics for Model 1 are larger than those for Model 2-cal, but smaller than those for Models 3-cal and 4-cal. Consequently, Models 3-cal and 4-cal are preferred over Models 1 and 2-cal.

The conclusions about the predictive power of Model 1, Models 2-cal--4-cal do not change if $S_L(Y,X)$ also contains an empirical estimate of ${\rm MCB}_L(Y,X)$.

All details concerning the estimation and the implementation of the four predictive models are available at \url{https://github.com/LukaszDelong/Miscalibration_Discrimination}.

\section{Proofs of the results}
\label{app_proofs}

\noindent\textbf{Theorem \ref{prop_murphy_new_decomp}:} 
We use the representation \eqref{bregman_representation_2} of a Bregman divergence $L$ with $\phi$. We compute
\beqo
{\rm DSC}_{L_\phi}(Y,X)&=& S_{L_\phi}(Y,\be[Y])-S_{L_\phi}(Y, \be[Y|X]) \\
&=&\be\Big[\phi(Y)-\phi(\be[Y])-(Y-\be[Y])\phi'(\be[Y])\\
&&-\phi(Y)+\phi(\be[Y|X])+(Y-\be[Y|X])\phi'(\be[Y|X])\Big]\\
&=&\be\Big[\phi(\be[Y|X])-\phi(\be[Y])-(\be[Y|X]-\be[Y])\phi'(\be[Y])\Big],
\eeqo
where we used the properties
\beqo
\be\Big[(Y-\be[Y|X])\phi'(\be[Y|X])\Big]&=&\be\Big[\be\Big[(Y-\be[Y|X])\phi'(\be[Y|X])\Big]\Big|X\Big]=0,\\
\be\Big[(Y-\be[Y])\phi'(\be[Y])\Big]&=&\be\Big[\be\Big[(Y-\be[Y])\phi'(\be[Y])\Big]\Big|X\Big].
\eeqo
\noindent The formula for ${\rm MCB}_L(Y,X)$ is derived in the same way. This completes the proof.
\qed

\

\noindent\textbf{Proposition \ref{prop_abc_sq}:}
Let $F_X(X)=U$. Since $F_X$ is assumed to be continuous on $[0,\infty)$, we know that $U\sim {\rm Unif}(0,1)$. We start as follows, using that the distribution $F_X$ is continuous and strictly increasing, we get
\beq\label{function_z}
\lefteqn{\be\Big[(Y-X)\mathbf{1}\{X\leq F_X^{-1}(p)\}\Big]=\be\Big[(Y-X)\mathbf{1}\{F_X(X)\leq p\}\Big]=\be\Big[(Y-X)\mathbf{1}\{U\leq p\}\Big]}\nonumber\\
&=&\be\Big[\be\Big[(Y-X)\mathbf{1}\{U\leq p\}\big|U\Big]\Big]=\int_0^{p}\be[(Y-X)|U=u]du=\int_0^{p}z(u) du,
\eeq
where we have introduced $z(u)=\be[(Y-X)|U=u]$. Let the apostrophe $'$ generically denotes an independent copy. Using the assumption of global unbiasedness and the independence between $Y-X$ and $Y'-X'$ given $(U,U')$, we deduce that
\beq\label{abc_sq_proof}
\be[Y]^2{\rm ABC}^2 (Y,X)&=&\int_0^1\Big(\int_0^{p}z(u)du\Big)^2dp=\int_0^1\int_0^p\int_0^pz(u)z(v) dudvdp\nonumber\\
&=&\int_0^1\int_0^1z(u)z(v)\big(1-\max\{u,v\}\big) dudv\nonumber\\
&=&\be\Big[\be\big[(Y-X)|U\big]\be\big[(Y'-X')|U'\big]\big(1-\max\{U,U'\}\big)\Big]\nonumber\\
&=&\be\Big[\be\big[(Y-X)(Y'-X')|U,U'\big]\big(1-\max\{U,U'\}\big)\Big]\nonumber\\
&=&\be\Big[\be\big[(Y-X)(Y'-X')\big(1-\max\{U,U'\}\big)\big | U,U'\big]\Big].
\eeq
The first representation follows from \eqref{abc_sq_proof} and the tower property. To prove the second representation, we notice that
\beqo
\lefteqn{\be\Big[\big(Y-X\big)\big(Y'-X'\big)\Big(1-\max\big\{F_X(X),F_X(X')\big\}\Big)\Big]}\\
&=&\be\Big[\big(Y-X\big)\be\Big[(Y'-X')\Big(1-\max\big\{F_X(X),F_X(X')\big\}\Big)\Big | X,Y\Big]\Big]=\be\Big[(Y-X)\mathcal{Q}(U)\Big].
\eeqo
In the definition of the mapping $z\mapsto\mathcal{Q}(z)$ we have used the property that $X$ and $U=F_X(X)$ generate the same $\sigma$-algebra, since $x\mapsto F_X(x)$ is strictly increasing. The third representation follows from the second and the assumption that $\be[Y]=\be[X]$. Finally, we calculate the value of $\mathcal{Q}$ at $x=0$
\beqo
\mathcal{Q}(0)=\frac{\be[(Y-X)(1-F_X(X))]}{\be[Y]}=-\frac{\be[(Y-X)F_X(X)]}{\be[Y]}=-{\rm ABC}(Y,X), 
\eeqo
and $\mathcal{Q}(1)=0$ is trivial. We prove the continuity of $z\mapsto \mathcal{Q}(z)$ on $[0,1]$ by the dominated convergence theorem. This completes the proof.
\qed

\

\noindent \textbf{Lemma \ref{lemma_abc_sq}:}
First, we define $\mathcal{H}(x)=\mathcal{Q}(F_X(x))$ for $x\in[0,\infty)$. Since the distribution $F_X$ is absolutely continuous and strictly increasing, we consider
\beqo
\mathcal{H}(x)\be[Y]&=&\be\Big[\big(Y-X\big)\Big(1-F_X(X)\Big)\mathbf{1}\{X>x\}\Big]+\Big(1-F_X(x)\Big)\be\Big[\big(Y-X\big)\mathbf{1}\{X\leq x\}\Big]\\
&=&\int_x^\infty\Big(1-F_X(z)\Big)\be\big[Y-X|X=z\big]f_X(z)dz
\\&&+~\Big(1-F_X(x)\Big)\int_0^x\be\big[Y-X|X=z\big]f_X(z)dz.
\eeqo
Taking the derivative w.r.t.~$x$ and using $F'_X(x)=f_X(x)$ for absolutely continuous $F_X$ give us
$$
\mathcal{H}'(x)\be[Y]=
-f_X(x)\int_0^x\be\big[Y-X|X=z\big]f_X(z)dz.
$$
Recalling the previous proof, for an absolutely continuous and strictly increasing distribution $F_X$ we have 
\beqo
\int_0^x\be\big[Y-X|X=z\big]f_X(z)dz&=&\int_0^x\be\big[Y-X|F_X(X)=F_X(z)\big]f_X(z)dz\\
&=&\int_0^{F_X(x)}\be\big[Y-X|U=u\big]du\\
&=&\be\Big[(Y-X)\mathbf{1}\{X\leq F_X^{-1}(F_X(x))\}\Big].
\eeqo
We deduce
\beqo
\mathcal{H}'(x)=-f_X(x)\Big(CC_{F_X(x)}(Y,X)-LC_{F_X(x)}(X)\Big),\qquad \text{for all $x\in(0,\infty)$}.
\eeqo
Next, for $\mathcal{Q}(x)=\mathcal{Q}(F_X(F_X^{-1}(x)))=\mathcal{H}(F_X^{-1}(x))$ we derive
\beqo
\mathcal{Q}'(x)=\frac{\mathcal{H}'\big(F_X^{-1}(x)\big)}{f_X\big(F_X^{-1}(x)\big)}=LC_{x}(X)-CC_{x}(Y,X), \qquad \text{for all $x\in(0,1)$}.
\eeqo
By representation \ref{function_z} for Lorenz curve and concentration curve, we deduce that $x\mapsto LC_x(X)$ and $x\mapsto CC_x(X)$ are continuous on $[0,1]$. Since $x\mapsto \mathcal{Q}(x)$ is continuous on $[0,1]$, and $\lim_{x\rightarrow 0^+}\mathcal{Q}'(x)=\lim_{x\rightarrow 1^-}\mathcal{Q}'(x)=0$, we establish the derivative $\mathcal{Q}'(x)$ on $[0,1]$.
\qed

\

\noindent \textbf{Proposition \ref{prop_abc_zero}:}
To prove (iii), we notice that ${\rm ABC}^2(Y,X)=0$ implies that $CC_p(Y,X)-LC_p(X)=0$ for a.e. $p\in[0,1]$. Recalling \eqref{function_z} for $CC_p(Y,X)-LC_p(X)$, we deduce that $\int_0^{p}z(u)du=0$ for a.e. $p\in[0,1]$. Hence, $z(u)=0$ for a.e. $u\in[0,1]$. As a result, $\be[Y-X|F_X(X)]=0$. Since we assume that $F_X$ is strictly increasing, we conclude $\be[Y|X]=X$. 
\qed

\

\noindent \textbf{Theorem \ref{main_thm_2}:}
We calculate the expected loss resulting from the mismatch between $\be[Y|X]$ and $X$ measured with the elementary loss function. By assumption (A) and formula \eqref{elementary Bregman loss} we get
\beqo
\lefteqn{M_\theta\big(\be[Y|X],X\big)=\be\big[L_\theta\big(\be[Y|X],X\big)\big]}\\
&=&\be\Big[\big(\be[Y|X]-\theta\big)^+-\mathbf{1}\{X>\theta\}\big(X-\theta\big)-\mathbf{1}\{X>\theta\}\big(\be[Y|X]-X\big)\Big]\\
&=&\be\big[\big(\be[Y|X]-\theta\big)^+\big]+\theta \mathbb{P}(X>\theta)-{\rm Cov}\big(\be[Y|X],\mathbf{1}\{X>\theta\}\big)-\be[X]\be[\mathbf{1}\{X>\theta\}].
\eeqo
We integrate each element above with respect to $H$. Recall that $H(0)=0$. For the first term we derive
\beqo
\int_0^\infty \be\big[\big(\be[Y|X]-\theta\big)^+\big]dH(\theta)&=&\int_0^\infty \int_0^\infty \big(\be[Y|X=x]-\theta\big)^+dF_X(x)dH(\theta)\\
&=&\int_0^\infty \int_0^\infty \big(\be[Y|X=x]-\theta\big)^+dH(\theta)dF_X(x)\\
&=&\int_0^\infty \int_0^{\be[Y|X=x]} \big(\be[Y|X=x]-\theta\big)dH(\theta)dF_X(x)\\
&=&\int_0^\infty \int_0^{\be[Y|X=x]} H(\theta)d\theta dF_X(x)\\
&=&\be\Big[\int_0^{\be[Y|X]} H(\theta)d\theta\Big],
\eeqo
and for the second term we have 
\beqo
\int_0^\infty\theta \mathbb{P}(X>\theta)dH(\theta)&=&\int_0^\infty\int_0^\infty\theta\mathbf{1}\{x>\theta\} dF_X(x)dH(\theta)\\
&=&\int_0^\infty\int_0^\infty \mathbf{1}\{\theta<x\}\theta dH(\theta) dF_X(x)
\\&=&\int_0^\infty\int_0^x \theta dH(\theta)dF_X(x)\\
&=&\be\Big[\int_0^X \theta dH(\theta)\Big].
\eeqo
For the third term we have 
\beqo
\int_0^\infty {\rm Cov}\big(\be[Y|X],\mathbf{1}\{X>\theta\}\big)dH(\theta)&=&{\rm Cov}\big(\be[Y|X],H(X)\big),
\eeqo
and for the last term we prove
\beqo
\int_0^\infty\be[\mathbf{1}\{X>\theta\}]dH(\theta)&=&\int_0^\infty\int_0^\infty \mathbf{1}\{x>\theta\} dF_X(x))dH(\theta)\\
&=&\int_0^\infty\int_0^\infty \mathbf{1}\{\theta<x\} dH(\theta) dF_X(x),\\
&=&\int_0^\infty H(x)dF_X(x)=\be[H(X)].
\eeqo
The first equality in \eqref{mcb_decomposition} follows from \eqref{murphy_sciore_3}. Collecting and re-arranging the terms, we show that
\beqo
\lefteqn{\int_0^\infty M_\theta\big(\be[Y|X],X\big)dH(\theta)}\\
&=&\be\Big[\int_0^{\be[Y|X]} H(\theta)d\theta\Big]+\be\Big[\int_0^X \theta dH(\theta)\Big]-{\rm Cov}\big(\be[Y|X],H(X)\big)-\be[X]\be[H(X)]\\
&=&\be\Big[\int_0^{\be[Y|X]} H(\theta)d\theta\Big]+\be\Big[\int_0^X \theta dH(\theta)\Big]+{\rm Cov}\big(X-\be[Y|X],H(X)\big)-\be[XH(X)]\\
&=&\be\Big[\int_0^{\be[Y|X]} H(\theta)d\theta\Big]+\be\Big[\int_0^X (\theta-X) dH(\theta)\Big]+{\rm Cov}\big(X-\be[Y|X],H(X)\big)\\
&=&\be\Big[\int_0^{\be[Y|X]} H(\theta)d\theta\Big]-\be\Big[\int_0^X H(\theta)d\theta\Big]+{\rm Cov}\big(X-\be[Y|X],H(X)\big).
\eeqo
To prove our representation, we just notice that
\beqo
\be\Big[\int_0^X H(\theta)d\theta\Big]&=&\int_0^\infty\int_0^\infty \mathbf{1}\{\theta<x\} H(\theta)d\theta dF_X(x)\\
&=&\int_0^\infty\int_0^\infty \mathbf{1}\{\theta<x\} dF_X(x)H(\theta)d\theta=\int_0^\infty(1-F_X(\theta)) H(\theta)d\theta.
\eeqo
This completes the proof.
\qed

\

\noindent \textbf{Corollary \ref{main_cor_2}:}
The results follow from Proposition \ref{prop_abc_sq}, Theorem \ref{main_thm_2}, equations \eqref{ABC_measure} and \eqref{ABC_measure_sq}. If the mapping $x\mapsto \mathcal{Q}(x)$ is non-decreasing on $[0,1]$, then ${\rm ABC}(Y,X)\geq 0$. Hence, $H(0)=\mathcal{Q}(0)+{\rm ABC}(Y,X)=0$ and $\theta\mapsto H(\theta)$ is non-decreasing on $[0,\infty)$. If the mapping $x\mapsto \mathcal{Q}(x)$ is non-increasing on $[0,1]$, then ${\rm ABC}(Y,X)\leq 0$. Hence, $H(0)=-\mathcal{Q}(0)-{\rm ABC}(Y,X)= 0$ and $\theta\mapsto H(\theta)$ is non-decreasing on $[0,\infty)$.
\qed

\

\noindent \textbf{Theorem \ref{thm_main_3}:}
Since $X\sim H$, from equation \eqref{score_H} we deduce that
\beqo
S_{L_H}(Y,X)&=&-\be\big[\big(X-X'\big)^+\big]+\be\big[\big(Y-X'\big)^+\big]\nonumber\\
&=&-\be[Y]{\rm Gini}(X)+\be\big[\big(Y-X'\big)^+\big],
\eeqo
where $X'\sim X$ is an independent copy of $X$, and we use representation \eqref{gini} of the Gini index. Next, we observe that
\beqo
\int_0^\infty\big(\be[Y]-\theta\big)^+dF_X(\theta)=\be\big[\big(\be[Y]-X\big)^+\big]=\frac{1}{2}\be\big[\big|X-\be[Y]\big|\big],
\eeqo
and we conclude from \eqref{discrimination_X_H} that
\beqo
{\rm DSC}_{L_H}(Y,X) = \be[Y]{\rm Gini}(X)-\frac{1}{2}\be\big[\big|X-\be[Y]\big|\big].
\eeqo
Finally, the third representation comes from equation (90) in \cite{mad}. This completes the proof.
\qed

\

\noindent \textbf{Proposition \ref{prop_murphy_gini_dominance}:}
Recall that the distributions $F_{X_1}$ and $F_{X_2}$ are assumed to be continuous and strictly increasing by assumption (C). Assertion (i) follows immediately from Propositions \ref{prop_murphy_dominance}. We prove assertion (ii). The Murphy's curve difference for two mean-calibrated predictors $X_1$ and $X_2$ with equal expected values is given by
\beqo
\Delta MC_{X_1,X_2}(\theta)=\be\left[(X_2-\theta)^+\right]- \be\left[(X_1-\theta)^+\right]=-\int_0^\theta\left(F_{X_1}(x)-F_{X_2}(x)\right)dx,
\eeqo
and the Lorenz curve difference for two mean-calibrated predictors $X_1$ and $X_2$ with equal expected values is given by
\beqo
\Delta {LC}_{X_1,X_2}(p)&=&\frac{1}{\be[Y]}\left(\be\big[X_1\mathbf{1}\{X_1\leq F_{X_1}^{-1}(p)\}\big]-\be\big[X_2\mathbf{1}\{X_2\leq F_{X_2}^{-1}(p)\}\big]\right)\\
&=&\frac{1}{\be[Y]}\left(\be\big[X_1\mathbf{1}\{F_{X_1}(X_1)\leq p\}\big]-\be\big[X_2\mathbf{1}\{F_{X_2}(X_2)\leq p\}\big]\right)\\
&=&\frac{1}{\be[Y]}\,\int_0^p\left(F_{X_1}^{-1}(x)-F_{X_2}^{-1}(x)\right)dx.
\eeqo
We are now in the framework of Lemma 1 and Theorem 2 of \cite{tryptych}, and the conclusion presented in (ii) follows. Finally, we prove assertion (iii). Assume that $p\mapsto LC_p(X_1)$ and $p\mapsto LC_p(X_2)$ cross once from above. Since $F_{X_1}$ and $F_{X_2}$ cross twice, we consider $[0,x_1]$ till the first crossing point. $F^{-1}_{X_1}(x)\geq F^{-1}_{X_2}(x)$ must hold for $x\in[0,F_{X_1}(x_1)]$ to guarantee the assumed inequality $LC_p(X_1)\geq LC_p(X_2)$ for all $p\in[0,p^*]$. The inequality for the quantiles is equivalent to 
$F_{X_1}(x)\leq F_{X_2}(x)$ for all $x\in[0,x_1]$. Hence, $\theta\mapsto M_\theta(Y,X_1)$ and $\theta\mapsto M_\theta(Y,X_2)$ cross once from above. The reverse relation is proved analogously.
\qed

\

\noindent \textbf{Proposition \ref{prop_2murphy}:}
By \eqref{m_s}, we recall the definition of the expected Bregman loss
\beqo
S_{L_H}(Y,X)=\int_0^\infty M_\theta(Y,X)dH(\theta),
\eeqo
and by Theorem \ref{prop_murphy_new_decomp}, its Murphy's decomposition for mean-calibrated predictors
\beqo
S_{L_H}(Y,X)={\rm UNC}_{L_H}(Y)-{\rm DSC}_{L_H}(Y,X),\quad M_\theta(Y,X)=M_\theta(Y,\be[Y])-M_\theta(X,\be[Y]).
\eeqo
We prove (i). ($\Longleftarrow$) The conclusion is trivial, we just set $p=\infty$. ($\Longrightarrow$) Let $\theta^*\in(0,\infty)$ denote the only crossing point of Murphy's curves. First, we can immediately deduce that
\beq
\int_0^pM_\theta(Y,X_1)dH(\theta)\geq \int_0^pM_\theta(Y,X_2)dH(\theta), \qquad \text{ for all } p\in[0,\theta^*].
\eeq
Second, we can also deduce that
\beq\label{disc_ineq}
\int_p^\infty M_\theta(Y,X_1)dH(\theta)\leq \int_p^\infty M_\theta(Y,X_2)dH(\theta), \qquad \text{ for all } p\in[\theta^*,\infty).
\eeq
At the same time, the relation between the discrimination statistics implies that
\beqo
\int_0^pM_\theta(Y,X_1)dH(\theta)+\int_p^\infty M_\theta(Y,X_1)dH(\theta)\geq \int_0^pM_\theta(Y,X_2)dH(\theta)+\int_p^\infty M_\theta(Y,X_2)dH(\theta),
\eeqo
for any $p\in(0,\infty)$. Hence, in order to have \eqref{disc_ineq} valid, we must have
\beqo
\int_0^pM_\theta(Y,X_1)dH(\theta)\geq \int_0^pM_\theta(Y,X_2)dH(\theta), \qquad \text{ for all } p\in[\theta^*,\infty).
\eeqo
The relation between $M_\theta(X_1,\be[Y])$ and $M_\theta(X_2,\be[Y])$ is proved immediately since the second component of $M_\theta(Y,X)$, which is $M_\theta(Y,\be[Y])$, does not depend on the predictor. The case (ii) is proved in the same way.
\qed

\

\noindent \textbf{Lemma \ref{lem_order_dist_quant_utility}:}
(i) We prove the limit
\beq\label{original_limit_0}
\lim_{x\rightarrow \infty} |U(x)|\big(1-F_{X}(x)\big)=0.
\eeq
Since $x\mapsto U(x)$ is piece-wise monotonic on $[0,\infty)$, the function $x\mapsto |U(x)|$ is non-decreasing or non-increasing on $[x_0,\infty)$, which is the last interval on which the function is monotonic. If $x\mapsto |U(x)|$ is non-increasing on $[x_0,\infty)$, then $\lim_{x\rightarrow \infty}|U(x)|<\infty$ must hold and the limit follows immediately. If $x\mapsto |U(x)|$ is non-decreasing on $[x_0,\infty)$, we observe that $0\leq |U(x)|(1-F(x))\leq \be[|U(X)|\mathbf{1}\{X>x\}]$ for any $x\in[x_0,\infty)$. Next, $\be[|U(X)|\mathbf{1}\{X>x\}]\rightarrow 0$ if $x\rightarrow\infty$, by the dominated convergence theorem.

\noindent We prove the limit
\beq\label{original_limit_00}
\lim_{x\rightarrow 0} |U(x)|F_{X}(x)=0.
\eeq
As above, the function $x\mapsto |U(x)|$ is non-decreasing or non-increasing on $[0,x^*]$, which is the first interval on which the function is monotonic. If $x\mapsto |U(x)|$ is non-decreasing on $[0,x^*]$, then $|U(0)|<\infty$ must hold and the limit follows immediately. If $x\mapsto |U(x)|$ is non-increasing on $[0,x^*]$, we observe that $0\leq |U(x)|F(x)\leq \be[|U(X)|\mathbf{1}\{X\leq x\}]$ for any $x\in(0,x^*]$. Next, $\be[|U(X)|\mathbf{1}\{X\leq x\}]\rightarrow 0$ if $x\rightarrow 0$, by the dominated convergence theorem and assumption (C) that $F_X(0)=0$.

\noindent (ii) We prove the limit
\beq\label{original_limit}
\lim_{x\rightarrow \infty} U'(x)\int_0^x\big(F_{X_1}(t)-F_{X_2}(t)\big)dt=0.
\eeq
If $\lim_{x\rightarrow \infty} |U'(x)|<\infty$, the limit follows since $X_1$ and $X_2$ have equal expected values and $\int_0^\infty\big(F_{X_1}(t)-F_{X_2}(t)\big)dt$ by assumption (A). We consider the case $\lim_{x\rightarrow \infty} |U'(x)|=+\infty$. The assumption that $U''(x)\geq 0$ for $x>0$ implies that $U'(x)>0$ for all $x\in[x_0,\infty)$ with sufficiently large $x_0$. Using integration by parts, we get
\beqo
\int_{x_0}^\infty U(x)dF_{X}(x)=-U(x)(1-F_X(x))\big|_{x=x_0}^{x=\infty}+\int_{x_0}^\infty U'(x)(1-F_X(x))dx.
\eeqo
Since $\int_0^\infty U(x)dF_{X}(x)<\infty$, by the assumption that $\be[|U(X)|]<\infty$, and the limits are finite, we conclude that $\int_{x_0}^\infty U'(x)(1-F_X(x))dx<\infty$. Next, we deduce that
\begin{equation}\label{prop_1}
\int_{x_0}^\infty U'(x)|F_{X_1}(x)-F_{X_2}(x)|dx\leq \int_{x_0}^\infty U'(x)(1-F_{X_1}(x))dx+\int_{x_0}^\infty U'(x)(1-F_{X_2}(x))dx<\infty.
\end{equation}
Recall that $X_1$ and $X_2$ have equal expected values, hence $\int_0^x\big(F_{X_1}(t)-F_{X_2}(t)\big)dt=-\int_x^\infty\big(F_{X_1}(t)-F_{X_2}(t)\big)dt$ for any $x\geq 0$, and instead of \eqref{original_limit} the goal is to prove the limit
\beq\label{new_limit}
\lim_{x\rightarrow \infty} U'(x)\int_x^\infty\big(F_{X_1}(t)-F_{X_2}(t)\big)dt=0.
\eeq
Since $x\mapsto U'(x)$ is non-decreasing and positive on $[x_0,\infty)$, we derive
\beq\label{prop_2}
0&\leq& \Big|U'(x)\int_x^\infty(F_{X_1}(t)-F_{X_2}(t))dt\Big| \leq U'(x)\int_x^\infty|F_{X_1}(t)-F_{X_2}(t)|dt\nonumber\\
&\leq& \int_x^\infty U'(t)|F_{X_1}(t)-F_{X_2}(t)|dt, \quad \text{for all } x\in[x_0,\infty).
\eeq
We take the limit $x\rightarrow\infty$. By \eqref{prop_1}, the right-hand side of \eqref{prop_2} converges to zero. We get the desired limit \eqref{new_limit}, and \eqref{original_limit}.

\noindent We prove the limit
\beq\label{original_limit_00}
\lim_{x\rightarrow 0} U'(x)\int_0^x\big(F_{X_1}(t)-F_{X_2}(t)\big)dt=0.
\eeq
If $\lim_{x\rightarrow 0} |U'(x)|<\infty$, the limit follows. We consider the case $\lim_{x\rightarrow 0} |U'(x)|=+\infty$. The assumption that $U''(x)\geq 0$ for $x>0$ implies that $U'(x)<0$ for all $x\in(0,x^*]$ with sufficiently small $x^*$. Let us switch to $\tilde{U}(x)=-U(x)$. Using integration by parts, we get
\beqo
\int_0^{x^*} \tilde{U}(x)dF_{X}(x)=\tilde{U}(x)F_X(x)\big|_{x=0}^{x=x^*}-\int_0^{x^*}\tilde{U}'(x)F_X(x)dx.
\eeqo
Since $\int_0^\infty \tilde{U}(x)dF_{X}(x)<\infty$, by the assumption that $\be[|U(X)|]<\infty$, and the limits are finite, we conclude that $\int_0^{x^*} \tilde{U}'(x)F_X(x)dx<\infty$. Next, we deduce that
\beq\label{prop_1_00}
\int_0^{x^*} \tilde{U}'(x)|F_{X_1}(x)-F_{X_2}(x)|dx\leq \int_0^{x^*} \tilde{U}'(x)F_{X_1}(x)dx+\int_0^{x^*}\tilde{U}'(x)F_{X_2}(x)dx<\infty.
\eeq
Since $x\mapsto \tilde{U}'(x)$ is non-increasing and positive on $(0,x^*]$, we derive
\beq\label{prop_2_00}
0 &\leq& \Big|\tilde{U}'(x)\int_0^x(F_{X_1}(t)-F_{X_2}(t))dt\Big|\leq \tilde{U}'(x)\int_0^x|F_{X_1}(t)-F_{X_2}(t)|dt\nonumber\\
&\leq& \int_0^x \tilde{U}'(t)|F_{X_1}(t)-F_{X_2}(t)|dt, \quad \text{for all } x\in(0,x^*].
\eeq
We take the limit $x\rightarrow 0$. By \eqref{prop_1_00}, the right-hand side of \eqref{prop_2_00} converges to zero and we get the desired limit \eqref{original_limit_00}. Since $X_1$ and $X_2$ have equal expected values, we also conclude that
\beq\label{new_limit_00}
\lim_{x\rightarrow 0} U'(x)\int_x^\infty\big(F_{X_1}(t)-F_{X_2}(t)\big)dt=0.
\eeq

\noindent (iii) We prove the limit
\beq\label{original_limit_000}
\lim_{x\rightarrow \infty} U''(x)\int_x^\infty \int_v^\infty (F_{X_2}(t)-F_{X_1}(t))dtdv=0,
\eeq
for $U\in\mathcal{V}$. Using integration by parts, and (i)-(ii), we derive
\beqo
\int_0^\infty U(x)dF_{X}(x) = \int_0^\infty U'(x)(1-F_{X}(x))dx = \int_0^\infty U''(x)\int_x^\infty(1-F_{X}(t))dtdx.
\eeqo
Since $\int_0^\infty U(x)dF_{X}(x)<\infty$, we conclude that $\int_0^\infty U''(x)\int_x^\infty(1-F_{X}(t))dtdx<\infty$. Next, we deduce that
\beq\label{prop_1_000}
\int_0^{\infty} U''(x)\int_x^\infty|F_{X_1}(t)-F_{X_2}(t)|dtdx&\leq& \int_0^{\infty} U''(x)\int_x^\infty(1-F_{X_1}(t))dtdx\nonumber\\
&&+\int_0^{\infty}U''(x)\int_x^\infty (1-F_{X_2}(t))dtdx<\infty.
\eeq
Since $x\mapsto U''(x)$ is non-decreasing and non-negative on $(0,\infty)$, we derive
\beqo\label{prop_2_000}
0 &\leq& \Big|U''(x)\int_x^\infty\int_v^\infty (F_{X_1}(t)-F_{X_2}(t))dtdv\Big| \\
&\leq& U''(x)\int_x^\infty\int_v^\infty |F_{X_1}(t)-F_{X_2}(t)|dtdv\leq \int_x^\infty U''(v)\int_v^\infty |F_{X_1}(t)-F_{X_2}(t)|dtdv, \ \text{for all } x>0.
\eeqo
We take the limit $x\rightarrow \infty$. By \eqref{prop_1_000}, the right-hand side of \eqref{prop_2_000} converges to zero and we get the desired limit \eqref{original_limit_000}.

\noindent (iv) We prove \eqref{order_dist_quant_up_utility}. Let $U\in\mathcal{U}$. We follow the arguments from \cite[Theorem 3.3]{dominance_book}. $(\Longrightarrow)$ Using integration by parts, we derive
\beqo
\lefteqn{\be[U(X_1)]-\be[U(X_2)]=\int_0^\infty U(x)d\big(F_{X_1}(x)-F_{X_2}(x)\big)}\\
&=&-U(x)(F_{X_2}(x)-F_{X_1}(x))\Big|_{x=0}^{x=\infty}+\int_0^\infty U'(x)(F_{X_2}(x)-F_{X_1}(x))dx\\
&=& U'(x)\int_0^x(F_{X_2}(t)-F_{X_1}(t))dt\Big|_{x=0}^{x=\infty}-\int_0^\infty U''(x)\int_0^x(F_{X_2}(t)-F_{X_1}(t))dtdx\\
&=& -U''(x)\int_0^x\int_0^v(F_{X_2}(t)-F_{X_1}(t))dtdv\Big|_{x=0}^{x=\infty}+\int_0^\infty U'''(x)\int_0^x\int_0^v(F_{X_2}(t)-F_{X_1}(t))dtdvdx\\
&=&-U''(\infty)\int_0^\infty\int_0^v(F_{X_2}(t)-F_{X_1}(t))dtdv+\int_0^\infty U'''(x)\int_0^x\int_0^v(F_{X_2}(t)-F_{X_1}(t))dtdvdx.
\eeqo
The first two limits at $x=0$ and $x=\infty$ vanish by (i)-(ii). $|U''(\infty)|<\infty$ and $U''(\infty)\geq 0$ since $x\mapsto U''(x)$ is non-negative, continuous and non-increasing on $(0,\infty)$. Finally, by equation (10) in \cite{role_of_variance} (their result can be extended to unbounded support which we consider here), the following formula holds for $X_1$ and $X_2$ with equal expected values and finite variances
\beq\label{variance_formula}
\int_0^\infty\int_0^v(F_{X_2}(t)-F_{X_1}(t))dtdv=\frac{1}{2}\big(Var[X_2]-Var[X_1]\big)<\infty,
\eeq
and we deduce that the term is finite since $U(x)=x^2\in\mathcal{U}$. If $\int_0^u\int_0^v(F_{X_2}(t)-F_{X_1}(t))dtdv\leq 0$ for all $u\geq 0$, and $\int_0^\infty\int_0^v(F_{X_2}(t)-F_{X_1}(t))dtdv\leq 0$ by continuity, then we conclude that $\be[U(X_1)]-\be[U(X_2)]\geq 0$ for any $U\in\mathcal{U}$. 

\noindent $(\Longleftarrow)$ Assume that there exists $x_0\in(0,\infty)$ such that $\int_0^{x_0}\int_0^v(F_{X_2}(t)-F_{X_1}(t))dtdv> 0$. By continuity, we observe that $\int_{0}^{x}\int_0^v(F_{X_2}(t)-F_{X_1}(t))dtdv\geq 0$ for all $x\in(x_0-\epsilon, x_0+\epsilon)$ if $\epsilon$ is chosen sufficiently small. We next define
\beqo
U'''(x)=
\begin{cases}
  0, & \mbox{if } x\leq x_0-\epsilon \\
  f(x), & \mbox{if } x\in(x_0-\epsilon, x_0+\epsilon) \\
  0, & \mbox{if } x\geq x_0+\epsilon.
\end{cases}
\eeqo
One can choose smooth function $x\mapsto f(x)\leq 0$ such that $U\in\mathcal{U}$ and $U''(\infty)=0$. Consequently, from (iv) we derive
\beqo
\be[U(X_1)]-\be[U(X_2)]=\int_{x_0-\epsilon}^{x_0+\epsilon} U'''(x)\int_0^x\int_0^v(F_{X_2}(t)-F_{X_1}(t))dtdvdx\leq 0,
\eeqo
and we end up with a contradiction.

\noindent (v) We prove \eqref{order_dist_quant_downn_utility}. Let $U\in\mathcal{V}$. We derive
\beqo
\lefteqn{\be[U(X_1)]-\be[U(X_2)]=\int_0^\infty U(x)d\big(F_{X_1}(x)-F_{X_2}(x)\big)}\\
&=&-U(x)(F_{X_2}(x)-F_{X_1}(x))\Big|_{x=0}^{x=\infty}+\int_0^\infty U'(x)(F_{X_2}(x)-F_{X_1}(x))dx\\
&=& -U'(x)\int_x^\infty(F_{X_2}(t)-F_{X_1}(t))dt\Big|_{x=0}^{x=\infty}+\int_0^\infty U''(x)\int_x^\infty(F_{X_2}(t)-F_{X_1}(t))dtdx\\
&=& -U''(x)\int_x^\infty\int_v^\infty(F_{X_2}(t)-F_{X_1}(t))dtdv\Big|_{x=0}^{x=\infty}+\int_0^\infty U'''(x)\int_x^\infty\int_v^\infty(F_{X_2}(t)-F_{X_1}(t))dtdvdx\\
&=&U''(0)\int_0^\infty\int_v^\infty(F_{X_2}(t)-F_{X_1}(t))dtdv+\int_0^\infty U'''(x)\int_x^\infty\int_v^\infty(F_{X_2}(t)-F_{X_1}(t))dtdvdx,
\eeqo
and we proceed as in (iv). By (iii) the third limit at $x=\infty$ vanishes.
\qed

\

\noindent \textbf{Theorem \ref{prop_gini_dominance_2nd}:}
(i) We prove \eqref{2lorenz_dominance}. We observe that the results of Lemma 4 and Lemma 9 from \cite{role_of_variance} do not depend on the authors' assumption that the distributions are supported on $[0,1]$ and they can be immediately extended to distributions supported on $[0,\infty)$. Let $p\mapsto LC_p(X_1)$ and $p\mapsto LC_p(X_2)$ cross once from above. The function
\beqo
S(v)=\int_0^v(F_{X_2}(t)-F_{X_1}(t))dt,\quad v\in[0,\infty),
\eeqo
satisfies $S(0)=0, S(\infty)=0$, by assumption (A), $S(v)\geq 0$ for $v\in[0,v^*]$ and $S(v)\leq 0$ for $v\in[v^*,\infty)$ for some $v^*\in(0,\infty)$, by the results from \cite{role_of_variance}. By the properties of $S$ and formula \eqref{variance_formula} for difference in variances, we now deduce the equivalence relation
\beq\label{variance_property}
\begin{aligned}
\int_0^uS(v)dv &= \int_0^u\int_0^v(F_{X_2}(t)-F_{X_1}(t))dtdv\geq 0, \quad \text{ for all } u\geq 0\\ \nonumber 
&\Longleftrightarrow \quad \int_0^\infty S(v)dv = \frac{1}{2}\big(Var(X_2)-Var(X_1)\big)\geq 0.
\end{aligned}
\eeq
Under assumption (D), the expected loss w.r.t.~a Bregman divergence is given with $S_{L_\phi}(Y,X)=\be[L_{\phi}(Y,X)]=\be[\phi(Y)]-\be[\phi(X)]$. By Lemma \ref{lem_order_dist_quant_utility} and \eqref{order_dist_quant_up_utility} we have the equivalence relations
\beqo
\lefteqn{\int_0^u\int_0^v(F_{X_2}(t)-F_{X_1}(t))dtdv\geq 0, \quad \text{ for all } u\geq 0}\nonumber\\
&\Longleftrightarrow& \quad \be[\phi(X_1)]\leq \be[\phi(X_2)], \quad \text{for all $\phi\in\mathcal{U}$} \nonumber\\
&\Longleftrightarrow& \quad \be[L_{\phi}(Y,X_1)]\geq \be[L_{\phi}(Y,X_2)], \quad \text{for all $L_\phi$ with $\phi\in\mathcal{U}$}. 
\eeqo
Since $S_L(Y,X)={\rm UNC}_L(Y)-{\rm DSC}_L(Y,X)$, we immediately deduce the last relation for DSC.

\noindent (ii) We prove \eqref{2lorenz_dominance_2}. We extend the results from \cite{role_of_variance} to cover the case $Var(X_2)\leq Var(X_1)$. We introduce the function
\beqo
\tilde{S}(v)=-S(v)=\int_v^\infty(F_{X_2}(t)-F_{X_1}(t))dt,\quad v\in[0,\infty).
\eeqo
In the view of the properties of $S$, the function $\tilde{S}$ satisfies $\tilde{S}(0)=0, \tilde{S}(\infty)=0, \tilde{S}(v)\leq 0$ for $v\in[0,v^*]$ and $\tilde{S}(v)\geq 0$ for $v\in[v^*,\infty)$ for some $v^*\in(0,\infty)$. Moreover, $\int_0^\infty \tilde{S}(v)dv=-\int_0^\infty S(v)dv=-\frac{1}{2}(Var(X_2)-Var(X_1))\geq 0$ by \eqref{variance_formula}. We deduce the equivalence relation
\beq\label{variance_property_2}
\begin{aligned}
\int_u^\infty\tilde{S}(v)dv&=\int_u^\infty\int_v^\infty(F_{X_2}(t)-F_{X_1}(t))dtdv\geq 0, \quad \text{ for all } u\geq 0  \\ \nonumber
&\Longleftrightarrow \quad \int_0^\infty\tilde{S}(v)dv=\frac{1}{2}\big(Var(X_1)- Var(X_2)\big)\geq 0.
\end{aligned}
\eeq
The remaining steps are as in (i).
\qed 

\end{appendices}

\end{document}